\documentclass[12pt]{article}
\usepackage[margin=1in]{geometry}

\usepackage{amssymb}
\usepackage{amsmath}
\usepackage{amsthm}
\usepackage{mathtools}
\usepackage{bbold}
\usepackage{bbm}
\usepackage{setspace}
\usepackage{booktabs}
\usepackage[authoryear]{natbib}
\usepackage{bibunits}
\usepackage{caption}
\usepackage{subcaption,tikz}
\usetikzlibrary{decorations.pathreplacing}

\newtheorem{example}{Example}

\newtheorem{lemma}{Lemma}
\newtheorem{theorem}{Theorem}

\newtheorem{prop}{Proposition}

\newtheorem{corollary}{Corollary}

\newcommand\norm[1]{\left\lVert#1\right\rVert}

\usepackage{esint}
\usepackage{color,hyperref}
\definecolor{darkblue}{rgb}{0.0,0.0,0.6}
\hypersetup{
	colorlinks=true, 
	linktoc=all,     
	linkcolor=darkblue,  
	citecolor=darkblue,
	urlcolor=darkblue,
	hypertexnames=false,
}
\makeatletter

\let\BU@natlinkstart\hyper@natlinkstart
\renewcommand{\hyper@natlinkstart}[1]{%
	\BU@natlinkstart{\@bibunitname.#1}%
}

\let\BU@natanchorstart\hyper@natanchorstart
\renewcommand{\hyper@natanchorstart}[1]{%
	\BU@natanchorstart{\@bibunitname.#1}%
}

\let\BU@natlinkbreak\hyper@natlinkbreak
\renewcommand{\hyper@natlinkbreak}[2]{%
	\BU@natlinkbreak{#1}{\@bibunitname.#2}%
}

\makeatother

\allowdisplaybreaks[1]
\usepackage{tikz}
\usepackage{graphicx}
\usetikzlibrary{arrows.meta,positioning}

\definecolor{ink}{HTML}{34383D}
\definecolor{accent}{HTML}{226E86}
\definecolor{accentfill}{HTML}{E8F1F5}

\title{Evolution of Behavioral Rules via Context-based Imitation}

\author{Enrique Urbano Arellano \and Xinyang Wang\footnote{Both Urbano (\href{mailto:enrique.urbano@itam.mx}{enrique.urbano@itam.mx}) and Wang (\href{mailto:xinyang.wang@itam.mx}{xinyang.wang@itam.mx}) are at ITAM. We are grateful to Victor Aguiar, Jaden Chen,  Mehmet Ekmekci, Fryderyk Falniowski, Ludovic Julien,  Edi Karni, Ali Khan, Jian Li, Richard McLean, Romans Pancs, Klaus Ritzberger, Ariel Rubinstein, Larry Samuelson, Karl Schlag, Tridib Sharma, Konstantin Sonin, Keisuke Teeple, Simone Tonin, Lin Tren, Carlos Urrutia, Levent \"{U}lk\"{u}, Rakesh Vohra, and Michael Zierhut, as well as seminar and conference participants at the Johns Hopkins University, University of Catania, University of Rochester, the SAET Conferences in 2022, 2023 and 2026, the 2023 Workshop of Strategic Interaction and General Equilibrium, the 2023 Central European Program in Economic Theory, the 2024 Southern Economic Association Conference, the 2023 Rethinking Economic Theory International Workshop, the 2023 Midwest Economic Theory Conference, the European Workshop on Economic Theory in 2023 and 2026 for their helpful comments.}}
\date{}

\usepackage{tgpagella}
\usepackage[T1]{fontenc}

\newcommand{\supp}{\operatorname{supp}}

\begin{document}
	\begin{bibunit}[abbrvnat]
		
\maketitle

\begin{abstract}
This paper studies the evolution of behavioral rules across multiple decision contexts. Agents imitate better-performing peers within a context and apply those rules across contexts. We show that the imitation dynamics admit a context-weighted social choice representation. This representation provides a unified approach to studying global discrete-time dynamics across all imitation protocols: global convergence to consensus occurs if and only if a Condorcet winner exists, whereas generic non-convergence occurs if and only if the essential set contains a Condorcet cycle. To illustrate how an environment can shape prevailing rules, we apply this framework to belief evolution in an insurance economy. We show that a large number of states can favor a rational-expectations consensus, whereas a large number of beliefs can make generic non-convergence likely. Imitation can also generate persistent waves of optimism and pessimism. We also discuss the robustness of ordinal imitation protocols and their dynamic implications.
\end{abstract}


\clearpage

\section{Introduction}
Decision makers often rely on behavioral rules, and the rules they adopt can shape aggregate economic outcomes. While a growing literature studies a wide range of behavioral rules, it often focuses on rationalizing their use in specific decision contexts. A major benefit of using a behavioral rule, however, is that it simplifies decision making when agents navigate across contexts that may differ in nature. Thus, understanding which rules agents adopt calls for studying complex environments with multiple decision contexts. 

Behavioral rules serve as general guiding principles for decision making. For instance, a rule might represent an agent's optimistic or pessimistic outlook, a preferred narrative or economic perspective for communication and analysis, or a practice of delegating decisions to an artificial intelligence tool.\footnote{Related examples appear in several strands of the recent literature. Biased beliefs are studied in \citet*{frick2024welfare}. The role of narratives in shaping behavior is discussed in \citet*{shiller2017narrative}, \citet*{eliaz2020model}, and \citet*{flynn2022macroeconomics}. Simplified economic models as general guides for reasoning are discussed in \citet*{gilboa2014economic} and \citet*{mailath2020learning}. Delegation of decisions to AI systems is studied in \citet*{fudenberg2025friend}.} In complex environments, a rule that outperforms another in one decision context may underperform it in another. Nevertheless, we assume that each agent applies a single rule across contexts; this rule may itself take a sophisticated form, such as a menu of context-specific strategies.

In this paper, we employ an evolutionary approach to study which behavioral rules agents adopt in complex environments. Rather than inventing new rules independently, agents learn socially by imitating more successful peers. We work with \emph{context-based imitation}: agents imitate better-performing peers within the same context and subsequently apply the adopted rule across contexts. In a complex environment, comparing outcomes across contexts can make it difficult to assess the relative performance of behavioral rules, as a peer's better performance may reflect a more favorable decision context. Learning from comparable situations is a common practice: in politics, \citet*{yeo2018forty} explains that shared Chinese origins made Singapore's economic success a relevant example for Deng Xiaoping as China opened its economy; in education, \citet*{arendale1994understanding} describes supplemental instruction programs that encourage students to learn study strategies from peers who have succeeded in comparable courses and apply these strategies in their subsequent courses.\\

\noindent {\bf The Imitation of Behavioral Rules.} The imitation of behavioral rules raises natural questions. Do agents converge to a common behavioral rule, which can be interpreted as a shared cultural trait as in \citet*{nunn2012culture}? Or does the distribution of behavioral rules remain non-convergent, sustaining fluctuations such as the waves of optimism and pessimism described by \citet{keynes1936general}? Furthermore, how do the characteristics of an environment shape the rules agents adopt? Does a more complex environment with more decision contexts destabilize aggregate behavior? Does a richer class of rules help to stabilize it?

To address these questions, we study an environment with finitely many contexts and finitely many behavioral rules. The set of contexts and their occurrence frequencies jointly describe the decision environment, and the available behavioral rules specify the class of heuristics agents may rely on.

A continuum of agents update their behavioral rules through context-based imitation in discrete time. In each period, each agent independently draws a context from a fixed distribution, applies their current rule, and realizes a performance. The agent then randomly samples a peer within the same context and, if the peer performs better, adopts the peer's rule with a probability determined by the imitation protocol. We do not restrict the imitation protocol to any specific form.\\

\noindent {\bf A Social Choice Representation and Characterizations of the Imitation Dynamics.} The driving observation of this paper is that multiple contexts turn the imitation dynamics into a context-weighted social choice problem. This representation follows from a social choice interpretation of the skew-symmetric pairwise comparison matrix governing the dynamics. Under the benchmark imitate-if-better protocol, agents imitate with probability one whenever they encounter a better-performing peer. The pairwise comparison between two rules then reduces to a voting margin across contexts: each context votes for the better-performing rule, with a voting weight given by its frequency of occurrence. Under more general imitation protocols, each context's voting weight is multiplied by the relevant imitation probability, which can be interpreted as a comparison-specific voter turnout rate.

This social choice perspective provides a unified treatment of global discrete-time imitation dynamics across all imitation protocols. First, the imitation dynamics converge from every interior initial distribution to a consensus on some rule if and only if that rule is a Condorcet winner, namely one that defeats every other rule in pairwise comparisons. In contrast, the dynamics fail to converge from almost every interior initial distribution if and only if the essential set of rules contains a Condorcet cycle. Unless pairwise comparisons tie, the absence of a Condorcet winner implies a Condorcet cycle within the essential set. To cover cases with ties, we also show that the absence of a Condorcet cycle within the essential set characterizes global convergence to a stationary distribution under sufficiently small imitation intensities.

These characterizations reveal that the imitation of behavioral rules can lead either to a cultural consensus, where all agents adopt the same rule, or to persistent fluctuations in behavior. The environment shapes which rules prevail. While obtaining specific features of prevailing rules usually requires a more complete description of the environment, we provide three further results. First, if the performance of behavioral rules satisfies a single-crossing property, the familiar median-voter theorem implies the imitation dynamics under imitate-if-better converge to a consensus on the best-performing rule for the "median" context. Second, non-convergent dynamics can also exhibit regularity: when there are three rules, every non-convergent trajectory approaches a heteroclinic cycle, with each rule's share repeatedly coming arbitrarily close to zero and one. Finally, we characterize the set of surviving rules under convergent dynamics and provide a necessary condition for surviving rules under non-convergent dynamics.\\

\noindent {\bf Belief Evolution, Once More.} To illustrate how imitation shapes prevailing behavioral rules in a specific environment, we apply our framework to the evolution of subjective beliefs in an insurance economy, revisiting the discussion of market beliefs through social imitation. Motivated by \citet*{friedman1953methodology}, classical market-selection models such as \citet*{sandroni2000markets} and \citet*{blume2006if} treat beliefs or belief-updating rules as fixed individual characteristics, with selection operating through the wealth or consumption shares of their users. In contrast, we let agents switch beliefs through imitation, so a belief can become extinct even when its former users remain active in the market.

In this application, each agent faces an idiosyncratic income state, which serves as their decision context, and relies on a subjective belief as their behavioral rule to make portfolio choices. We assume that a subjective belief performs better in a realized income state if it assigns a higher probability to that state, reflecting that agents hold more contingent claims for states they view as more likely.

We first show a tradeoff between the complexity of the decision environment and the richness of available beliefs. Suppose rational expectations is available and the remaining beliefs are drawn uniformly at random. For a fixed number of beliefs, the probability of global convergence to a rational-expectations consensus approaches one as the number of income states grows to infinity. In contrast, under some regularity conditions, for a fixed number of income states, the probability of generic non-convergence approaches one as the number of beliefs grows to infinity. Thus, a complex decision environment helps select rational expectations, whereas a rich set of beliefs can induce persistent belief fluctuations.

In addition, we show that waves of optimism and pessimism described by \citet*{keynes1936general} can emerge endogenously from belief imitation. The ranking criterion matters: when beliefs are ranked by the monotone likelihood ratio, under some regularity conditions, the imitation dynamics always converge to a consensus on the belief that assigns the highest probability to the median income state; in contrast, beliefs ranked by first-order stochastic dominance can exhibit persistent non-convergence.\\

\noindent {\bf Ordinal Imitation.} The choice of imitation protocol can substantially influence the resulting imitation dynamics. To illustrate, we discuss ordinal imitation, where imitation intensity depends on only performance rankings and is thus discontinuous in performance profiles. We start by showing that ordinal imitation must be a version of imitate-if-better, up to context-specific rescalings of imitation intensities. In addition, the ordinality of imitation can cause persistent non-convergent dynamics: we prove that the only class of protocols guaranteeing global convergence across all environments is proportional imitation, up to context-specific performance rescalings. To motivate why agents might adopt ordinal imitation, we characterize the imitate-if-better protocol by a myopic envy-minimizing criterion. Finally, we remark that ordinal imitation can be helpful for analysts: it is robust to context-specific performance rescalings and certain imitation errors. For this reason, ordinal imitation is appropriate for modeling context-based imitation, acknowledging that performance in different contexts may be measured in different units.

We conclude with additional discussion of our framework. We explain how to accommodate extended model assumptions such as stochastic performance, universal sampling, and across-context imitation, and we discuss the role that a continuum of agents and discrete time play in our analysis.\\

\noindent {\bf Related Literature.} Our treatment of a behavioral rule as a guiding principle across decision contexts is related to
\citet*{samuelson2001analogies}, who studies how agents apply  analogies across strategic interactions, and
\citet*{mengel2012learning}, who studies the joint learning of actions
and partitions of games. Under this cross-context interpretation,
\citet*{heller2016rule} and \citet*{aumann2019synthesis} provide rationales for adopting the rule that maximizes some measure of its average utility across decision contexts. 

Building on \citet*{ellison1995word} and
\citet*{schlag1998imitate}, we model imitation through
word-of-mouth communication of individual experiences with
behavioral rules. Comparisons are restricted to agents facing
the same realized context. Learning from similar but different
situations, which \citet*{schlag2012imitation} identifies as an
open question, therefore lies outside our baseline model.

Our global analysis relates long-run behavioral-rule adoption
to environmental primitives without relying on a specific
initial population distribution. 
Our perspective differs from local stability analysis:
\citet*{cubitt1998selection} characterize Lyapunov stability
and rule out asymptotically stable interior states under
imitation, while \citet{ritzberger1995evolutionary} characterize
asymptotic stability under a broader class of evolutionary
dynamics. It also differs from
analyses of stable states resisting small invasions or mutations in the evolution of preferences in \citet*{dekel2007evolution},
misspecifications in
\citet*{fudenberg2023misspecifications} and \citet*{he2025misspecified}, moral principles in \citet*{avataneo2025evolutionary}, and the study of
stochastic stability in \citet*{hwang2026pairwise}. Related to ours, \citet*{cerreia2024dynamic} study the global dynamics of opinion aggregation through non-Bayesian updating in a social network. Surveys of global evolutionary dynamics include
\citet*{alos2009imitation} and \citet*{sandholm2010}. 

As \citet*{alos2003comment} point out, the single-sample
imitation dynamics are a form of replicator dynamics for a symmetric zero-sum game. Multiple contexts preserve this skew-symmetric
structure. Earlier work connects this structure to social
choice through results on convergence in
\citet*{AkinLosert1984}, asymptotic stability in
\citet*{ritzberger1995evolutionary}, and Lyapunov stability in
\citet*{cubitt1998selection}. Section~\ref{subsec:relation-to-literature} provides a
detailed comparison.

We study belief evolution in an insurance application,
providing an imitation-based reinterpretation of the
market-selection argument in
\citet*{friedman1953methodology}. Building on \citet*{sandroni2000markets} and \citet*{blume2006if},
subsequent studies such as \citet*{yan2008natural}, \citet*{kogan2017market}, and \citet*{beker2023if} typically examine the selection of beliefs through
the survival of their holders or their long-run influence
on prices.
In our framework, beliefs are transferable behavioral
rules, so selection operates directly on belief shares as
agents abandon poorly performing beliefs and imitate
better-performing ones. This approach pursues a direction
opened by \citet*{blume1992evolution}.

In our insurance application, we derive three results showing how the richness of contexts and beliefs shapes the stability of belief dynamics. First, sufficiently many contexts make consensus on the rational-expectations belief arbitrarily likely. This result identifies a different source of stability from
\citet*{hommes2010does}, who provide conditions under which
more agent types help to enforce the rational-expectations steady state. Second, a sufficiently large set of candidate beliefs makes generic non-convergence
arbitrarily likely. This result complements \citet*{brock2005evolutionary}, who
show through examples that greater belief dispersion can destabilize belief dynamics; here, we vary the number of candidate beliefs rather than their dispersion. Third, we show that imitation can endogenously generate the ``waves of optimistic and pessimistic sentiment'' described by
\citet*{keynes1936general}. Our imitation approach differs from Bayesian learning from market outcomes in
\citet*{townsend1978market} and \citet*{feldman1987example}, inference
from others' actions in \citet*{vives1996social}, and adaptive
statistical learning in \citet*{evans1994local} and \citet*{evans1999learning}.

The imitation protocol influences the behavioral-rule dynamics. \citet*{schlag1998imitate} justifies proportional imitation
through maximal expected payoff improvement. We complement this
justification by showing that only protocols of the
proportional-imitation form always yield convergence in our
framework, and that the imitate-if-better protocol
minimizes a myopic measure of envy. Details of protocol
specification matter: \citet*{falniowski2025discrete} show that
different discrete-time evolutionary processes sharing the
same continuous-time limit can exhibit different global
dynamics. \citet*{mertikopoulos2022survival}
work with payoff-continuous protocols, whereas our analysis
also accommodates payoff-discontinuous ordinal imitation and shows that its dynamical properties can be more robust.

\section{Model}\label{sec:model}
\subsection{Environment}
We study a dynamic environment with a unit mass of agents in discrete time $t=1,2,\ldots$, described by the primitives $(\mathcal{K},\pi, \mathcal{R}, (u_k)_{k\in\mathcal{K}}, \mu_1).$ Agents face decision problems across various contexts, and use behavioral rules to make decisions. All agents are ex ante identical except that they may use different behavioral rules.

Let $\mathcal{K}=\{1,\dots,K\}$ denote the finite set of contexts, with their frequencies given by a full-support distribution $\pi\in\Delta_+(\mathcal{K})$.\footnote{Throughout the paper, for any finite set $A$, we use $\Delta_+(A)$ to denote the set of full-support probability distributions on $A$.} Contexts are the basic units through which agents observe and learn from one another: when evaluating outcomes, agents treat outcomes within the same context as comparable and those in different contexts as incomparable. In each period, every agent independently encounters a context drawn according to the stationary distribution $\pi$. For notational convenience, we write $\pi(A)=\sum_{k\in A}\pi_k$ for the objective probability of any subset of contexts $A \subseteq \mathcal{K}$.

Let $\mathcal{R}=\{R_1,\dots,R_N\}$ denote the set of behavioral rules agents apply to make decisions. Each rule is applicable across every context. For example, a rule might represent an optimistic or pessimistic attitude an agent uses to face all problems, or a theoretical approach an academic uses for all her research. The description of rules is comprehensive enough that agents use the same rule across all contexts.

For each context $k\in\mathcal{K}$, there exists a real-valued performance function $u_k$ defined over the set of rules $\mathcal{R}$. The profile $(u_k(R))_{R\in \mathcal{R}}$ represents the ranking of the outcomes of behavioral rules in context $k$. We assume that these performance functions are common to all agents, and that the performance ranking of behavioral rules within each context is deterministic.\footnote{Section \ref{sec:ordinal} explains why the specific choice of cardinal representation $u_k$, as well as the commonality assumption, does not matter for ordinal imitation (though it can matter for non-ordinal imitation), and Section~\ref{subsec:stochasticachievement} discusses stochastic within-context performances.}

At the beginning of period $1$, each agent is endowed with an initial rule. Let $\mu_1 \in \Delta_+(\mathcal{R})$ denote the initial distribution of rules across the population, where $\mu_1(i)>0$ represents the initial mass of agents using rule $R_i$. 

\subsection{Context-Based Imitation Protocols}

Agents update their behavioral rules through context-based imitation: they compare outcomes with peers only within the same context and then apply imitated rules across all contexts in their subsequent decision making. Performance across contexts is treated as incomparable for the purpose of imitation. One interpretation is that agents attribute the better performance of others in different contexts to the luck of being assigned a more favorable context rather than to better behavioral rules. Word-of-mouth communication drives imitation: agents share general behavioral heuristics rather than context-specific strategies.

An \emph{imitation protocol} is denoted by a tuple of functions $F=(F_1,\dots,F_K)$, where each $F_k$ maps $\mathbb{R}\times \mathbb{R}$ to $[0,1]$. For a given context $k \in \mathcal{K}$, the value $F_k(u,v)$ represents the probability that an agent with performance $u$ imitates the rule of a sampled peer with performance $v$. To avoid trivial discussions, we suppose $F_k$ is not identically zero for any $k$.

We say an imitation protocol $F$ is \emph{improving} if, for each $k$, $F_k(u,v)=0$ whenever $v\le u$. Under an improving protocol, an agent never switches rules after sampling a peer with weakly worse performance; in particular, sampling a peer who uses the same rule naturally results in the agent retaining her current rule. We denote the set of all improving imitation protocols by $\mathcal{F}$.

We specify two classes of imitation protocols. First, the \emph{imitate-if-better} protocol is given by $F_k(u,v)=\mathbf{1}\{v>u\}$ for all $u,v\in\mathbb{R}$ and $k\in \mathcal{K}$. Second, for any compact utility interval $U\subset \mathbb{R}$, we say $F$ exhibits \emph{proportional imitation} on $U$ if $F_k(u,v)=\sigma (v-u)^+$ holds for all $u,v\in U$ and some $\sigma>0$. Here, $x^+=\max\{x,0\}$ denotes the positive part of $x$, and $\sigma>0$ is an imitation intensity parameter chosen such that $F_k(u,v) \in [0,1]$ for all bounded performances in $U$. This boundedness condition on $U$ is necessary: without it, ensuring imitation probabilities remain bounded in $[0,1]$ globally over $\mathbb{R}$ would require $\sigma=0$, making the protocol degenerate.

All agents use the same imitation protocol $F$. The timeline of imitation for a given agent is illustrated in Figure~\ref{fig:context_imitation_timeline} and proceeds as follows. In each period $t\ge 1$, the agent enters with a behavioral rule $R$, and a context $k$ is drawn according to $\pi$, independently across agents and over time. The agent applies rule $R$ and realizes a performance $u_k(R)$. She then samples a peer from the population of agents who faced the same context $k$.\footnote{Sections \ref{subsec:universalsampling} and \ref{subsec:acrosscontextimitation} discuss  alternative sampling and comparison methods.} If the sampled peer uses rule $R'$, the agent updates her rule to $R'$ with probability $F_k(u_k(R),u_k(R'))$ and retains her rule as $R$ with probability $1-F_k(u_k(R),u_k(R'))$. The rule held after this update is carried into the next period.

\begin{figure}[t]
	\centering
	\begin{tikzpicture}[
		x=0.95cm,y=0.95cm,
		>=Latex,
		timeline/.style={thick},
		stepbox/.style={
			align=center,
			font=\footnotesize,
			inner sep=3pt,
			text width=2.1cm,
			minimum height=0.95cm
		}
		]
		
		\draw[timeline,->] (0,0) -- (11.2,0);
		
		\foreach \x in {0.8,2.6,4.4,6.2,8.0,9.8} {
			\draw[thick] (\x,0.12) -- (\x,-0.12);
		}
		
		\node[font=\footnotesize] at (0.8,-0.35) {$t$ begins};
		\node[font=\footnotesize] at (9.8,0.55) {$t+1$ begins};
		
		\node[stepbox] (s1) at (0.8,1.45) {Rule-$R$ agent enters};
		\draw[->] (0.8,0.12) -- (s1.south);
		
		\node[stepbox] (s2) at (2.6,-1.45) {Draws $k\sim\pi$\\Gets $u_k(R)$};
		\draw[->] (2.6,-0.12) -- (s2.north);
		
		\node[stepbox] (s3) at (4.4,1.45) {Samples a peer in the same context};
		\draw[->] (4.4,0.12) -- (s3.south);
		
		\node[stepbox] (s4) at (6.2,-1.45) {Observes $R'$\\and $u_k(R')$};
		\draw[->] (6.2,-0.12) -- (s4.north);
		
		\node[stepbox] (s5) at (8.0,1.45) {Updates via\\$F_k(u_k(R),u_k(R'))$};
		\draw[->] (8.0,0.12) -- (s5.south);
		
		\node[stepbox] (s6) at (9.8,-1.45) {Carries rule\\to $t+1$};
		\draw[->] (9.8,-0.12) -- (s6.north);
		
	\end{tikzpicture}
	\caption{Timeline of context-based imitation}
	\label{fig:context_imitation_timeline}
\end{figure}

\subsection{Evolution of Behavioral Rules}\label{subsec:evolution}

We study the evolution of behavioral rules induced by the imitation protocol $F$. Let $\mu_t \in \Delta(\mathcal{R})$ denote the distribution of behavioral rules at the beginning of period $t$.

Since context assignments are independent, for each context $k$ and time $t$, the distribution of rules among the group of agents assigned to context $k$ is $\mu_t$. Within this group, an $R_i$-user samples an $R_j$-user with probability $\mu_t(j)$ and, conditional on such sampling, switches from $R_i$ to $R_j$ with probability $F_k(u_k(R_i),u_k(R_j))$. Thus, the mass of agents in context $k$ who switch from $R_i$ to $R_j$ in period $t$ is $\pi_k \mu_t(i)\mu_t(j)F_k(u_k(R_i),u_k(R_j))$. Aggregating over all rules, the total mass of agents in context $k$ who switch away from $R_i$ is $\pi_k \sum_{j=1}^N \mu_t(i)\mu_t(j)F_k(u_k(R_i),u_k(R_j))$, and the total mass of agents in context $k$ who switch to $R_i$ is $\pi_k \sum_{j=1}^N \mu_t(j)\mu_t(i)F_k(u_k(R_j),u_k(R_i))$. Summing across all contexts, the evolution of the mass of $R_i$-users is therefore given by
\[
\mu_{t+1}(i) = \mu_{t}(i) + \sum_{k=1}^K \pi_k \sum_{j=1}^N \mu_t(i)\mu_t(j)\big[F_k(u_k(R_j),u_k(R_i))-F_k(u_k(R_i),u_k(R_j))\big], \qquad t\ge 1.
\]

This evolution can be represented in matrix form as a system of difference equations that is quadratic in $\mu_t$. For each imitation protocol $F$, we define an $N\times N$ \emph{pairwise comparison matrix} $W^F$, where, for each $1\le i,j\le N$, 
\begin{equation}
W_{ij}^F=\sum_{k=1}^K \pi_k \big[F_k(u_k(R_j),u_k(R_i))-F_k(u_k(R_i),u_k(R_j))\big].
\label{eqn:pairwisecomparisonmatrix}
\end{equation} 
By construction, $W^F$ is antisymmetric. Using this matrix, the evolution equation for the behavioral rule distribution $\mu_t$ can be represented by
\begin{equation}
	\mu_{t+1}(i) = \mu_t(i) + \mu_t(i)\,(W^F\mu_t)(i), \qquad i=1,\dots,N, \; t\ge 1.
	\label{eqn:evolution}
\end{equation}

We establish two immediate properties of these imitation dynamics in the following lemma. First, under any imitation protocol, the dynamics are always well-defined, and the population share of any rule never vanishes in finite time. Second, the evolution of the population ratios between any pair of behavioral rules admits a clean representation.

\begin{lemma}\label{lem:relative-growth}
	For any trajectory $(\mu_t)_{t\ge 1}$ generated by \eqref{eqn:evolution} under imitation protocol $F$, we have the following two statements.
	\begin{enumerate}
		\item For any $t\ge 1$, we have that $\mu_t\in\Delta(\mathcal{R})$ and that $\mu_t$ remains an interior distribution.
		\item For any pair of rules $R_i,R_j$ and any $t\geq 1$, we have
		\begin{equation}\label{eq:ratio}
			\frac{\mu_{t+1}(i)}{\mu_{t+1}(j)}
			=\frac{\mu_t(i)}{\mu_t(j)}
			\frac{1+(W^F\mu_t)(i)}{1+(W^F\mu_t)(j)}.
		\end{equation}
		Consequently, the population ratio $\mu_t(i)/\mu_t(j)$ increases, remains constant, or decreases depending on whether $(W^F\mu_t)(i)-(W^F\mu_t)(j)$ is positive, zero, or negative, respectively.
	\end{enumerate}
\end{lemma}

\section{A Social Choice Representation of Imitation Dynamics}\label{sec:socialchoice}

Throughout this section, we fix an imitation protocol $F\in\mathcal{F}$ and develop a social choice interpretation of the pairwise comparison matrix $W^F$. In the next section, we use this interpretation to provide a unified analysis of the imitation dynamics in \eqref{eqn:evolution}.

We say that rule $R_i$ \emph{defeats} rule $R_j$ under $F$ if $W_{ij}^F>0$, and that the two rules \emph{tie}  under $F$ if $W_{ij}^F=0$. The pairwise comparison matrix $W^F$ in \eqref{eqn:pairwisecomparisonmatrix} induces a directed weighted graph $\mathcal G(F)=(\mathcal{R},W^F)$, where the vertices are given by the set of rules $\mathcal{R}$, and there exists an edge from $R_i$ to $R_j$ with weight $W_{ij}^F$ whenever $R_i$ defeats $R_j$. Since $W^F$ is antisymmetric, $\mathcal G(F)$ is a tournament whenever no two distinct rules tie. Figure~\ref{fig:social-choice}(a) illustrates an induced graph with three rules.

\begin{figure}[t] 
	\centering
	\resizebox{\textwidth}{!}{%
		\begin{tikzpicture}[
			rule/.style={
				circle, draw=ink, fill=white, line width=.65pt,
				minimum size=7.8mm, inner sep=0pt
			},
			selected/.style={
				rule, draw=accent, fill=accentfill, line width=1pt
			},
			edge/.style={
				-{Latex[length=2mm,width=1.45mm]},
				draw=ink, line width=.7pt,
				shorten >=1pt, shorten <=1pt
			},
			strong/.style={edge, draw=accent, line width=1pt},
			weight/.style={fill=white, inner sep=1.5pt, font=\small},
			text=ink
			]
			
			\begin{scope}
				\node[rule] (a1) at (.7,2.15) {$R_1$};
				\node[rule] (a2) at (3.6,2.15) {$R_2$};
				\node[rule] (a3) at (2.15,.15) {$R_3$};
				
				\draw[edge] (a1) -- node[weight,above] {$a$} (a2);
				\draw[edge] (a2) -- node[weight,right] {$b$} (a3);
				\draw[edge] (a3) -- node[weight,left]  {$c$} (a1);
				
				\node at (2.15,-.85) {(a)};
			\end{scope}
			
			\begin{scope}[xshift=4.95cm]
				\node[selected] (b1) at (2.05,2.55) {$R_1$};
				\node[rule] (b2) at (.65,.15) {$R_2$};
				\node[rule] (b3) at (3.45,.15) {$R_3$};
				\node[rule] (b4) at (2.05,.95) {$R_4$};
				
				\draw[strong] (b1) -- (b2);
				\draw[strong] (b1) -- (b3);
				\draw[strong] (b1) -- (b4);
				\draw[edge] (b2) -- (b3);
				\draw[edge] (b3) -- (b4);
				\draw[edge] (b4) -- (b2);
				
				\node at (2.05,-.85) {(b)};
			\end{scope}
			
			\begin{scope}[xshift=10cm]
				\draw[gray,dashed,line width=.7pt,rounded corners=9pt]
				(-.05,-.45) rectangle (3.95,2.95);
				
				\node[selected] (c1) at (1.95,2.28) {$R_1$};
				\node[selected] (c2) at (.55,.15) {$R_2$};
				\node[selected] (c3) at (3.35,.15) {$R_3$};
				\node[rule] (c4) at (1.95,.92) {$R_4$};
				\node[rule] (c5) at (5.95,1.25) {$R_5$};
				
				\draw[strong] (c1) -- (c2);
				\draw[strong] (c2) -- (c3);
				\draw[strong] (c3) --
				node[weight,right] {$1/2$} (c1);
				
				\draw[edge] (c4) -- (c1);
				\draw[edge] (c2) -- (c4);
				\draw[edge] (c3) -- (c4);
				
				\draw[edge] (3.95,1.25) --
				node[midway,above=3pt,font=\small,align=center]
				{each\\defeats}
				(c5);
				
				\node at (3.1,-.85) {(c)};
			\end{scope}
		\end{tikzpicture}%
	}
	\caption{Social choice representations.
		Panel (a) illustrates a Condorcet cycle, with $W_{12}^F=a$,
		$W_{23}^F=b$, and $W_{31}^F=c$, where $a,b,c\in(0,1]$.
		In panel (b), $R_1$ is a Condorcet winner.
		In panel (c),  every unlabeled
		edge has weight $1/4$. The blue nodes form the support of the unique
		maximal lottery $p^F=(1/4,1/2,1/4,0,0)$, and the top cycle $\mathcal{S}=
		\{R_1,R_2,R_3,R_4\}$ is enclosed by the dashed box. }
	\label{fig:social-choice}
\end{figure}

This induced graph admits a social choice interpretation. When $F$ is the imitate-if-better protocol, we have $W_{ij}^{F}= \pi(\{k: u_k(R_j)<u_k(R_i)\}) - \pi(\{k: u_k(R_j)>u_k(R_i)\})$. In this case, $W_{ij}^F$ can be interpreted as the aggregate voting margin in favor of $R_i$ over $R_j$: each context serves as a voter with a weight given by its frequency $\pi_k$. For any pair of candidate rules, each context votes for the better-performing rule in the context and abstains if the two rules are equal in performance. Consequently, a rule defeats another if it outperforms the latter more frequently across contexts.

This social choice interpretation extends to any improving imitation protocol. For any $F\in\mathcal{F}$, we have $W_{ij}^F=\sum_{k\in\mathcal{K}}\pi_k \big[F_k(u_k(R_j),u_k(R_i))-F_k(u_k(R_i),u_k(R_j))\big]$. In this general setting, $W_{ij}^F$ can still be interpreted as the aggregate voting margin in favor of $R_i$ over $R_j$, with each context acting as a voter with weight $\pi_k$. However, each voter now has a turnout rate specific to the pair $(R_i, R_j)$: when comparing these rules, voter (context) $k$ casts a fraction $F_k(u_k(R_j),u_k(R_i))$ of her ballots for $R_i$, a fraction $F_k(u_k(R_i),u_k(R_j))$ of her ballots for $R_j$, and abstains with the remainder. Because $F$ is an improving protocol, $F_k(u_k(R_j),u_k(R_i))$ and $F_k(u_k(R_i),u_k(R_j))$ cannot be simultaneously strictly positive; thus, voters either vote for the better-performing rule in their own context or abstain. 

This social choice representation suggests that the context distribution determines the imitation dynamics. Although each context has a transitive performance ranking over the rules, the aggregation of these rankings depends on the context frequencies and need not be transitive.

\subsection{Social Choice Notions}

To facilitate our subsequent analysis, we introduce five notions from social choice theory and mention their basic properties. 

A \emph{Condorcet winner} is a rule $R_i$ that defeats every other rule $R_j$. Formally, $R_i$ is a Condorcet winner if $W_{ij}^F>0$ for all $j\neq i$. A Condorcet winner, if one exists, is necessarily unique. Figure~\ref{fig:social-choice}(b) illustrates a Condorcet winner.

A \emph{Condorcet cycle} is a sequence of rules $(R_{i_1},\ldots,R_{i_m})$ such that each rule defeats the subsequent one, and the final rule defeats the first. Formally, $(R_{i_1},\ldots,R_{i_m})$ is a Condorcet cycle if $W_{i_1 i_2}^F>0, \dots, W_{i_{m-1} i_m}^F>0,$ and $W_{i_m i_1}^F>0$. Since there is at most one edge between any pair of rules, a Condorcet cycle must consist of at least three rules. Moreover, when the induced graph $\mathcal G(F)$ is a tournament, the absence of a Condorcet winner implies the existence of a Condorcet cycle. Figure~\ref{fig:social-choice}(a) illustrates a Condorcet cycle.

A \emph{maximal lottery} is a probability distribution $p\in\Delta(\mathcal R)$ satisfying $p^\top W^F q \geq 0$ for every $q\in\Delta(\mathcal R)$. Here, the product $p^\top W^F q$ can be interpreted as the expected comparison margin between rules drawn independently from $p$ and $q$. Thus, intuitively, a distribution $p$ is a maximal lottery if it weakly defeats any other distribution $q$. If a Condorcet winner exists, the unique maximal lottery assigns probability one to that rule. Moreover, by the antisymmetry of $W^F$, one can show that $p\in\Delta(\mathcal R)$ is a maximal lottery if and only if $W^Fp\leq0$. \citet*{fishburn1984probabilistic} proved that a maximal lottery always exists, and \citet*{brandl2016consistent} and \citet{brandl2017distribution} showed that it is generically unique. We denote the set of maximal lotteries by $\mathcal M^F=\{p\in\Delta(\mathcal R): W^Fp\leq0\}$. Figure~\ref{fig:social-choice}(c) illustrates a graph with a unique maximal lottery.

The \emph{essential set} $E^F=\bigcup_{p\in\mathcal M^F}\operatorname{supp}(p)$ is defined as the union of the supports of all maximal lotteries. We adopt this terminology from \citet{dutta1999comparison}.\footnote{\citet*{brandt2018structure} refer to the essential set as the bipartisan set.} Intuitively, the essential set consists of rules that are used to construct maximal lotteries. In Figure~\ref{fig:social-choice}(c), the essential set is the support of the unique maximal lottery.

The \emph{top cycle} $S^F= \bigcap\big\{S\subseteq\mathcal R: S \neq \emptyset \text{ and } W_{ij}^F>0 \text{ for every } i\in S, j\notin S\big\}$ is the smallest nonempty subset of rules that defeats its complement. The top cycle is uniquely defined and always contains the essential set; see \citet*{brandt2023characterizing}. Figure~\ref{fig:social-choice}(c) demonstrates that the top cycle can strictly contain the essential set.

We note that these notions are invariant under the rescaling of imitation intensities. For any improving protocol $F\in\mathcal{F}$ and any scalar $\lambda\in(0,1)$, the scaled protocol $\lambda F$ is also an improving imitation protocol. The pairwise comparison matrix satisfies $W^{\lambda F} = \lambda W^{F}$. Thus, all social choice notions defined in this subsection remain invariant under such rescalings.

\section{Imitation Dynamics}\label{sec:evolution}

In this section, we use the social choice representation developed in the previous section to characterize the global imitation dynamics and establish their implications for the survival of rules. For ease of exposition, we defer most discussion of our contributions to Subsection~\ref{subsec:relation-to-literature}. For notational clarity, we identify rule $R_i$ with its index $i$ in equations.

\subsection{Convergence to a Consensus}\label{subsec:consensus}

When the mass of agents using a rule $R_i$ converges to one, so that $\lim_{t\to\infty}\mu_t(i)=1$, we say that the imitation dynamics converge to a \emph{consensus} on $R_i$. Our first theorem characterizes global convergence to a consensus.

\begin{theorem}\label{thm:consensus-iff-condorcet}
	The imitation dynamics converge to a consensus on $R_i$ from every initial distribution if and only if $R_i$ is a Condorcet winner.
\end{theorem}

Global convergence to a consensus requires some rule $R_i$ to defeat every other rule. When this condition holds, the weakest pairwise advantage of $R_i$ over other rules determines the rate of convergence. The following proposition describes this rate. Throughout the paper, we use the convention that $\log 0=-\infty$.

\begin{prop}\label{prop:consensus-convergencerate}
	If $R_i$ is a Condorcet winner, then $\mu_t(i)$ strictly increases to one from every initial distribution, and the rate of convergence satisfies
	\[
	\lim_{t\to\infty}\frac{1}{t}\log[1-\mu_t(i)]
	=\log\left(1-\min_{j\neq i}W^F_{ij}\right).
	\]
\end{prop}

When the imitation dynamics converge to a consensus, the convergence is exponential when $\min_{j\neq i}W^F_{ij}<1$ and is faster than exponential when $\min_{j\neq i}W^F_{ij}=1$. The latter case is extreme: every other rule imitates $R_i$ with probability one in every context, while $R_i$ never imitates any other rule.

Theorem~\ref{thm:consensus-iff-condorcet} reduces the analysis of imitation dynamics leading to a consensus to the study of Condorcet winners. This connection allows us to leverage classical results from social choice theory to analyze imitation dynamics. To illustrate, we apply the representative-voter argument for single-crossing preferences developed in \citet{gans1996majority}.

We say the performance profile $(u_k)_{k\in\mathcal{K}}$ satisfies \emph{single-crossing} if the rules and contexts can be ordered such that, for every $i<j$ and $k<k'$, $u_k(R_j) \ge u_k(R_i)$ implies $u_{k'}(R_j) \ge u_{k'}(R_i)$. When single-crossing holds, the performance ranking of any pair of rules switches at most once across contexts. The following proposition shows that, subject to regularity conditions, single-crossing ensures that the imitation dynamics globally converge to a consensus under the imitate-if-better protocol.

\begin{prop}\label{prop:singlecrossing-existence-condorcet}
	Suppose the performance profile $(u_k)_{k\in\mathcal{K}}$ satisfies single-crossing with $u_k(R_i)\neq u_k(R_j)$ for all $i\neq j$, and suppose there exists a median context $M\in\mathcal{K}$ satisfying $\sum_{k<M}\pi_k<\frac12$ and $\sum_{k>M}\pi_k<\frac12$. Under the imitate-if-better protocol, the imitation dynamics globally converge to a consensus on the rule having the highest performance in the median context $M$.
\end{prop}

\subsection{Convergence to a Stationary Distribution}\label{subsec:stationary-convergence}

The imitation dynamics may converge to a stationary distribution that is not a consensus. We call a distribution $p\in\Delta(\mathcal{R})$ \emph{stationary} if $\mu_t = p$ implies $\mu_{t+1} = p$ under \eqref{eqn:evolution}, or equivalently, if $p_i (W^F p)_i = 0$ for all $i = 1,\dots, N$. The following theorem characterizes global convergence for sufficiently small imitation intensities.\footnote{Although, for sufficiently small imitation intensities, the discrete-time dynamics approximate their continuous-time counterparts arbitrarily well on finite time horizons, this theorem does not automatically characterize long-run convergence in continuous time. }

\begin{theorem}\label{thm:weak-imitation-convergence}
	For any imitation protocol $F\in \mathcal F$, the following statements are equivalent:
	\begin{enumerate}
		\item The essential set $E^F$ contains no Condorcet cycle. Equivalently, $W^F_{E^F, E^F}=0$.\footnote{Here, $W^F_{S_1,S_2}$ denotes the submatrix of $W^F$ having rows in $S_1$ and columns in $S_2$.}
		\item There exists $\bar\lambda\in(0,1]$ such that, for every $\lambda\in(0,\bar\lambda)$, the imitation dynamics under $\lambda F$ converge from every initial distribution.
		\item The set of initial distributions leading to convergent dynamics under $F$, 
		$\big\{\mu_1\in\Delta_+(\mathcal{R}): \text{the sequence } (\mu_t)_{t\ge 1} \text{ generated by } \eqref{eqn:evolution} \text{ converges}\big\}$, 
		has strictly positive Lebesgue measure.\footnote{Here, Lebesgue measure refers to the $(N-1)$-dimensional Lebesgue measure on the simplex $\Delta(\mathcal{R})$.}
	\end{enumerate}
\end{theorem}

This theorem paves the way for characterizing generically non-convergent dynamics in Theorem~\ref{thm:generic-nonconvergence}. In general, non-convergent imitation dynamics are unavoidable: in Theorem~\ref{thm:universalconvergencecharacterization}, we show that, up to context-specific performance scalings, proportional imitation is the only protocol that guarantees convergence in all environments. 

The condition in Theorem~\ref{thm:weak-imitation-convergence} is knife-edge when there is no Condorcet winner: the essential set $E^F$ contains multiple elements, and the condition $W^F_{E^F,E^F}=0$ requires all rules in the essential set to tie pairwise. While this feature reveals the general fragility of global convergence to non-consensus stationary distributions, specific imitation protocols can induce this nongeneric structure: for instance, under proportional imitation, the pairwise comparison matrix is of rank at most two.\footnote{To see this, define the expected performance of rule $R_i$ across contexts as $\bar u_i=\sum_{k\in\mathcal K}\pi_k u_k(R_i)$. Then $W^F_{ij}=\sigma(\bar u_i-\bar u_j)$, so $W^F=\sigma(\bar u\mathbf1^\top-\mathbf1\bar u^\top)$, where $\mathbf1$ is the vector of ones.}

The following proposition characterizes the set of possible limiting distributions and the rate of convergence when the dynamics converge.

\begin{prop}\label{prop:stationary-convergence}
	For each imitation protocol $F$, let $L^F$ be the set of maximal lotteries $p$ satisfying $\supp(p)=E^F$ and $(W^F p)_j<0$ for every $j\notin E^F$. The following statements hold:
	\begin{enumerate}
		\item The set $L^F$ is nonempty and convex. 
		\item The imitation dynamics converge to $p$ from some initial distribution if and only if $p\in L^F$.
		\item Suppose the trajectory $(\mu_t)_{t\ge 1}$ converges to $p$. If $E^F=\mathcal R$, the imitation dynamics must be stationary. If $E^F\neq \mathcal R$, the trajectory $(\mu_t)_{t\ge 1}$ satisfies	
		\[
		\lim_{t\to\infty}
		\frac{1}{t}\log\|\mu_t-p\|_1
		=
		\log\left(
		1-\min_{j\notin E^F}[-(W^F p)_j]
		\right).
		\]
	\end{enumerate}
\end{prop}

As the following example illustrates, imitation intensity can affect long-run convergence even though it leaves the essential set unchanged. Thus, the small-intensity requirement in Theorem \ref{thm:weak-imitation-convergence} cannot be dropped.

\begin{example}\label{example:largeintensitybreakconvergence}
	Suppose there are five rules, and the pairwise comparison matrix is given by
	\[
	W^F
	=
	\begin{pmatrix}
		0&0&0.51&-0.49&0.01\\
		0&0&-0.49&0.51&0.01\\
		-0.51&0.49&0&0.1&-0.1\\
		0.49&-0.51&-0.1&0&0.1\\
		-0.01&-0.01&0.1&-0.1&0
	\end{pmatrix}.\footnote{$W^F$ is a pairwise comparison matrix generated by the following imitation protocol $F\in\mathcal F$: consider two contexts with $\pi_1=0.51$ and $\pi_2=0.49$, performance profile $(u_1(R_i))_{i=1}^5=(5,3,4,2,1)$, and $u_2=-u_1$. For each pair $(i,j)$ such that $W_{ij}^F>0$, set $F_k(u_k(R_j),u_k(R_i))=W_{ij}^F/\pi_k$ in the unique context $k$ where $u_k(R_i)>u_k(R_j)$, and set $F_k=0$ elsewhere.}
	\]
	
	One can show that the set of maximal lotteries consists of distributions $(\alpha,1-\alpha,0,0,0)$ for $\alpha\in [0.49,0.51]$. 
	Hence, the essential set is $E^{F}=\{R_1,R_2\}$. Since $W^F_{E^F, E^F}=0$, Theorem~\ref{thm:weak-imitation-convergence} implies that every trajectory under $\lambda F$ converges whenever $\lambda$ is sufficiently small. Nevertheless, in Online Appendix \ref{appendix:largeintensitybreakconvergence}, we show that when $\lambda=1$, there exists a nonempty, relatively open set of initial distributions from which the imitation dynamics do not converge. 
\end{example}

\subsection{Generic Non-convergence}\label{subsec:nonconverge}

The imitation of behavioral rules can lead to persistent non-convergence. Negating statements 1 and 3 in Theorem~\ref{thm:weak-imitation-convergence}, we obtain the following characterization of non-convergent dynamics.

\begin{theorem}\label{thm:generic-nonconvergence}
	The imitation dynamics fail to converge from almost every initial distribution\footnote{The qualification ``almost every'' allows for an exceptional set of initial distributions of zero $(N-1)$-dimensional Lebesgue measure. This qualification is necessary, as Proposition~\ref{prop:stationary-convergence} implies that there always exists an initial distribution from which the imitation dynamics converge.} if and only if the essential set $E^F$ contains a Condorcet cycle.
\end{theorem}

When no pair of rules ties, either a Condorcet winner exists or the essential set contains a Condorcet cycle. Thus, in such cases, Theorems~\ref{thm:consensus-iff-condorcet} and \ref{thm:generic-nonconvergence} fully characterize the global imitation dynamics: they either converge to a consensus from every initial distribution or fail to converge from almost every initial distribution. When some rules tie, Theorem~\ref{thm:weak-imitation-convergence} bridges the remaining gap for sufficiently small imitation intensities.

The distribution of contexts shapes which behavioral rules prevail in an environment. We illustrate this with the following example involving three rules and three contexts.

\begin{figure}[htbp]
	\centering
	\begin{subfigure}[t]{0.48\textwidth}
		\centering
		\includegraphics[width=\linewidth]{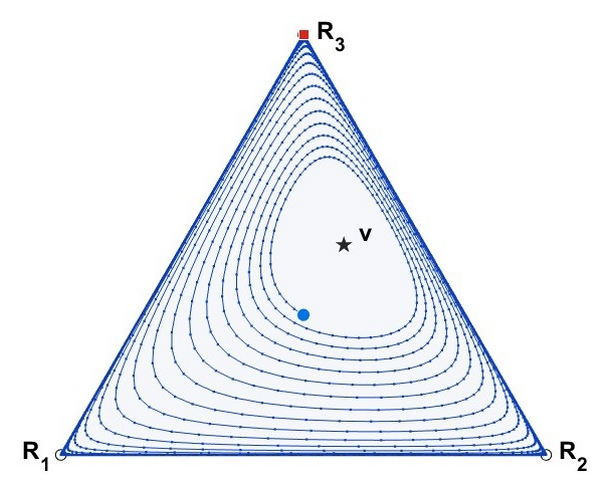}
		\caption{}
		\label{fig:simplex-three}
	\end{subfigure}\hfill
	\begin{subfigure}[t]{0.48\textwidth}
		\centering
		\includegraphics[width=\linewidth]{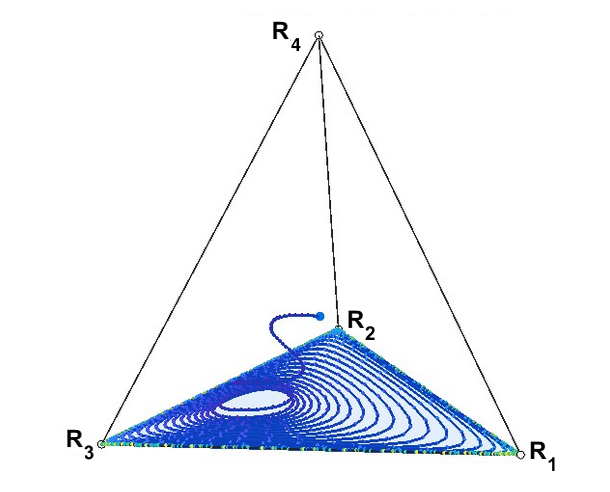}
		\caption{}
		\label{fig:simplex-four}
	\end{subfigure}
	\caption{The trajectory in the three-rule simplex (a) is generated by the non-convergent environment in Example \ref{example:condorcet-contexts} and Proposition \ref{prop:heteroclinic-cycle}, with $W_{12}^F=0.12, W_{23}^F=0.04, W_{31}^F=0.08$, and an initial distribution $\mu_1=(1/3,1/3,1/3)$. The trajectory in the four-rule simplex (b) illustrates that a dominated rule does not survive the imitation dynamics. It is generated by the example in Appendix \ref{appendix:dominated-rule-extinction}, with an initial distribution $\mu_1=(1/4,1/4,1/4,1/4)$.}
\end{figure}

\begin{example}\label{example:condorcet-contexts}
	Consider an environment with three rules and three contexts. Suppose the performance profile satisfies $u_1(R_1)>u_1(R_2)>u_1(R_3)$, $u_2(R_2)>u_2(R_3)>u_2(R_1)$, and $u_3(R_3)>u_3(R_1)>u_3(R_2)$. Under the imitate-if-better protocol $F$, the pairwise comparison margins are $W^F_{12}=1-2\pi_2$, $W^F_{23}=1-2\pi_3$, and $W^F_{31}=1-2\pi_1$. By relabeling the rules and contexts, we may assume that $\pi_1\geq\max\{\pi_2,\pi_3\}$.
	
	If a context occurs with majority frequency ($\pi_1>1/2$), then $W^F_{1j}\geq 2\pi_1-1>0$ for every $j\neq 1$. Thus, $R_1$ is a Condorcet winner, and Theorem~\ref{thm:consensus-iff-condorcet} implies global convergence to a consensus on $R_1$. In this case, the weakest pairwise advantage of $R_1$ is $2\pi_1-1$. Proposition~\ref{prop:consensus-convergencerate} therefore implies that convergence accelerates as the frequency $\pi_1$ increases. 
	
	If $\pi_1=1/2$, we have $W^F_{31}=0$, $W^F_{12}>0$, and $W^F_{23}>0$. The essential set is $E^F=\{R_1,R_3\}$, and the maximal lotteries are of the form $(p_1,0,1-p_1)$ with $p_1\in[2\pi_2,1]$. Theorem~\ref{thm:weak-imitation-convergence} implies that the imitation dynamics under $\lambda F$ globally converge whenever the imitation intensity $\lambda$ is sufficiently small, and Proposition~\ref{prop:stationary-convergence} implies that a distribution $(p_1,0,1-p_1)$ is attainable as a limiting distribution from some initial distribution if and only if $p_1\in (2\pi_2,1)$. Since $(W^Fp)_2=2\pi_2-p_1$, the same proposition gives the asymptotic convergence factor $1-\lambda(p_1-2\pi_2)$. Thus, for a fixed imitation intensity $\lambda$, trajectories that converge to a larger limiting share $p_1$ do so asymptotically faster.
	
	If $\pi_1<1/2$, then there exists a Condorcet cycle with $W^F_{12}, W^F_{23}, W^F_{31}>0$. The distribution $v=(1-2\pi_3,1-2\pi_1,1-2\pi_2)\in\Delta_+(\mathcal R)$ is the unique maximal lottery, and it satisfies $W^F v=0$. Thus, the essential set is $E^F=\mathcal{R}$. Theorem~\ref{thm:generic-nonconvergence} then implies that the imitation dynamics under $\lambda F$ for any $\lambda \in (0,1]$ fail to converge from almost every initial distribution. Figure~\ref{fig:simplex-three} illustrates the non-convergent dynamics under the scaled imitate-if-better protocol $\lambda F$ with $\lambda=6/25$ and context distribution $\pi=(1/3,1/4,5/12)$ (without relabeling).
\end{example}

The following proposition describes the asymptotic behavior of non-convergent imitation dynamics. Here, $\partial\Delta(S^F)$ denotes the relative boundary of the simplex $\Delta(S^F)$ spanned by the top cycle $S^F$, that is, the set of distributions supported on proper subsets of $S^F$.

\begin{prop}\label{prop:boundary-approach}
	For any imitation protocol $F$, let $(\mu_t)_{t \ge 1}$ be a trajectory generated by the associated imitation dynamics. The following statements hold:
	\begin{enumerate}
		\item If $S^F=\mathcal R$, then $\mu_t(S^F)=1$ for all $t \ge 1$. Otherwise, $\mu_t(S^F)$ strictly increases to $1$ and satisfies
		\[
		\limsup_{t\to\infty}\frac{1}{t}\log[1-\mu_t(S^F)]
		\leq
		\log\left(1-\min_{i\in S^F, j\notin S^F}W^F_{ij}\right).
		\]
	\item \sloppy If $z\in\Delta_+(S^F)$ is an accumulation point\footnote{A point $z$ is an accumulation point of $(\mu_t)_{t\ge 1}$ if some subsequence converges to $z$. Here, $\Delta_+(S^F)$ is the set of probability vectors in $\Delta(\mathcal R)$ with support equal to $S^F$.} of the trajectory $(\mu_t)_{t\ge 1}$, then the trajectory converges to $z$. Consequently, if the imitation dynamics do not converge, then $\operatorname{dist}\bigl(\mu_t,\partial\Delta(S^F)\bigr)\to 0$ as $t\to\infty$.
	\end{enumerate}
\end{prop}

This proposition serves as the counterpart to Proposition~\ref{prop:stationary-convergence} for non-convergent dynamics. The first statement shows that every rule outside the top cycle vanishes at least exponentially fast from any initial distribution, strengthening the observation in \citet[Remark~2]{ritzberger1995evolutionary} that such rules vanish when their combined initial share is sufficiently small.\footnote{ \citet[Remark~2]{ritzberger1995evolutionary} show the top cycle face is asymptotically stable. The evolutionary dynamics studied there are more general than imitation dynamics and operate in continuous rather than discrete time; nevertheless, analogous arguments apply.} The second statement reflects the familiar ``overshooting'' phenomenon of discrete-time dynamics: in each period, the population distribution moves slightly further along the tangent direction of the continuous-time trajectories, pushing the dynamics toward the relative boundary. In particular, when the dynamics fail to converge, the smallest population share among top-cycle rules must converge to zero.\footnote{The same result does not hold in continuous time: under certain conditions, \citet[Theorem~7]{boone2019darwin} show that almost every interior trajectory returns arbitrarily close to its initial state infinitely often.}

The following proposition provides sufficient conditions under which the dynamics fail to converge from a given initial distribution. It relies on a strong assumption: the maximal lottery is unique and its support (the essential set) equals the top cycle.\footnote{With three rules and no pairwise ties, this assumption holds automatically. With more than three rules, the assumption is substantive: even when the maximal lottery is unique, the essential set can be strictly smaller than the top cycle, as illustrated in \citet*[Example~2]{de2000choosing}. Special classes of environments satisfying this assumption have been identified by, for instance, \citet*{brandt2018structure}, who show that the condition holds when all pairwise comparisons within the top cycle share a common margin in absolute value and each top-cycle rule defeats exactly half of the others.} This assumption ensures a unique interior stationary distribution within the top cycle. Under this condition, the dynamics are non-convergent whenever the initial distribution is sufficiently far from this unique stationary distribution, as measured by the Kullback--Leibler divergence: for $p\in\Delta(\mathcal R)$ and $q\in\Delta_+(\mathcal R)$,
$D_{\mathrm{KL}}(p\|q)=\sum_{i\in \supp(p)}p_i\log (p_i/q_i).$

\begin{prop}\label{prop:nonconvergencesufficient}
	For any imitation protocol $F$, suppose the maximal lottery $v$ is unique and has support $S^F$. Starting from an initial distribution $\mu_1\in\Delta_+(\mathcal R)$, the imitation dynamics fail to converge under either of the following conditions:
	\begin{enumerate}
		\item $S^F=\mathcal R$ and $\mu_1\neq v$;
		\item \sloppy $S^F\neq\mathcal R$ and 
		$D_{\mathrm{KL}}(v\|\mu_1)
		>-\frac{\beta}{\delta}\log \mu_1(S^F)$ where $\beta=\max_{j\notin S^F}((W^F)^\top v)_j$ and $\delta=\min_{i\in S^F,j\notin S^F}W^F_{ij}$.\footnote{When a Condorcet winner exists, neither condition holds, because $S^F$ is a singleton and thus $-\log \mu_1(S^F) = D_{\mathrm{KL}}(v\|\mu_1)$.}
	\end{enumerate}
\end{prop}

Non-convergent dynamics may exhibit structured long-run behavior. While a general characterization is difficult to obtain, the following proposition shows that when there are three rules, non-convergent imitation dynamics always approach a heteroclinic cycle, strengthening Proposition~\ref{prop:boundary-approach}(2). In this case, each rule alternately comes arbitrarily close to consensus and to extinction, which may create a false impression of social consensus or rule extinction. We refer to Figure~\ref{fig:simplex-three} again for an illustration of a three-rule heteroclinic cycle.

\begin{prop}\label{prop:heteroclinic-cycle}
	Suppose there are three rules with $a=W^F_{12}>0$, $b=W^F_{23}>0$, and $c=W^F_{31}>0$. The unique interior stationary distribution is $v=(b,c,a)/(a+b+c)$. From every initial distribution $\mu_1\neq v$, the imitation dynamics fail to converge and satisfy
	\[
	\liminf_{t\to\infty}\mu_t(i)=0 \quad \text{and} \quad \limsup_{t\to\infty}\mu_t(i)=1 \qquad \text{for all } i\in\{1,2,3\}.
	\]
	Moreover, the dynamics approach a boundary heteroclinic cycle along $e_1\to e_3\to e_2\to e_1$.
\end{prop}

As a final remark, social welfare under non-convergent dynamics can be difficult to evaluate, as the time averages of population shares may fail to converge. Even in this three-rule environment with a limiting heteroclinic cycle, \citet*{ganikhodzhaev2004necessary} show that $(\frac{1}{T}\sum_{t=1}^T \mu_t)_{T=1}^{\infty}$ fails to converge from every initial distribution $\mu_1\neq v$.

\subsection{Surviving Behavioral Rules}\label{subsec:survival}

\sloppy Now, we summarize the implications of the preceding discussion for surviving rules. We say a rule $R_i$ \emph{survives} if its population share does not converge to zero; that is, $\limsup_{t\to\infty}\mu_t(i) > 0$. As demonstrated in Proposition~\ref{prop:heteroclinic-cycle}, a surviving rule may still become arbitrarily rare along a subsequence of periods.

For convergent dynamics, Proposition~\ref{prop:stationary-convergence} shows that the surviving rules coincide with the essential set. For non-convergent dynamics, Proposition~\ref{prop:boundary-approach} shows that only rules in the top cycle can survive.\footnote{The same necessary condition for survival under continuous-time dynamics is given in \citet[Theorem~4.3]{biggar2024attractor}.} Example~\ref{example:largeintensitybreakconvergence} shows that this necessary condition cannot be strengthened to membership in the essential set. Below, we show that top-cycle membership is not sufficient for survival: a dominated rule can never survive, even though it may belong to the top cycle. Figure~\ref{fig:simplex-four} illustrates that a dominated rule can vanish while the dynamics remain non-convergent.

\begin{prop}\label{prop:dominated-rule-extinction}
	For any imitation protocol $F$, suppose rule $R_j$ is dominated by rule $R_i$, in the sense that $W^F_{i\ell} \geq W^F_{j\ell}$ for every $\ell\in \{1,2,\dots, N\}$ and $W^F_{ij} > 0$. Then:
	\begin{enumerate}
		\item Rule $R_j$ cannot survive; that is, $\lim_{t\to\infty} \mu_t(j) = 0$.
		\item There exist environments in which a dominated rule belongs to the top cycle $S^F$.
	\end{enumerate}
\end{prop}

\subsection{Relation to the Literature}
\label{subsec:relation-to-literature}

The imitation dynamics studied in this paper are discrete-time replicator dynamics with an antisymmetric payoff matrix, also known as Volterra quadratic stochastic operators. Our main contribution is to show that linking imitation dynamics to a social-choice representation yields a unified characterization of global dynamics. Below, we discuss how our results relate to and strengthen the literature.

Theorem~\ref{thm:consensus-iff-condorcet} establishes that global convergence to consensus is equivalent to the existence of a Condorcet winner, without imposing any regularity assumptions on $W^F$. Prior work captures only one direction or relies on local or nondegeneracy restrictions: \citet*[Principle~1]{nagylaki1983evolution} and \citet*[Proposition~1]{bilancini2021pairwise} establish sufficiency, while \citet[Proposition~3]{nachbar1990evolutionary} shows that convergence from an interior population state to consensus implies a weak Condorcet condition. For local stability, \citet[Theorem~2.4(c) and Corollary~2.9]{AkinLosert1984} characterize the local asymptotic stability of consensus under the assumption that all even-dimensional principal minors of $W^F$ are nonzero,\footnote{Corollary 2.9 also establishes that the Condorcet winner condition serves as a sufficient condition for global convergence.} and for continuous-time multipopulation dynamics, \citet[Proposition~2]{ritzberger1995evolutionary} characterize local asymptotic stability via a strict Condorcet condition without imposing nondegeneracy.\footnote{\citet{ritzberger1995evolutionary} analyze a multipopulation environment with two pools of agents that update separately. Their dynamics reduce to the continuous-time counterpart of ours when specialized to the symmetric zero-sum game induced by $W^F$ with identical initial distributions in both pools.} In addition, the convergence rate characterized in Proposition~\ref{prop:consensus-convergencerate} sharpens the bound specified in \citet[Theorem~7]{ganikhodzhaev1993quadratic}.

Theorem~\ref{thm:weak-imitation-convergence} characterizes global convergence under sufficiently weak imitation. For any imitation intensity, Proposition~\ref{prop:stationary-convergence} identifies which stationary distributions can arise as limits, including mixed distributions in which multiple behavioral rules retain positive shares in the long run, and gives the associated convergence rates.
Existing results provide only a partial account of convergence to mixed distributions: \citet[Theorem~1]{saburov2013some} characterizes global convergence from all initial distributions (including boundary distributions) via a transitivity condition. However, under the nondegeneracy assumption imposed there, transitivity implies a Condorcet winner and thus rules out convergence to nontrivial mixtures.\footnote{Their result also does not cover Theorem~\ref{thm:consensus-iff-condorcet}, which does not require $W^F$ to be nondegenerate.} Regarding the surviving rules under convergent dynamics (equivalently, the support of limiting distributions), \citet[Theorem~10]{AkinLosert1984} show that under nondegeneracy and a sufficiently small step size, every accumulation point of an interior trajectory is supported on the essential set. \citet[Theorem~3.4]{piliouras2014optimization} provide a continuous-time counterpart to this necessary condition. Also in continuous time, \citet[Theorem~1]{ritzberger1995evolutionary} characterize which faces of the state space are asymptotically stable; however, face stability does not imply pointwise convergence to a mixed stationary distribution.\footnote{More specifically, when a unique mixed stationary distribution exists, asymptotic stability of the face spanned by its support does not guarantee convergence to that distribution; when multiple limiting distributions exist, the convexity of the limiting distribution set (Proposition~\ref{prop:stationary-convergence}) implies that no mixed limiting distribution is asymptotically stable, as every neighborhood contains another stationary point.}

Theorem~\ref{thm:generic-nonconvergence} gives a clean characterization of generic non-convergence, and Proposition \ref{prop:nonconvergencesufficient} provides sufficient conditions for non-convergence from a given initial distribution. While \citet{AkinLosert1984} show that discrete-time replicator dynamics can fail to converge even with three strategies, prior general non-convergence results rely on restrictive structural assumptions: \citet[Theorem~12]{ganikhodzhaev2006quadratic} show that the set of initial distributions leading to convergent dynamics is lower-dimensional under nontransitivity, homogeneity, and nondegeneracy,\footnote{Their paper uses the term ``typical'' to describe this nondegeneracy condition.} and \citet[Theorem~4]{ganikhodzhaev1993quadratic} identifies sufficient conditions ensuring non-convergence by assuming an isolated interior stationary distribution. In the continuous-time counterpart of our model, \citet[Theorem 1 and Remark 2]{ritzberger1995evolutionary} allow for non-convergent dynamics and show that only rules in the top cycle can survive when the initial distribution is sufficiently close to the top-cycle face. This necessary condition for survival leaves open whether the population distribution converges. Finally, \citet{falniowski2025discrete} show that discrete-time models with the same continuous-time replicator limit can exhibit different long-run behavior.

\section{Belief Evolution Through Imitation}
\label{sec:insurance_belief_evolution}

Section \ref{sec:evolution} characterizes how an environment can shape the long-run behavioral rule adoption. Obtaining more specific predictions about which rules prevail, however, requires specifying the environment in more detail. In this section, we apply our general imitation framework to study belief evolution in an insurance economy, following \citet{blume1992evolution}. Agents hold subjective beliefs and imitate the beliefs of peers in the same income state. We show how the distribution of income states and the set of initial beliefs can lead either to a consensus on rational expectations or to persistent non-convergent belief dynamics, including waves of optimism and pessimism.  

\subsection{Environment}\label{subsec:insuranceenvironment}

There is a unit mass of agents facing idiosyncratic income shocks. The set of income states is $\mathcal K=\{1,\dots,K\}$. In state $k\in\mathcal K$, an agent has an income $y_k > 0$. The objective distribution of income states is given by $\pi\in\Delta_+(\mathcal K)$. In each period, each agent's income state is drawn according to $\pi$, independently across agents and over time. We order the income states such that $y_1<\cdots<y_K$.

Agents hold subjective beliefs about their income states. The set of subjective beliefs is given by $\mathcal R=\{\hat\pi^1,\dots,\hat\pi^N\}$, where each subjective belief $\hat\pi^i$ has full support. Let $\pi^{\mathrm{RE}}=\pi$ denote the rational-expectations belief. We do not require $\pi^{\mathrm{RE}}\in\mathcal R$, thereby allowing for an economy in which no agent holds rational expectations.

Each agent maintains the same subjective belief regardless of their realized income state. Their beliefs determine how they allocate wealth across state-contingent claims in the insurance market, and the realized income state determines the performance of these investment decisions. Suppressing explicit portfolio choice problems, we represent the performance ranking of subjective beliefs in income state $k$ as a function
\begin{equation}
	u_k(\hat\pi)=\frac{\hat\pi_k-\pi_k}{\pi_k}.
	\label{eq:Uk_formula_ins}
\end{equation}
This performance function measures the relative over- or underestimation of the probability of state $k$ under subjective belief $\hat\pi$, and captures the intuition that an agent who assigns a higher subjective probability to state $k$ allocates more wealth to that state, and therefore performs better when state $k$ is realized.\footnote{For a more explicit justification of this performance function, consider agents endowed with logarithmic Bernoulli utility and wealth $w>0$ who purchase state-contingent securities. Suppose the market price of the security paying one unit in state $k$ is $q_k(\mu^t)>0$, which may depend on the history of population states $\mu^t=(\mu_s)_{s=1}^t$. An agent with belief $\hat\pi$ chooses her security portfolio to maximize the subjective expected utility of tomorrow's wealth, $\sum_{k=1}^K\hat\pi_k\log c_k$, subject to the budget constraint $\sum_{k=1}^Kq_k(\mu^t)c_k\leq w$, yielding state-contingent wealth $c_k(\hat\pi)=w\hat\pi_k/q_k(\mu^t)$. In this case, we have $c_k(\hat\pi^i)/w > c_k(\hat\pi^j)/w \Longleftrightarrow \hat\pi^i_k > \hat\pi^j_k \Longleftrightarrow u_k(\hat\pi^i) > u_k(\hat\pi^j)$. That is, the performance function $u_k$ represents the ranking over realized gross returns in state $k$. Moreover, when all prices $q_k(\mu^t)=\pi_k$ are actuarially fair, $u_k(\hat\pi)$ coincides with the realized net return on wealth.}  The rational-expectations belief yields zero performance in every state.

We suppose agents update via the imitate-if-better protocol, using \eqref{eq:Uk_formula_ins} only as an ordinal representation of within-state performance. In particular, agents do not observe its cardinal value, which depends on the objective probabilities. Imposing ordinal imitation\footnote{Ordinal imitation is defined and discussed in Section~\ref{sec:ordinal}.} is important for this application: if an agent could observe the cardinal performance of her subjective belief $\hat\pi$, she could infer the objective income distribution as soon as she had experienced every state at least once. In contrast, when imitation depends on only ordinal comparisons, the state-specific rankings do not identify the objective distribution: for any counterfactual distribution $\tilde\pi\in\Delta_+(\mathcal{K})$, the function $\tilde u_k(\hat\pi)=(\hat\pi_k-\tilde\pi_k)/\tilde\pi_k$ induces the same performance ranking of beliefs in state $k$.
	
The timeline of the imitation dynamics is as follows. Time starts in period $0$ and imitation starts in period $1$. At the beginning of period $t\ge 0$, each agent is randomly assigned an income state according to $\pi$ and uses her subjective belief to choose her portfolio. Her performance is realized when her income state in period $t+1$ is drawn. Then, each agent samples a peer who has realized the same income state, and imitates the peer's belief if the sampled peer has higher performance.  Belief evolution is governed by \eqref{eqn:evolution}, with $W^F$ given by 
\[W^F_{ij}=\pi(\{k\in \mathcal K: \hat\pi_k^i>\hat\pi_k^j\})-\pi(\{k\in \mathcal K: \hat\pi_k^i<\hat\pi_k^j\}),\qquad i,j=1,2,\dots, N.\]
	
\subsection{Convergence to Rational Expectations with a Large Number of States} \label{subsec:convergentbeliefdynamics}

To illustrate how the decision environment can shape market beliefs, we first show that a richer set of income states can help the economy converge to rationality. We employ a probabilistic argument: fixing the number of available beliefs, setting one of them to be the rational-expectations belief, and drawing the remaining beliefs independently and uniformly at random, we show that the rational-expectations belief becomes the consensus with arbitrarily large probability when there are sufficiently many income states. 

\begin{prop}\label{prop:REwins}
	Fix the number of beliefs $N\ge 2$. For any given number of income states $K \ge 1$, suppose the belief set $\mathcal R$ consists of the rational-expectations belief $\pi^{\mathrm{RE}}=\pi$ and $N-1$ other beliefs drawn independently and uniformly from $\Delta_+(\mathcal K)$, where $\mathcal K = \{1, \dots, K\}$. Then, for any $\varepsilon>0$, there exists an integer $K_0$ such that for all $K\ge K_0$ and all objective distributions $\pi\in\Delta_{+}(\mathcal K)$, the probability that $\pi^{\mathrm{RE}}$ is the Condorcet winner (and hence that the belief dynamics globally converge to a consensus on rational expectations) is greater than $1-\varepsilon$.
\end{prop}
 
The intuition for this result relies on the social-choice representation: under imitate-if-better, a belief is a Condorcet winner if it is more likely than any other belief to assign a higher probability to the realized state. When the number of states is large, most states necessarily have small objective probabilities. Thus, a uniformly drawn belief is likely to assign disproportionately
high probabilities to some of these states, in which case these
overestimates must be offset by many smaller underestimates or
by larger underestimates of more probable states. As a consequence, the rational-expectations belief defeats a randomly drawn belief with high probability when there are many states. This intuition relies on the uniform randomization, and the same conclusion need not hold for other belief selections.

Proposition~\ref{prop:REwins} is robust to small perturbations to the rational-expectations belief. In Online Appendix \ref{appendix:REsharp}, we show that the same conclusion holds if the fixed rational-expectations belief $\pi^{\mathrm{RE}}$ is replaced by any belief $\hat\pi^*$ that is sufficiently close to $\pi$ in total variation distance, and we identify the sharp threshold for this perturbation.

\subsection{Generic Non-convergence with a Large Number of Beliefs}
\label{subsec:nonconvergence_large_beliefs}

Whereas Proposition \ref{prop:REwins} shows that a rich state space favors rational expectations for a fixed number of beliefs, fixing the state space and expanding the set of beliefs reverses the conclusion.
Using the same probabilistic framework (where the number of income states is fixed and beliefs are drawn independently and uniformly at random), we show that, under mild regularity conditions on the objective distribution, the belief dynamics generically fail to converge with arbitrarily large probability when there are sufficiently many beliefs.

\begin{prop}\label{prop:NoCondorcet}
	Fix the number of income states $K \ge 3$. For any given number of beliefs $N \ge 2$, suppose the belief set $\mathcal R$ consists of $N$ beliefs drawn independently and uniformly from $\Delta_+(\mathcal K)$. Then, for any $\varepsilon>0$, there exists an integer $N_0$ such that for all $N\ge N_0$ and all objective distributions $\pi\in\Delta_{+}(\mathcal K)$ satisfying $\max_{1\le k\le K}\pi_k < 1/2$ and $\pi(A)\neq1/2$ for every $A\subseteq\mathcal K$, the probability that there exists a Condorcet cycle within the essential set (and hence that the belief dynamics are generically non-convergent) is greater than $1-\varepsilon$. The same statement holds if one of the beliefs is fixed as the rational-expectations belief.
\end{prop}

When the objective frequency of every state is below $1/2$,
no single state commands a majority in the social-choice
representation. The condition $\pi(A)\neq1/2$
further rules out pairwise ties almost surely.
Drawing candidate beliefs uniformly at random generates uniformly random ordinal rankings of beliefs within each state. These rankings are dependent across income states because no belief can assign a higher probability than another to every state.
Our result thus provides, to the best of our knowledge, a new counterpart to the probabilistic voting paradox of \citet{may1971some}, who considers independently and uniformly generated voter rankings.\footnote{Relatedly,
	\citet{laslier2010silico} provides simulation results showing that
	Condorcet winners become rare as the number of candidates
	grows in a similar allocation model, using a different
	sampling distribution.}
As the proposition shows, the same conclusion holds even if
one of the candidate beliefs is the rational-expectations
belief. Thus, the presence of rational expectations in the economy does not prevent generic non-convergence when there are many subjective beliefs.

\subsection{Waves of Optimistic and Pessimistic Sentiment}
\label{sec:optimism_pessimism}

While the previous subsection establishes that non-convergent belief dynamics can occur, it does not imply that the waves of optimistic and pessimistic sentiment described by \citet{keynes1936general} may arise. In particular, such waves require candidate beliefs to be ordered in terms of some measure of optimism. In this subsection, we show that Keynesian sentiment waves can indeed be generated by belief imitation.

There exist multiple ways to rank beliefs according to their degree of optimism or pessimism. We focus on two standard criteria. We say the beliefs in $\mathcal R$ are ranked by \emph{first-order stochastic dominance} (FOSD) if, for every $i<j$, we have $\hat\pi^j\succeq_{\mathrm{FOSD}}\hat\pi^i$; that is, $\sum_{k=1}^{r}\hat\pi_k^j \leq \sum_{k=1}^{r}\hat\pi_k^i$ for every $r=1,\ldots,K-1$. Furthermore, we say the beliefs in $\mathcal R$ are ranked by the \emph{monotone likelihood ratio} (MLR) if, for every $i<j$, the likelihood ratio $\hat\pi_k^j/\hat\pi_k^i$ is weakly increasing in $k$. Under both rankings, beliefs with larger indices can be interpreted as more optimistic. 

Ranking candidate beliefs by MLR is a stronger assumption than ranking them by FOSD. Recall that the belief profile satisfies single-crossing if, for every $i<j$ and $k<k'$, $\hat\pi_k^j\geq\hat\pi_k^i$ implies $\hat\pi_{k'}^j\geq\hat\pi_{k'}^i$. The MLR ranking implies this single-crossing property, which in turn implies the FOSD ranking. Neither converse holds in general.

Consequently, for MLR-ranked beliefs, Proposition \ref{prop:singlecrossing-existence-condorcet} implies that a Condorcet-winning belief exists under mild nondegeneracy conditions. Under these conditions, the imitation dynamics globally converge to a consensus on the subjective belief that assigns the highest probability to the median income state.

\begin{corollary}\label{cor:ordered_belief_evolution}
	Suppose candidate beliefs in $\mathcal R$ are ranked by monotone likelihood ratio. Suppose further that the environment is nondegenerate: no two beliefs assign the same probability to any state ($\hat\pi_k^i\neq \hat\pi_k^j$ for all $i\neq j$ and $k\in \mathcal{K}$), and no subset of states has objective probability $1/2$ ($\pi(A)\neq 1/2$ for any $A\subseteq \mathcal K$). Then, there exists a unique $\pi$-median state $M\in\mathcal K$ (a state satisfying $\sum_{k<M}\pi_k<1/2$ and $\sum_{k>M}\pi_k<1/2$), and the imitation dynamics globally converge to a consensus on the belief assigning the largest probability to the median income state $M$.
\end{corollary}

When beliefs satisfy the stronger MLR ranking, the imitation dynamics cannot sustain persistent waves of optimism and pessimism; instead, they necessarily converge to a consensus belief. Moreover, this corollary shows that imitation favors extreme sentiments whenever the median state is extreme (e.g., in a two-state environment or when an extreme state occurs with probability exceeding $1/2$) and moderate sentiments otherwise. Rational expectations survive if and only if all other subjective beliefs underestimate the likelihood of the median income state.

If beliefs satisfy only the weaker FOSD ranking, the following proposition shows that non-convergent belief dynamics can arise whenever there are four or more states.\footnote{Under the stated nondegeneracy condition, the inequality $\pi_M+\min\{\pi_1,\pi_K\}>1/2$ necessarily holds when there are only two or three states, but it can fail when there are four or more. Indeed, for objective income distributions $\pi$ drawn uniformly at random, one can show that $\pi_M+\min\{\pi_1,\pi_K\}<1/2$ holds with probability arbitrarily close to one when there are sufficiently many income states.} Moreover, it shows that belief imitation can generate persistent waves of optimistic and pessimistic sentiment even when the available beliefs consist only of rational expectations and arbitrarily small optimistic and pessimistic misspecifications. Here, we use a total variation distance to measure the distance between two beliefs $\pi^1,\pi^2\in\Delta(\mathcal K)$:
\[d(\pi^1,\pi^2)=\frac{1}{2}\sum_{k=1}^K|\pi^1_k-\pi^2_k|.\]

\begin{prop}\label{prop:FOSD_belief_evolution}
	Fix the number of states $K\geq 2$ and an objective distribution
	$\pi\in\Delta_+(\mathcal K)$ such that
	$\pi(A)\neq 1/2$ for every $A\subseteq\mathcal K$.
	Let $M$ be the $\pi$-median state. The following statements hold:
	\begin{enumerate}
		\item Suppose $\pi_M+\min\{\pi_1,\pi_K\}>1/2$.
		If the belief set $\mathcal R$ is FOSD-ranked and no two
		beliefs assign the same probability to any state, then
		the belief assigning the highest probability to the
		median state $M$ is the unique Condorcet winner, and the imitation dynamics globally converge to a consensus
		on this belief.
		
		\item Suppose $\pi_M+\min\{\pi_1,\pi_K\}<1/2$.
		For every $\delta>0$, there exist beliefs
		$\hat\pi^P,\hat\pi^O\in\Delta_+(\mathcal K)$ satisfying
		$\hat\pi^O\succeq_{\mathrm{FOSD}}\pi
		\succeq_{\mathrm{FOSD}}\hat\pi^P$,
		$d(\hat\pi^P,\pi)<\delta$, and
		$d(\hat\pi^O,\pi)<\delta$,
		such that no two beliefs in
		$\mathcal R=\{\hat\pi^P,\pi,\hat\pi^O\}$
		assign the same probability to any state and
		the three beliefs form a Condorcet cycle.
		For this belief set, the imitation dynamics do not
		converge from any interior initial distribution other than
		the unique interior stationary distribution.
	\end{enumerate}
\end{prop}

Intuitively, FOSD pins down belief rankings at only the two extreme states. The belief that assigns the highest probability to the median state $M$ is therefore guaranteed to outperform any rival at both $M$ and one of the extreme states, which secures a majority whenever $\pi_M+\min\{\pi_1,\pi_K\}>1/2$. When this sum falls below $1/2$, the lack of single-crossing on intermediate states leaves enough room for creating a Condorcet cycle. 

\section{Ordinal Imitation}\label{sec:ordinal}

The imitate-if-better protocol plays a crucial role in linking imitation dynamics to classical social choice theory, where ordinal preferences are aggregated. Imitate-if-better is a form of ordinal imitation, in which imitation probabilities depend on only ordinal performance rankings rather than cardinal performance levels. Because ordinal imitation demands less information, it is better suited for modeling imitation in environments where performances are incomparable across contexts, or where agents do not infer the underlying environment parameters, as in Section~\ref{subsec:insuranceenvironment}.

In this section, we study ordinal imitation. We show that it opens up the possibility of non-convergent dynamics and admits an envy-based justification. Although ordinal imitation protocols are discontinuous in the payoff space, they can generate dynamics whose qualitative properties are robust to perturbations of the protocol.

\subsection{Ordinal Imitation and Imitate-if-better}

An imitation protocol $F$ is \emph{ordinal} if, for every context $k\in\mathcal{K}$ and every strictly increasing function $\varphi:\mathbb{R}\rightarrow\mathbb{R}$, we have $F_k(u,v)=F_k(\varphi(u),\varphi(v))$ for all $u,v\in\mathbb{R}$. That is, a protocol is ordinal when the imitation probability is invariant under strictly increasing transformations of performances. It is straightforward to verify that the imitate-if-better protocol is ordinal.

Ordinal imitation substantially weakens the informational requirements of our framework. It does not require agents to coordinate on a shared cardinal representation of performances: as long as agents rank rules within each context in the same way, each agent can use her own cardinal scale to evaluate both the performance of her own rule and that of her sampled peer. Furthermore, when no across-context imitation occurs, performances in different contexts can be measured in different units without affecting the dynamics.

The following proposition shows that any improving ordinal imitation protocol must take the form of imitate-if-better, up to context-specific scalings of imitation intensities. As a consequence of this proposition, any improving ordinal imitation protocol must be discontinuous in payoffs. 

\begin{prop}\label{prop:ordinalimitations}
	An improving imitation protocol $F\in\mathcal{F}$ is ordinal if and only if $F_k(u,v)=a_k\mathbf{1}\{v>u\}$ holds for all $u,v\in\mathbb{R}$ and $k\in\mathcal{K}$, where $a_k\in (0,1]$.
\end{prop}

\subsection{Guaranteed Convergence Requires Proportional Imitation}

Our earlier results in Propositions~\ref{prop:NoCondorcet} and~\ref{prop:FOSD_belief_evolution} show that the imitate-if-better protocol can generate non-convergent dynamics. A natural follow-up question is whether this non-convergence is an artifact of the discontinuity inherent in ordinal imitation, or whether it arises under a broader class of imitation protocols.

In this subsection, we show that an imitation protocol always leads to convergent dynamics across all environments if and only if it takes the form of proportional imitation over the entire range of performance levels. Thus, guaranteed convergence relies on a strong behavioral assumption: each agent must imitate with a probability tailored to cardinal performances in every decision context.

\begin{theorem}\label{thm:universalconvergencecharacterization}
	Fix a compact interval
	$U=[\underline u,\overline u]$ with $\underline u\neq \overline u$, a finite context set
	$\mathcal K$ with $|\mathcal K|\geq3$, and an improving imitation
	protocol $F\in \mathcal{F}$ such that, for every $k\in\mathcal{K}$, $F_k$ is not identically zero on $U\times U$. The following statements are equivalent:
	\begin{enumerate}
		\item For every distribution $\pi\in\Delta_+(\mathcal K)$, every
	 set of rules $\mathcal R$, and every
		performance profile $(u_k)_{k\in\mathcal K}$ with
		$u_k:\mathcal R\to U$, the imitation dynamics under $F$ 
		globally converge.
				
		\item For every context $k\in\mathcal K$, there exists a weakly
		increasing function $g_k: U\to[0,1]$ such that
		$F_k(u,v)=(g_k(v)-g_k(u))^+$ for all $u,v\in U$.
		
		\item For each $k\in\mathcal K$, the imitation probabilities
		satisfy additivity: 
		$F_k(u,w)=F_k(u,v)+F_k(v,w)$
		whenever $u,v,w\in U$ and $u<v<w$.
	\end{enumerate}
\end{theorem}

Theorem~\ref{thm:universalconvergencecharacterization} establishes that guaranteed convergence requires the protocol to take a proportional form, up to context-specific performance rescalings. The theorem also shows that this representation is equivalent to additivity, which rules out improving ordinal imitation: by Proposition~\ref{prop:ordinalimitations}, any improving ordinal protocol takes the form $F_k(u,v)=a_k\mathbf{1}\{v>u\}$ with $a_k\in(0,1]$, so additivity for $u<v<w$ would require $a_k = a_k + a_k$, contradicting $a_k > 0$. Consequently, every improving ordinal imitation protocol generates non-convergent dynamics in some environment.

\subsection{Envy-Minimizing Imitation}
\label{sec:comparison_loss_foundation}

As \citet{coricelli2010counterfactual} note, envy arising from social comparisons can influence behavioral choices in adaptive learning. We now motivate the imitate-if-better protocol via a myopic envy-minimization criterion. Suppose that agents incur an envy-driven disutility from being outperformed by a randomly sampled peer in a subsequent comparison, and receive no corresponding utility gain from outperforming others. We represent this disutility by a weakly increasing loss function $\ell:\mathbb{R}\to[0,\infty)$ satisfying $\ell(x)=0$ for $x\leq 0$ and $\ell(x)>0$ for $x>0$. The value $\ell(x)$ captures the utility loss an agent suffers when encountering a peer whose performance is $x$ units higher than her own. This formulation accommodates various forms of envy, including the binary loss $\ell(x)=\mathbf{1}\{x>0\}$ and the proportional loss $\ell(x)=\max\{x,0\}$.

Agents evaluate an imitation protocol $F$ after their contexts are assigned but before they sample a peer. At a population state $\mu$, an agent who uses rule $R$ and faces context $k$ anticipates an expected utility loss from a subsequent peer comparison given by $L(R \mid k,\mu)=\sum_{i=1}^N\mu(i)\ell(u_k(R_i)-u_k(R))$. Her expected utility loss from adopting $F$ is therefore
\[
L(F \mid R,k,\mu)=\sum_{j=1}^N\mu(j)\Big[F_k(u_k(R),u_k(R_j))L(R_j \mid k,\mu)+\bigl(1-F_k(u_k(R),u_k(R_j))\bigr)L(R \mid k,\mu)\Big].
\]
The first term in brackets corresponds to the expected loss if the agent adopts $R_j$, and the second to the expected loss if she retains $R$. The subsequent comparison is with a new peer drawn independently from the current population distribution $\mu$. This evaluation is myopic: agents assume they remain in the same context and ignore simultaneous rule updating by other agents.

The following proposition establishes that imitate-if-better is the unique imitation protocol that minimizes the aggregate expected utility loss $L(F\mid \mu)=\sum_{k=1}^K\pi_k\sum_{i=1}^N\mu(i)L(F\mid R_i,k,\mu)$ at every population state and across all environments.

\begin{prop}\label{prop:myopic_comparison_loss}
	A protocol $F\in\mathcal{F}$ satisfies $L(F\mid\mu) \leq L(F'\mid\mu) $ for all $F'\in\mathcal{F}$ in every environment and at every population state $\mu$ if and only if $F$ is imitate-if-better.
\end{prop}

Intuitively, whenever an agent encounters a peer using a better-performing rule, switching to that rule strictly reduces her expected envy in a subsequent same-context comparison. Thus, the expected utility loss is minimized by switching to a better-performing rule with maximal probability. This result complements the expected payoff-maximizing characterization of proportional imitation established by \citet{schlag1998imitate}. In Sections~\ref{subsec:stochasticachievement} and~\ref{subsec:acrosscontextimitation}, we explain why our envy-based characterization does not extend to settings with stochastic performances or across-context comparisons.

\subsection{Robust Ordinal Imitation}

Ordinal imitation is a useful modeling tool. In addition to reducing the informational demands of imitation, it exhibits several robustness properties, which we discuss below. 

\subsubsection*{Invariance to Performance Rescalings}

Under ordinal imitation, the behavioral-rule dynamics are
invariant to context-specific strictly increasing transformations
of performance. This feature is useful when outcomes across
contexts are incomparable, or when agents differ in their
cardinal evaluations while agreeing on performance rankings
within each context.

In contrast, non-ordinal protocols such as proportional imitation can be sensitive to such performance rescalings, as illustrated in the following example.

\begin{example}\label{ex:rescaling_sensitivity}
	Suppose agents use context-based proportional imitation in the belief-evolution setting of Section~\ref{sec:insurance_belief_evolution}. By Theorem~\ref{thm:universalconvergencecharacterization}, the dynamics always converge; moreover, under proportional imitation, they concentrate on the belief(s) with the highest expected performance and yield a consensus when the maximizer is unique. However, which belief prevails depends on the chosen performance scale:
	\begin{enumerate}
		\item Original scale \eqref{eq:Uk_formula_ins}: every belief has the same expected performance $\sum_{k=1}^K \pi_k u_k(\hat\pi^i)=0$, so $W^F = 0$ and the imitation dynamics are always stationary.
		\item \sloppy Concave rescaling $\varphi(x)=\log(1+x)$: belief $\hat\pi^i$ has expected performance $\sum_{k=1}^K \pi_k \log\frac{\hat\pi_k^i}{\pi_k} = -D_{\mathrm{KL}}(\pi\|\hat\pi^i)$, so the dynamics converge to the belief(s) minimizing misspecification in Kullback--Leibler divergence.
		\item Convex rescaling $\varphi(x)=(1+x)^2$: belief $\hat\pi^i$ has expected performance $\sum_{k=1}^K \pi_k (\frac{\hat\pi_k^i}{\pi_k})^2 = 1 + \chi^2(\hat\pi^i\|\pi)$, so the dynamics converge to the belief(s) maximizing misspecification in chi-square divergence.\footnote{For distributions $p,q$ on $\{1,\dots,K\}$ with $q$ having full support, $\chi^2(p\|q)=\sum_{k=1}^K p_k^2/q_k - 1$.}
	\end{enumerate}
\end{example}

\subsubsection*{Robustness to Imitation Errors}

Under any improving ordinal protocol, the probability of imitating a better-performing rule remains bounded away from zero as the performance gap vanishes. In contrast, under non-ordinal imitation, this probability can shrink to zero as the gap diminishes. This difference can make ordinal imitation more robust to small errors when performance differences are small.

To illustrate this robustness, we adapt an example from \citet{mertikopoulos2022survival} to our multi-context environment by introducing a perturbation to agents' sampling: when an agent samples a peer using the same rule, she samples once more. Under this perturbation, a strictly dominated rule still goes extinct under ordinal imitation, yet can survive under proportional imitation.

\begin{example}\label{example:robustperturbation}
	Consider three rules $\mathcal{R}=\{R_1,R_2,R_3\}$ and two
	equiprobable contexts in which the performance profiles for
	$(R_1,R_2,R_3)$ are $(1,0,-\delta)$ and $(0,1,1-\delta)$, with
	$\delta=0.001$. Note that $R_2$ strictly dominates $R_3$ by a
	small margin. Suppose imitation is perturbed as follows: whenever an agent samples a peer using the same rule, she resamples once. Under this perturbation, users of more prevalent rules are more likely
	to resample.
	
	We compare two baseline protocols: proportional imitation (with imitation intensity $\sigma=1/2$) and imitate-if-better. Without perturbations, Proposition~\ref{prop:dominated-rule-extinction} shows that the dominated rule $R_3$ goes extinct under both protocols. With this perturbation, because of redrawing, the induced evolutionary dynamics under either protocol are no longer quadratic, so the same proposition does not apply. In Online Appendix~\ref{appendix:robustperturbation}, we prove that $R_3$ survives under perturbed proportional imitation, yet is driven to extinction under perturbed imitate-if-better.
	
	Intuitively, because users of prevalent rules resample more often
	than users of rare ones, the perturbation generates a net inflow
	from $R_1$ into $R_3$ whenever $R_3$ is rare. Under proportional
	imitation, this net inflow offsets the outflow to $R_2$, because
	$R_3$-users who sample $R_2$ switch with a very small probability.
	Under imitate-if-better, by contrast, $R_3$-users who sample $R_2$
	switch with probability one, generating a large outflow that
	cannot be offset.
\end{example}

This example confirms the remark in \citet{mertikopoulos2022survival} that their resampling argument does not extend to discontinuous imitation protocols. The innovative dynamics studied by
\citet{hofbauer2011survival}, which allow agents to adopt
currently unused strategies, involve a different mechanism
that we do not analyze here.

\section{Discussion of Modeling Choices}\label{sec:discussions}

\subsection{Stochastic Performances}\label{subsec:stochasticachievement}

So far, we have assumed that the performance of each behavioral rule is deterministic. A natural generalization is to allow \emph{stochastic performances} by treating each $u_k(R_i)$ as an independent random variable. In this case, agents imitate based on realized performances. The evolution equation~\eqref{eqn:evolution} continues to govern the aggregate dynamics, provided the pairwise comparison matrix is redefined as
\[
W^{F}_{ij}=\sum_{k=1}^K \pi_k \mathbb{E}\big[F_k(u_k(R_j),u_k(R_i))-F_k(u_k(R_i),u_k(R_j))\big].
\]
All characterization results in Section~\ref{sec:evolution} continue to hold under this redefined matrix. Nevertheless, with stochastic performances, imitate-if-better need not minimize expected envy as in Proposition~\ref{prop:myopic_comparison_loss}: as illustrated below, a lucky performance realization can induce agents to adopt a rule that generates greater expected loss in the subsequent comparison. 

\begin{example}
	Consider a single context $k$ with two rules and the binary
	loss function $\ell(x)=\mathbf{1}\{x>0\}$. Suppose $R_1$ has a
	deterministic performance of $1$, whereas $R_2$ yields $0$ or
	$2$ with equal probability $1/2$. At the population state
	$\mu=(1/2,1/2)$, the expected loss of using $R_1$ is
	$L(R_1\mid k,\mu)=\mu(2)\mathbb{P}\big(1<u_k(R_2)\big)=1/4$,
	while the expected loss of using $R_2$ is
	$L(R_2\mid k,\mu)=\mathbb{P}\big(u_k(R_2)=0\big)\big[\mu(1)\mathbb{P}\big(0<u_k(R_1)\big)+\mu(2)\mathbb{P}\big(0<u_k(R_2)\big)\big]=3/8$.
	Thus, $R_2$ generates a strictly greater expected loss than $R_1$.
	
	Under imitate-if-better, an agent who samples a peer using the
	other rule switches with probability $1/2$, leaving the
	population shares unchanged. Consider instead the alternative
	improving protocol $\widetilde F_k(u,v)=\mathbf{1}\{u\le 0 <v\}$,
	under which an agent imitates a better-performing peer if and
	only if her own performance is zero. Under $\widetilde F$, no
	$R_1$-user switches, whereas an $R_2$-user switches to $R_1$
	whenever her realized performance is $0$ and she samples an
	$R_1$-user. This strictly reduces the share of $R_2$-users, so
	$\widetilde F$ achieves a strictly smaller aggregate expected
	loss than imitate-if-better. 
\end{example}

\subsection{Universal Sampling}\label{subsec:universalsampling}

Another assumption we have maintained is that each agent samples a peer from her own context. An alternative is \emph{universal sampling}, under which each agent samples a peer uniformly at random from the entire population. Maintaining the assumption that outcomes are not compared across contexts, an agent imitates a sampled peer only if that peer faces the same context.

This modification changes only the probability of encountering a comparable peer. In context $k$, an $R_i$-user now samples a comparable $R_j$-user with probability $\pi_k\mu_t(j)$ rather than $\mu_t(j)$. Thus, the evolution equation~\eqref{eqn:evolution} continues to govern the imitation dynamics, provided the pairwise comparison matrix is redefined as
\[
W^{F}_{ij}=\sum_{k=1}^K \pi_k^2 \big[F_k(u_k(R_j),u_k(R_i))-F_k(u_k(R_i),u_k(R_j))\big].
\]

Universal sampling effectively reweights each context from $\pi_k$ to $\pi_k^2$. In the social choice interpretation, while every context's voting weight decreases (slowing the overall pace of evolution), less frequent contexts lose more weight than more frequent ones. Consequently, universal sampling amplifies the relative influence of more frequent contexts and favors rules that perform well in them.

\subsection{Across-Context Imitation}\label{subsec:acrosscontextimitation}

Treating performances as comparable across contexts, we now discuss across-context imitation under universal sampling. Such imitation can arise when contexts are interpreted as unobservable fundamentals that drive the performance of each decision.

Across-context imitation reduces to the study of stochastic performances of Section~\ref{subsec:stochasticachievement}. When context labels are ignored and the imitation protocol is context-independent with identical $F_k$ over $k$, the comparison between rules $R_i$ and $R_j$ is governed by the random performance pair $(u_k(R_i), u_{k'}(R_j))$, where $(k,k')$ is drawn from the product distribution $\pi\otimes\pi$. The resulting imitation dynamics are therefore still governed by a redefined pairwise comparison matrix. Thus, for the same reason, imitate-if-better no longer minimizes expected envy as in Proposition~\ref{prop:myopic_comparison_loss}.

Across-context imitation can substantially change the imitation dynamics, as the following example illustrates.

\begin{example}
	Consider two rules, $\alpha$ and $\beta$, and three
	equiprobable contexts in which the performance pairs for
	$\alpha$ and $\beta$ are $(0,5)$, $(3,1)$, and $(4,2)$.
	Under imitate-if-better with within-context comparisons,
	$\alpha$ outperforms in two of the three contexts, so the dynamics globally
	converge to a consensus on $\alpha$. With across-context
	comparisons, $\beta$ performs better in five of the nine equally likely comparisons, among $(0,5)$, $(0,1)$, $(0,2)$, $(3,5)$, $(3,1)$, $(3,2)$, $(4,5)$, $(4,1)$, and $(4,2)$, so the
	dynamics globally converge to a consensus on $\beta$ instead.
\end{example}

\subsection{Continuum versus Discrete}\label{subsec:continuumpopulation}

Our analysis relies on a continuum population and discrete time. The continuum-population formulation is critical for obtaining a credible analysis of global dynamics. With finitely many agents, there are only finitely many population states, and the imitation process evolves as an absorbing Markov chain. Consequently, the dynamics must converge almost surely. However, such convergence is often driven by extremely rare events, such as all agents being assigned to the same context. The continuum-population formulation allows us to abstract away from the influence of these rare events and conduct a more robust long-run analysis.

The discrete-time formulation prevents agents from continuously updating their behavioral rules, making it more biologically plausible \citep{AkinLosert1984}, albeit at the cost of analytical tractability \citep{sandholm2005negative}. In the short run, with sufficiently small imitation intensities, the discrete-time dynamics closely approximate their continuous-time counterparts. Nevertheless, their long-run behavior can differ substantially. For instance, Proposition~\ref{prop:boundary-approach} shows that non-convergent discrete-time trajectories must approach the boundary of the population simplex, whereas their continuous-time counterparts can remain bounded away from it.

\section{Concluding Remarks}\label{sec:concluding}

Employing an imitation framework, we study which behavioral rules prevail in an environment with multiple decision contexts. The key restriction is that the performances of behavioral rules are compared within each decision context and the learned rules are applied across contexts. We observe that multiple contexts turn the imitation dynamics into a context-weighted social choice problem. Building on this observation, we characterize the global imitation dynamics, illustrate the framework through an application to belief evolution, and examine the role of ordinal imitation in modeling and its impact on long-run dynamics.

Two directions merit further exploration. First, our analysis assumes that all agents share the same imitation protocol and the same perception of decision contexts. Allowing agents to differ in how they imitate peers or judge comparability would capture environments with heterogeneous imitative traditions and cognitive abilities. Second, our general analysis does not directly characterize the substantive features of prevailing behavioral rules. Addressing this question requires studying more structured environments, as we illustrate in our belief-evolution application.

\addcontentsline{toc}{section}{References}

\begin{singlespace}
	\putbib[RS]
\end{singlespace}

\appendix

\section{Proofs of Main Results}\label{appendix:main-proofs}

Throughout this appendix, we fix $F$ and write $W=W^F$, $E=E^F$, and $J=\mathcal R\setminus E$. Some of the proofs below rely on auxiliary results in the Online Appendix. 

\subsection{Proof of Theorem~\ref{thm:consensus-iff-condorcet}} \label{appendix:consensus-iff-condorcet}

Suppose first that $R_i$ is a Condorcet winner, and let $\delta=\min_{j\neq i}W_{ij}>0$. Then, by \eqref{eqn:evolution},
$\mu_{t+1}(i)-\mu_t(i)=\mu_t(i) \sum_{j\neq i} W_{ij}\mu_t(j)\geq\delta\mu_t(i)(1-\mu_t(i))>0.$ 
Thus, $\mu_t(i)$ strictly increases to some limit $\bar{\mu}\in(0,1]$. Taking $t\rightarrow\infty$, we have $0\geq\delta\bar{\mu}(1-\bar{\mu})$, which implies $\bar{\mu}=1$.

Conversely, suppose $\mu_t$ converges to $e_i$, the $i$-th standard basis vector. If $W_{ji}>0$ for some $j$, then $(W\mu_t)(j)-(W\mu_t)(i)$ converges to $W_{ji}>0$. Lemma~\ref{lem:relative-growth}(2) then implies that the ratio $\mu_t(i)/\mu_t(j)$ eventually decreases, contradicting the assumption that $\mu_t \to e_i$. Thus, we must have $W_{ij}\geq0$ for every $j\neq i$. Partition the remaining rules into $A=\{j\neq i: W_{ij}>0\}$ and $B=\{j\neq i: W_{ij}=0\}$. 

For each $j\in A$, \eqref{eqn:evolution} and $\mu_t\to e_i$ imply $\mu_{t+1}(j)/\mu_t(j)\to 1+W_{ji}=1-W_{ij}<1$. Thus, $\mu_t(j)$ is eventually bounded above by a geometrically decreasing sequence, implying $\sum_{t=1}^{\infty}\mu_t(j)<\infty$. Summing over all $j\in A$, we have $\sum_{t=1}^{\infty}\mu_t(A)<\infty$.

Suppose $B$ is nonempty, and let $b_t=\mu_t(B)$. By Lemma \ref{lem:relative-growth}, $b_t>0$. We sum \eqref{eqn:evolution} over $j\in B$. Because $W_{ji}=0$ for $j \in B$, we have
\[b_{t+1}-b_t =\sum_{j\in B}\sum_{\ell\neq i}W_{j\ell}\mu_t(j)\mu_t(\ell)=\sum_{j\in B}\sum_{\ell\in A}W_{j\ell}\mu_t(j)\mu_t(\ell) \geq-\sum_{j\in B}\sum_{\ell\in A}\mu_t(j)\mu_t(\ell) =-b_t\mu_t(A),\]
where the second equality relies on the antisymmetry of $W$, and the inequality follows from each $W_{j\ell}\ge -1$. Thus, $b_{t+1}/b_t\ge 1-\mu_t(A)$ for all $t$. Since $\mu_t(A)$ is summable over $t$, we can choose $T$ such that $\mu_t(A)\leq 1/2$ for all $t\geq T$. Using the inequality $\log(1-x)\geq-2x$ for $0\leq x\leq 1/2$, we obtain, for any $t\geq T$, $b_t \geq b_T\prod_{s=T}^{t-1}[1-\mu_s(A)] \geq b_T\exp\!\left(-2\sum_{s=T}^{\infty}\mu_s(A)\right)>0.$ Thus, the total share of rules tied with $R_i$ remains bounded away from zero, contradicting $\mu_t\to e_i$. Hence, $B$ must be empty, meaning $R_i$ is a Condorcet winner.

\subsection{Proof of Proposition~\ref{prop:stationary-convergence}}\label{appendix:stationary-convergence}

Fix $F$ and write $W=W^F$, $E=E^F$, and $J=\mathcal R\setminus E$.
As in the model, initial distributions belong to $\Delta_+(\mathcal R)$.
Recall that $p$ is a maximal lottery if and only if $Wp\leq0$ and
$p\in\Delta(\mathcal R)$. Antisymmetry gives $p^\top Wp=0$, so
$(Wp)(i)=0$ whenever $p(i)>0$.

\emph{Statement 1.}
For each $j\in J$, we first construct a maximal lottery $p^j$ satisfying $(Wp^j)(j)<0$. Every $x\geq0$ satisfying $Wx\leq0$ has $x(j)=0$,
since otherwise normalizing $x$ would give a maximal lottery assigning
positive mass to $j$. Apply Farkas' lemma to the $N$-by-$2N$ matrix $(I,W)$
and the vector $-e_j$. The inequalities $(I,W)^\top x\geq0$ and
$(-e_j)^\top x<0$ would give $x\geq0$, $Wx\leq0$, and $x(j)>0$, which we argued is impossible. Hence, for some $u,v\geq0$, we have $u+Wv=-e_j$. 

Since $Wv=-e_j-u$, we have $Wv\leq0$ and $(Wv)(j)\leq-1$.
In particular, $v\neq0$, so normalizing $v$ gives a maximal
lottery $p^j$ satisfying $(Wp^j)(j)<0$. For each $i\in E$, choose a maximal lottery $q^i$ with $q^i(i)>0$.
Set $p=N^{-1}\bigl(\sum_{i\in E}q^i+\sum_{j\in J}p^j\bigr)$,
where the second sum is zero if $J$ is empty. As an average of
maximal lotteries, $p$ is a probability vector satisfying $Wp\leq0$.
Every lottery in the average is supported on $E$, and each $i\in E$
receives positive mass from $q^i$. Hence $\supp(p)=E$.
For each $j\in J$, every lottery in the average contributes a
nonpositive term to $(Wp)(j)$, and $p^j$ contributes a strictly
negative term. Thus $p\in L^F$.

\emph{Statement 2.}
Suppose $\mu_t\to p$. Write $S=\supp(p)$ and
$V=\{j:(Wp)(j)<0\}$. Lemma~\ref{lem:estimateofD} in the online
appendix gives $V\subseteq S^c$. For any $j\in S^c\setminus V$, as $(Wp)(j)\ge 0$ because $j\notin V$ and $(Wp)(j)\le 0$ by Lemma \ref{lem:estimateofD}, we have 
$(Wp)(j)=0$. Fix $j\in S^c\setminus V$. For every $k\in V$, by Lemma \ref{lem:relative-growth},
\[
\frac{\mu_{t+1}(k)/\mu_{t+1}(j)}{\mu_{t}(k)/\mu_{t}(j)}
=\frac{1+(W\mu_t)(k)}{1+(W\mu_t)(j)}
\longrightarrow 1+(Wp)(k)<1.
\]
Hence, $\mu_t(k)/\mu_t(j)\to0$, which implies $\mu_t(V)/\mu_t(j)\to 0$. But Lemma~\ref{lem:estimateofD} gives
$\mu_t(j)\le \mu_t(S^c)\leq2\mu_t(V)$
for all sufficiently large $t$. Contradiction. Thus, we must have $V=S^c$. For any maximal lottery $q$, we have
$0\geq p^\top Wq=-(Wp)^\top q\geq0$, so $q(j)=0$ for every
$j\in V=S^c$. Since $p$ is itself a maximal lottery, $S=E$ and $p\in L^F$.

Conversely, fix $p\in L^F$. If $J$ is empty, then $\mu_1=p$ is
interior and generates a stationary trajectory. If $J$ is nonempty,
Lemma~\ref{lem:attainability} in the online appendix gives an interior initial distribution
whose trajectory converges to $p$.

\emph{Statement 3.}
Suppose $\mu_t\to p$ and write $D_t=D_{\mathrm{KL}}(p\|\mu_t)$.
If $J$ is empty, then $Wp=0$. By \eqref{eqn:differenceD} in the online appendix, $D_{t+1}\geq D_t$ for every $t$. Since $D_t\geq0$ and $D_t\to0$,
we have $\mu_t=p$ for every $t$.

Now suppose $J$ is nonempty. Write $b_t=\mu_t(J)$ and
$\alpha=\max_{j\in J}[1+(Wp)(j)]<1$. By \eqref{eqn:evolution}, we have
$\mu_{t+1}(j)/\mu_t(j)=1+(W\mu_t)(j)\rightarrow 1+(Wp)(j)$.
Thus, $\mu_t(j)^{1/t}\to1+(Wp)(j)$. As $J$ is finite, the asymptotic rate is governed by the largest element. That is, $b_t^{1/t}\to \alpha$. For any fixed $\rho\in(\alpha,1)$, there exists a $T$ such that for any $t\ge T$ $\mu_{t+1}(j)/\mu_t(j)<\rho$ holds for all $j\in J$. Therefore, 
$b_{t+1}/b_t<\rho$ for any $t\ge T$. Consequently, $\sum_{t=T}^{\infty}b_t\leq b_T/(1-\rho)$.

It remains to show that the error on $E$ is bounded by a constant
times $b_t=\mu_t(J)$. Write $r_t=\mu_t-p$, $h_t=(r_t)_E$, and
$D_t=D_{\mathrm{KL}}(p\|\mu_t)$. Pinsker's inequality gives
$\norm{r_t}_1^2\leq 2D_t$, and Lemma~\ref{lem:estimateofD} in
the online appendix gives $D_t\leq 2b_t$ for all sufficiently
large $t$. Hence $\norm{r_t}_1^2\leq 4b_t$ for all sufficiently large $t$.

Adopting the convention in Lemma \ref{lem:limit-linearization} in the online appendix, we let $P=\operatorname{diag}(p(i):i\in E)$, $Q=I+PW_{E,E}$, and
$\norm{h}_P=\norm{P^{-1/2}h}_2$. For each $i\in E$, we have $(Wp)(i)=0$ and hence
$(W\mu_t)(i)=(Wr_t)(i)$. Using $Q=I+PW_{E,E}$, one can show
\[
	h_{t+1}(i)-(Qh_t)(i)= p(i)\sum_{j\in J}W_{ij}\mu_t(j)+r_t(i)(Wr_t)(i).
\]

Since $|W_{ij}|\leq 1$, the first term has absolute value at most
$p(i)b_t$ and the second at most $|r_t(i)|\norm{r_t}_1$.
Summing over $i\in E$ gives
$\norm{h_{t+1}-Qh_t}_1\leq b_t+\norm{r_t}_1^2\leq 5b_t$ for all sufficiently large $t$. Since finite-dimensional norms are equivalent, $\norm{h_{t+1}-Qh_t}_P\leq Cb_t$
for some $C>0$ and all sufficiently large $t$. Statement~1 of Lemma~\ref{lem:limit-linearization} in the online
appendix gives $\norm{h_t}_P\leq\norm{Qh_t}_P$. Hence
$\norm{h_t}_P\leq\norm{h_{t+1}}_P+Cb_t$. Iterating this inequality
and using $h_{t+n}\to0$ as $n\to\infty$ gives
$\norm{h_t}_P\leq C\sum_{s=t}^\infty b_s\leq Cb_t/(1-\rho)$, for sufficiently large $t$. 

Since $p(j)=0$ for $j\in J$, $\norm{\mu_t-p}_1=b_t+\norm{h_t}_1\geq b_t$.
For the upper bound, Cauchy--Schwarz and $\sum_{i\in E}p(i)=1$ give
$$\norm{h_t}_1
=\sum_{i\in E}\sqrt{p(i)} \frac{|h_t(i)|}{\sqrt{p(i)}}
\leq \left(\sum_{i\in E}p(i)\right)^{1/2}
\left(\sum_{i\in E}\frac{h_t(i)^2}{p(i)}\right)^{1/2}
=\norm{h_t}_P.$$
Using the preceding estimate $\norm{h_t}_P\leq Cb_t/(1-\rho)$,
we obtain $b_t\leq\norm{\mu_t-p}_1\leq [1+C/(1-\rho)]b_t$
for all sufficiently large $t$. Taking $t$-th roots and using
$b_t^{1/t}\to\alpha$, we obtain $\norm{\mu_t-p}_1^{1/t}\to\alpha$.
Taking logarithms gives the stated rate, with $\log 0=-\infty$.

\subsection{Proof of Theorem~\ref{thm:weak-imitation-convergence}}\label{appendix:weak-imitation-convergence}

By Proposition~\ref{prop:stationary-convergence}, choose $p\in L^F$.
By definition, $p$ has support $E$, $(Wp)(i)=0$ for $i\in E$, and
$(Wp)(j)<0$ for $j\in J$.

\emph{The equivalence within Statement 1.}
If $W_{E,E}\neq0$, take $i,j\in E$ with $W_{ij}>0$.
Since $p(i)>0$ and $0=(Wp)(j)=\sum_{k=1}^N W_{jk}p(k)=\sum_{k\in E} W_{jk}p(k)$, the negative term $W_{ji}p(i)$
must be offset by a positive term $W_{jk}p(k)$ for some $k\in E$.
Thus every strict comparison in $E$ can be continued by another
strict comparison. Finiteness of $E$ gives a Condorcet cycle.
Conversely, $W_{E,E}=0$ rules out any such cycle.

\emph{Statement 1 implies Statements 2 and 3.}
Suppose $W_{E,E}=0$. If $J$ is empty, then $W=0$ and every trajectory
is stationary. Otherwise, set $\bar\lambda=\frac12\min_{j\in J}[-(Wp)(j)]$,
so $0<\bar\lambda\leq1/2$. Consider a trajectory under $\lambda F$,
where $0<\lambda\leq1$, and write $b_t=\mu_t(J)$ and
$D_t=D_{\mathrm{KL}}(p\|\mu_t)$. Since $b_t=\mu_t(J)=1-\mu_t(E)$ and $1-u\leq-\log u$ for $u>0$,
we have $b_t\leq-\log\mu_t(E)$.  By Jensen's
inequality, $D_t=-\sum_{i\in E}p(i)\log\frac{\mu_t(i)}{p(i)}
\geq-\log\left(\sum_{i\in E}p(i)\frac{\mu_t(i)}{p(i)}\right)
=-\log\mu_t(E)\ge b_t.$ Since $W_{E,E}=0$ and $|W_{ij}|\leq1$,
we have $|(W\mu_t)(i)|\leq b_t$ for $i\in E$. Thus,
$p^\top W\mu_t=-(Wp)^\top\mu_t\geq2\bar\lambda b_t$.

Whenever $\lambda b_t\leq\bar\lambda$, we can apply
$\log(1+u)\geq u-u^2$ for $|u|\leq1/2$ to $D_{t+1}-D_t=-\sum_{i\in E}p(i)\log[1+\lambda(W\mu_t)(i)]$ and obtain
\begin{equation}\label{eqn:Dincrement}
	D_{t+1}-D_t\leq
	-\lambda\sum_{i\in E}p(i)(W\mu_t)(i)+\lambda^2\sum_{i\in E}p(i)[(W\mu_t)(i)]^2\leq-2\lambda\bar\lambda b_t+\lambda^2b_t^2\leq-\lambda\bar\lambda b_t.
\end{equation}
For Statement 2, fix $\lambda<\bar\lambda$. Since $b_t\leq1$, we
have $\lambda b_t\leq\lambda<\bar\lambda$ at every date. The bound \eqref{eqn:Dincrement} therefore holds from every interior initial distribution. For Statement 3, set $\lambda=1$ and consider an initial
distribution with $D_1<\bar\lambda$. Whenever $D_t<\bar\lambda$,
we have $b_t\leq D_t<\bar\lambda$, so the bound \eqref{eqn:Dincrement} applies and gives
$D_{t+1}\leq D_t<\bar\lambda$. By induction, the same bound holds at every date. In both cases, \eqref{eqn:Dincrement} holds at every date. Summing
from $t=1$ to $T$ gives
$\lambda\bar\lambda\sum_{t=1}^{T}b_t\leq D_1-D_{T+1}\leq D_1$.
Since this holds for every $T$, we have $\sum_{t\geq1}b_t<\infty$.

We now show that each rule's share converges. For $i\in J$,
we have $\mu_t(i)\leq b_t\to0$. For $i\in E$, the evolution
equation gives $|\mu_{t+1}(i)-\mu_t(i)|
=\lambda\mu_t(i)|(W\mu_t)(i)|\leq\lambda b_t$.
Hence, for any $s>t$, we have
$|\mu_s(i)-\mu_t(i)|\leq\lambda\sum_{r=t}^{s-1}b_r
\leq\lambda\sum_{r=t}^{\infty}b_r$.
The last bound tends to zero as $t\to\infty$, so $(\mu_t(i))_{t\geq1}$
is Cauchy and converges. Therefore, $\mu_t$ converges in both cases.

It remains to show that a positive-measure set of
interior initial distributions satisfies $D_1<\bar\lambda$ for Statement 3. As $\mu_1$ approaches $p$ from the interior, each term
$p(i)\log[p(i)/\mu_1(i)]$ with $p(i)>0$ tends to zero.
Thus $D_1=D_{\mathrm{KL}}(p\|\mu_1)<\bar\lambda$ for every
interior $\mu_1$ sufficiently close to $p$. These initial
distributions contain a nonempty open ball in the affine hull
of the simplex and hence have positive Lebesgue measure.

\emph{Statement 3 implies Statement 1.}
Suppose $W_{E,E}\neq0$. For each $p\in L^F$, Lemma~\ref{lem:appendix-local-null} gives an open
neighborhood $\mathcal U_p$ such that the interior initial distributions
whose trajectories remain in $\mathcal U_p$ from date $m+1$ onward
form a null set for each $m\geq0$. Choose a countable subcover
$\{\mathcal U_{p_k}\}_{k\geq1}$ of $L^F$. By
Proposition~\ref{prop:stationary-convergence}, every convergent interior
trajectory has a limit in $L^F$. The limit belongs to some
$\mathcal U_{p_k}$, so the trajectory remains in that neighborhood from
some finite date onward. The set of interior initial distributions
leading to convergence is therefore contained in a countable union
of null sets, contradicting Statement 3.

\emph{Statement 2 implies Statement 1.}
Choose $\lambda\in(0,\bar\lambda)$ as in Statement 2. Statement 3
holds for $\lambda F$ because every interior initial distribution
leads to convergence under this protocol. Applying the implication
just proved to $\lambda F$ gives
$W^{\lambda F}_{E^{\lambda F},E^{\lambda F}}=0$.
Since $W^{\lambda F}=\lambda W$, the maximal lotteries and their
essential set are the same under $F$ and $\lambda F$.
Thus $\lambda W_{E,E}=0$, which proves Statement 1.

\subsection{Proof of Proposition~\ref{prop:REwins}}\label{appendix:REwins}

In Online Appendix~\ref{appendix:REsharp}, we prove a stronger result from which Proposition~\ref{prop:REwins} follows as a corollary. For completeness, we provide a proof sketch for Proposition \ref{prop:REwins} here.

For one uniformly drawn competing belief $\hat\pi$, let
$\phi(\hat\pi)=\sum_{k=1}^K\pi_k\mathbf1\{\hat\pi_k>\pi_k\}$.
	Since coordinate ties have probability zero, $\pi$ defeats
	$\hat\pi$ if and only if $\phi(\hat\pi)<1/2$.
	Lemma~\ref{lem:uniform-belief-moments} in the Online Appendix, applied with
	$\hat\pi^*=\pi$ and $b=\pi$, gives
	$\mathbb E[\phi(\hat\pi)]\leq(1-1/K)^{K-1}$ and 
	$\operatorname{Var}(\phi(\hat\pi))\leq 1/K.$
	Since $(1-1/K)^{K-1}<1/2$ for $K\geq3$, Chebyshev's
	inequality implies
	\[\Pr(\phi(\hat\pi)\ge 1/2)\le \frac{\operatorname{Var}(\phi(\hat\pi))}{(1/2-E[\phi(\hat\pi)])^2}\le \frac{1}{K\bigl(1/2-(1-1/K)^{K-1}\bigr)^2}.\]
	Since $\pi$ is not a Condorcet winner if and only if $\pi$ fails to defeat at least one of the $N-1$ randomly drawn beliefs, the union bound implies the probability that $\pi$ is not a Condorcet winner is bounded from above by $\frac{N-1}{K\bigl(1/2-(1-1/K)^{K-1}\bigr)^2}$. Since $(1-1/K)^{K-1}\to e^{-1}<1/2$ as $K$ goes to infinity, this probability tends to zero under a bound that is independent of the choice of objective distribution $\pi$.

\subsection{Proof of Proposition~\ref{prop:NoCondorcet}}\label{appendix:NoCondorcet}

We first state a lemma establishing a uniform lower bound on the probability that a random belief defeats a given candidate. Its proof is given in Online Appendix~\ref{appendix:uniform-defeat}.

\begin{lemma}\label{lem:appendix:uniform-defeat}
		Fix $K\geq3$ and $\pi\in\Delta_+(\mathcal K)$ with $\max_k\pi_k<1/2$. For any fixed belief $\hat\pi^*\in\Delta(\mathcal K)$ and $c_K=K^{2-K}$, a uniformly drawn belief $\hat\pi$ satisfies
		$\Pr\bigl(\hat\pi\text{ defeats }\hat\pi^*\bigr)\geq c_K.$
\end{lemma}
Now, we fix $K$ and $\pi$ as in the proposition. First, suppose that all $N\geq2$ beliefs
$\hat\pi^1,\ldots,\hat\pi^N$ are drawn independently and
uniformly from $\Delta_+(\mathcal K)$, and we call the belief set $\mathcal{R}_N$. By Lemma \ref{lem:appendix:uniform-defeat}, for any fixed $\hat\pi^i$, each of the other
$N-1$ beliefs defeats it with probability at least $c_K$. Thus,
the probability that $\hat \pi^i$ is the Condorcet winner of $\mathcal{R}_N$ is no larger than $(1-c_K)^{N-1}$. Applying the union bound across all $i=1,\dots,N$, we have that the probability that the belief set $\mathcal{R}_N$ contains a Condorcet winner is bounded from above by $N(1-c_K)^{N-1}$, which converges to zero as $N$ goes to infinity.

Next, suppose
$\mathcal R_N=\{\pi,\hat\pi^1,\ldots,\hat\pi^{N-1}\}$,
where the $N-1$ other beliefs are drawn independently
and uniformly from $\Delta_+(\mathcal K)$. Again by Lemma \ref{lem:appendix:uniform-defeat}, each random belief defeats $\pi$ with
probability at least $c_K$. Independence implies that the probability that $\pi$ is the Condorcet winner of $\mathcal{R}_N$ is no larger than $(1-c_K)^{N-1}$. Moreover, for each fixed $\hat\pi^i$ with $i=1,\ldots,N-1$, each other random belief defeats $\hat\pi^i$ with
probability at least $c_K$. Thus, the probability that $\hat\pi^i$ is the Condorcet winner of $\mathcal{R}_N$ is no larger than $(1-c_K)^{N-2}$. Applying the union bound, we know the probability  that the belief set $\mathcal{R}_N$ contains a Condorcet winner is bounded from above by $(1-c_K)^{N-1}+(N-1)(1-c_K)^{N-2}$, which also converges to zero as $N$ goes to infinity.

In both cases, with probability one, no pair of beliefs assigns
the same probability to any state. A pairwise tie would
therefore require the states in which one belief outperforms
the other to have total objective probability $1/2$,
which is excluded by assumption. Hence, with probability
one, there are no pairwise ties. In such cases, the nonexistence of a Condorcet winner is equivalent to the existence of a Condorcet cycle within the essential set. Thus, in both cases, the probability that a Condorcet cycle exists within the essential set is arbitrarily close to one when there are sufficiently many beliefs, and the bound is independent of the choice of the true distribution $\pi$.

\subsection{Proof of Proposition~\ref{prop:FOSD_belief_evolution}}
\label{appendix:FOSD_belief_evolution}

\emph{Statement 1.}
Suppose $\pi_M+\min\{\pi_1,\pi_K\}>1/2$, and let
$\hat\pi^*$ be the unique belief assigning the highest
probability to state $M$. If $\pi_M>1/2$, then $\hat\pi^*$
is the Condorcet winner because it outperforms every other
belief in state $M$.

Otherwise, $M$ is an interior state. Fix any other belief
$\hat\pi^j\in\mathcal R$. Since the beliefs are FOSD-ranked and no two beliefs assign the same probability to any state,
either $\hat\pi^*\succeq_{\mathrm{FOSD}}\hat\pi^j$, which
implies $\hat\pi_K^*>\hat\pi_K^j$, or
$\hat\pi^j\succeq_{\mathrm{FOSD}}\hat\pi^*$, which implies
$\hat\pi_1^*>\hat\pi_1^j$. Thus, $\hat\pi^*$ outperforms $\hat\pi^j$
in state $M$ and at least one extreme state, whose combined
objective probability is at least
$\pi_M+\min\{\pi_1,\pi_K\}>1/2$.
Therefore, $\hat\pi^*$ is the unique Condorcet winner.
Theorem~\ref{thm:consensus-iff-condorcet} implies convergence
to consensus on this belief.

\noindent \emph{Statement 2.} The proof uses the following lemma, proved in Online Appendix~\ref{appendix:FOSD-local-construction}.

\begin{lemma}\label{lem:FOSD-local-construction}
	Let $\pi\in\Delta_+(\mathcal K)$ and $1<M<K$.
	For every $\delta>0$, there exist beliefs
	$\hat\pi^P,\hat\pi^O\in\Delta_+(\mathcal K)$ satisfying
	$\hat\pi^O\succeq_{\mathrm{FOSD}}\pi
	\succeq_{\mathrm{FOSD}}\hat\pi^P$,
	$d(\hat\pi^P,\pi)<\delta$, and
	$d(\hat\pi^O,\pi)<\delta$, with the following
	state-by-state rankings: for $k<M$, $\hat\pi_k^P>\pi_k>\hat\pi_k^O$; for $k=M$, $\pi_k>\hat\pi_k^O>\hat\pi_k^P$; for $M<k<K$, $\hat\pi_k^O>\hat\pi_k^P>\pi_k$; for $k=K$, $\hat\pi_k^O>\pi_k>\hat\pi_k^P$.
\end{lemma}

Suppose $\pi_M+\min\{\pi_1,\pi_K\}<1/2$, and fix
$\delta>0$. By reversing the state indices and
interchanging the labels $P$ and $O$ if necessary,
we may assume $\pi_K\leq\pi_1$. Then
$\pi_M+\pi_K<1/2$, and $1<M<K$.
Choose $\hat\pi^P$ and $\hat\pi^O$ as in
Lemma~\ref{lem:FOSD-local-construction}. By the rankings in the lemma, $\hat\pi^P$ defeats $\pi$
because
$\sum_{k:\hat\pi_k^P>\pi_k}\pi_k
=1-\pi_M-\pi_K>1/2$;
$\pi$ defeats $\hat\pi^O$ because
$\sum_{k:\pi_k>\hat\pi_k^O}\pi_k
=\sum_{k=1}^{M}\pi_k>1/2$;
and $\hat\pi^O$ defeats $\hat\pi^P$ because
$\sum_{k:\hat\pi_k^O>\hat\pi_k^P}\pi_k
=\sum_{k=M}^{K}\pi_k>1/2$.
The last two inequalities follow from the definition
of the median state. Hence, these beliefs form a
Condorcet cycle. Thus, Proposition~\ref{prop:heteroclinic-cycle}
implies that the imitation dynamics do not converge from
any interior initial distribution other than the unique
interior stationary distribution.

\subsection{Proof of Theorem \ref{thm:universalconvergencecharacterization}} \label{appendix:universalconvergencecharacterization}

We use the following lemma for the necessity of additivity. Its proof is given in Online Appendix~\ref{appendix:additivity-cycle}.

\begin{lemma}\label{lem:additivity-cycle}
		Fix a compact interval $U=[\underline u,\overline u]$ containing more than one point, a finite context set $\mathcal K$ with $|\mathcal K|\geq3$, and an improving protocol $F$ such that each $F_k$ is not identically zero
		on $U\times U$. If additivity fails in some context, there exist three rules with performance functions taking values in $U$ and a distribution $\pi\in\Delta_+(\mathcal K)$ such that $W^F_{12}=W^F_{23}=W^F_{31}\neq0$. Consequently, $E^F=\mathcal R$.
\end{lemma}

Now, we prove the theorem. \emph{Statements 2 and 3 are equivalent.} It is evident that Statement 2  implies Statement 3. Conversely, suppose Statement 3 holds. Fix a context $k$
and define $g_k(u)=F_k(\underline u,u)\in[0,1]$ for each $u\in U$.
Since $F$ is improving, we have $g_k(\underline u)=0$. For $\underline u<u<v$, the additivity of $F_k$ implies 
$F_k(\underline u,v)=F_k(\underline u,u)+F_k(u,v)$. Thus, we have $F_k(u,v)=g_k(v)-g_k(u)$.
If instead $u=\underline u$, the same conclusion follows
because $F_k(\underline u,v)=g_k(v)$ by definition and
$g_k(\underline u)=0$.
Hence, $F_k(u,v)=g_k(v)-g_k(u)$ whenever $u<v$ in $U$. Because $F_k(u,v)\geq0$, this equality implies $g_k$ is weakly
increasing. Since $F_k(u,v)=0$ for
$v\leq u$, we obtain
$F_k(u,v)=(g_k(v)-g_k(u))^+$ for all $u,v\in U$.

\emph{Statement 2 implies Statement 1.}
Suppose $F_{k}(u,v)=(g_k(v)-g_k(u))^+$ for some weakly increasing function $g_k$. For each rule $R_i$, we define a score
$s_i=\sum_{k\in\mathcal K}\pi_k g_k(u_k(R_i))$.
Statement 2 implies $W_{ij}=s_i-s_j$.
Relabel the rules so that $s_1\geq\cdots\geq s_N$.
Thus, summing \eqref{eqn:evolution} for $i=1,\cdots,m$ gives
$\sum_{i=1}^m[\mu_{t+1}(i)-\mu_t(i)]
=\sum_{i=1}^m\sum_{j=1}^N
	(s_i-s_j)\mu_t(i)\mu_t(j)
= \sum_{i=1}^m\sum_{j=m+1}^N
(s_i-s_j)\mu_t(i)\mu_t(j)\ge 0.$
Hence, every cumulative share $\sum_{i\leq m}\mu_t(i)$
is nondecreasing and bounded above by one, so it converges. Since the total population share is always one, taking
differences of consecutive cumulative shares shows that
every individual share converges.

\emph{Statement 1 implies Statement 3.} We prove by contradiction. Suppose Statement 3 fails. Then, Lemma~\ref{lem:additivity-cycle} gives an environment with a Condorcet cycle within the essential set $E=\mathcal{R}$. By Theorem~\ref{thm:generic-nonconvergence}, the dynamics in this environment fail to converge generically, contradicting the global convergence required by Statement 1. Thus, Statement 1 implies Statement 3.

\end{bibunit}

\clearpage
\pagenumbering{arabic}
\renewcommand{\thepage}{OA-\arabic{page}}

\begin{bibunit}[abbrvnat]
	
\begin{center}
	\fontfamily{cmr}\selectfont
	{\Large\fontseries{bx}\selectfont
		Online Appendix to
		``Evolution of Behavioral Rules
		via Context-based Imitation''\par}
	\vspace{14pt}
	{\large Enrique Urbano Arellano and Xinyang Wang\par}
\end{center}

\section{Supplementary Proofs and Results}
\label{appendix:online-proofs}

This appendix contains the proofs of the auxiliary lemmas stated in Appendix~\ref{appendix:main-proofs} and the remaining omitted proofs. Throughout the online appendix, we fix $F$, and write $W=W^F$, $E=E^F$, $S=S^F$, and $J=\mathcal R\setminus E$.

\subsection{Proof of Lemma \ref{lem:relative-growth}}\label{appendix:relative-growth}

We prove $\mu_t\in\Delta(\mathcal{R})$ by induction. First, by assumption, $\mu_t$ is a probability vector for $t=1$. Suppose $\mu_t$ is a probability vector for some $t\ge 1$. Then, we have
\begin{align*}
	\sum_{i=1}^N \mu_{t+1}(i)
	&=
	\sum_{i=1}^N \mu_t(i)
	+
	\sum_{i=1}^N \mu_t(i)(W\mu_t)(i)
	=
	1 + \sum_{i,j=1}^N W_{ij}\mu_t(i)\mu_t(j)\\
	=&	1 + \frac{1}{2}\sum_{i,j=1}^N (W_{ij}+W_{ji})\mu_t(i)\mu_t(j)
	=
	1,
\end{align*}
where the last equation uses that $W$ is antisymmetric. Hence, $\mu_{t+1}$ is a probability vector. We next verify interiority of $\mu_t$. Note for any $i,j$, we have 
$|W_{ij}| \le \sum_{k=1}^K \pi_k|F_k(u_k(R_i),u_k(R_j))-F_k(u_k(R_j),u_k(R_i))|\le 1.$  Using $W_{ii}=0$, for any $i$ and $\mu_t\in\Delta_+(\mathcal{R})$, we have $1+(W\mu_t)(i)\geq1-\sum_{j\neq i} |W_{ij}|\mu_t(j)
\;\ge\;
1-\sum_{j\neq i} \mu_t(j)
=
\mu_t(i)>0.$ The evolution equation therefore gives $\mu_{t+1}(i)>0$ for every $i$. Thus, we have $\mu_t\in\Delta_+(\mathcal R)$ at every $t\ge 1$. Last, dividing the evolution equations for $i$ and $j$ gives \eqref{eq:ratio}, proving the second statement.

\subsection{Proof of Proposition \ref{prop:consensus-convergencerate}}\label{appendix:consensus-convergencerate}

Suppose $R_i$ is a Condorcet winner. Let $\delta=\min_{j\neq i}W_{ij}>0$. Then, $\mu_{t+1}(i)-\mu_t(i)\geq\delta\mu_t(i)(1-\mu_t(i))>0$. Thus, $\mu_t(i)$ strictly increases to one. For the rate of convergence, Lemma~\ref{lem:relative-growth} and the fact that $\mu_t\to e_i$ yield, for every $j\neq i$,
\[
\frac{\mu_{t+1}(j)/\mu_{t+1}(i)}
{\mu_t(j)/\mu_t(i)}
=\frac{1+(W\mu_t)(j)}{1+(W\mu_t)(i)}
\rightarrow \frac{1+W_{ji}}{1+W_{ii}}=1-W_{ij}.
\]
Therefore, $(\mu_{t}(j)/\mu_{t}(i))^{1/t}\to 1-W_{ij}$. Since $\mu_t(i)\to 1$, we have $(\mu_{t}(j))^{1/t}\to 1-W_{ij}$. Moreover, because $1-\mu_t(i)=\sum_{j\neq i}\mu_t(j)$, the asymptotic behavior of this sum is governed by its largest term. That is, $[1-\mu_t(i)]^{1/t}\to\max_{j\neq i}(1-W_{ij})=1-\delta$. Taking logarithms yields the stated rate of convergence.

\subsection{Lemmas for the Proof of Proposition \ref{prop:stationary-convergence}}

Recall that the Kullback--Leibler divergence from $p\in\Delta(\mathcal R)$ to $q\in\Delta_+(\mathcal R)$ is given by
$$D_{\mathrm{KL}}(p\|q)=\sum_{i\in \supp(p)}p(i)\log \frac{p(i)}{q(i)}.$$
The following lemma estimates the Kullback--Leibler divergence from the limiting distribution $p$ to the period-$t$ distribution $\mu_t$.

\begin{lemma}\label{lem:estimateofD}
	Fix an imitation protocol $F\in\mathcal F$ and write $W=W^F$. Suppose an interior trajectory converges
	to $p$. Let $S=\supp(p)$, $V=\{j:(Wp)(j)<0\}$, and
	$D_t=D_{\mathrm{KL}}(p\|\mu_t)$. Then $Wp\leq0$ and $(Wp)(i)=0$
	for every $i\in S$. Moreover, $\mu_t(S^c)\leq D_t$ for every $t$,
	and $D_t\leq2\mu_t(V)$ for all sufficiently large $t$.
\end{lemma}

\begin{proof}
	Taking limits in \eqref{eqn:evolution} gives $(Wp)(i)=0$ for every
	$i\in S$. If $(Wp)(j)>0$ for some $j\notin S$, then
	$\mu_{t+1}(j)/\mu_t(j)\geq1+(Wp)(j)/2$ for all sufficiently large $t$, contradicting
	$\mu_t(j)\to0$. Thus $Wp\leq0$.
	By Jensen's inequality, we have
	\[-D_t\leq\log\left(\sum_{i\in S}p(i)\frac{\mu_t(i)}{p(i)}\right)
	=\log\mu_t(S)\leq-\mu_t(S^c).\]
	That is, we have $\mu_t(S^c)\leq D_t$ for every $t$. To show the last statement, we note $D_t\to0$ because $\mu_t(i)\to p(i)>0$ for every $i\in S$. Using $\log(1+u)\leq u$ and the antisymmetry of $W$, we have
	\begin{equation}\label{eqn:differenceD}
		D_{t+1}-D_t
		=-\sum_{i\in S}p(i)\log[1+(W\mu_t)(i)]
		\geq-p^\top W\mu_t= (Wp)\cdot \mu_t 
		=\sum_{j\in V}(Wp)(j)\mu_t(j).
	\end{equation}
	For every $j\in V$, since $(W\mu_t)(j)\rightarrow (Wp)(j)<0$, we have $(W\mu_t)(j)\leq(Wp)(j)/2$ for all sufficiently large $t$. For all such sufficiently large $t$, we have
	$-(Wp)(j)\mu_t(j)\leq - 2 (W\mu_t)(j)\mu_t(j)=2[\mu_t(j)-\mu_{t+1}(j)],$ where the equality follows from the evolution equation \eqref{eqn:evolution}.
	By \eqref{eqn:differenceD},
	$D_t-D_{t+1}\leq 2[\mu_t(V)-\mu_{t+1}(V)]$.
	Summing from $t$ onward and using $D_s\to0$ and $\mu_s(V)\to0$
	gives $D_t\leq2\mu_t(V)$.
\end{proof}

Now, we extend the evolution map to $\mathbb R^N$ by defining a function $\Phi:\mathbb R^N\rightarrow \mathbb R^N$, where
\begin{equation}\label{eqn:Phi}
	\Phi(x)(i)=x(i)[1+(Wx)(i)].
\end{equation}
The antisymmetry of $W$ gives
$\sum_i\Phi(x)(i)=\sum_i x(i)$ for every $x\in\mathbb R^N$. The following lemma presents properties of the local linearization of $\Phi$ at the limit distribution $p$.
\begin{lemma}\label{lem:limit-linearization}
	Fix $p\in L^F$. Order the coordinates as $(E,J)$ and define
	$P=\operatorname{diag}(p(i):i\in E)$,
	$Q=I+PW_{E,E}$, and
	$U=\operatorname{diag}(1+(Wp)(j):j\in J)$. Then
	\[
	A:=D\Phi(p)=
	\begin{pmatrix}
		Q & PW_{E,J}\\
		0 & U
	\end{pmatrix}.
	\]
	The following statements hold.
	\begin{enumerate}
		\item Every diagonal entry of $U$ belongs to $[0,1)$, and $Q$
		is invertible. Moreover, $\norm{Q^{-1}h}_P\leq\norm{h}_P$ for
		every $h\in\mathbb R^{|E|}$, where
		$\norm{h}_P=\norm{P^{-1/2}h}_2$.
		
		\item Let $X_1=\{v:A^nv\to0\}$ and $X_2=\{v:v_J=0\}$.
		Then $\mathbb R^N=X_1\oplus X_2$, and both spaces are invariant
		under $A$. For every $z\in\mathbb R^{|J|}$, there is a unique $v\in X_1$ such that $v_J=z$. Moreover, $(Av)_J=Uv_J$ for every $v\in X_1$.
		
		\item There is a decomposition $X_2=X_2^c\oplus X_2^u$ where both spaces are invariant under $A$. The restriction of $A$ to $X_2^c$ is the identity. If $W_{E,E}\neq0$, then $\dim X_2^u\geq2$, and there exists $b>1$ such that $\norm{(Au)_E}_P\geq b\norm{u_E}_P$ for every $u\in X_2^u$. If $W_{E,E}=0$, then $X_2^u=\{0\}$.
	\end{enumerate}
\end{lemma}

\begin{proof}
	Since $p\in L^F$, we have $Wp\leq0$ and $\supp(p)=E$.
	Antisymmetry gives $p^\top Wp=0$. Since $p(i)>0$ for every
	$i\in E$, it follows that $(Wp)(i)=0$ for every $i\in E$.
	Note that $\Phi(x)(i)=x(i)[1+(Wx)(i)]=x(i)[1+\sum_{k} W_{ik}x(k)]$. Differentiating it with respect to $x(j)$ gives
	$$[D\Phi(x)]_{ij}=\delta_{ij}(1+(Wx)(i))+x(i)W_{ij}.$$
	Thus, $D\Phi(p)=I+\operatorname{diag}(Wp)+\operatorname{diag}(p)W$. Since $(Wp)_E=0$, the diagonal entries of $I+\operatorname{diag}(Wp)$ equal one on $E$. On $J$, they are $1+(Wp)(j)$, which gives the diagonal entries of the diagonal matrix $U$. For $\operatorname{diag}(p)W$, row $i$ of $W$ is multiplied by $p(i)$. Since $p_J=0$, every row indexed by $J$ becomes zero. The rows indexed by $E$ are multiplied by $P=\operatorname{diag}(p(i):i\in E)$. Adding the two matrices gives
	\[
	D\Phi(p)
	=
	\begin{pmatrix} I&0\\0&U\end{pmatrix}
	+
	\begin{pmatrix} PW_{E,E}&PW_{E,J}\\0&0\end{pmatrix}
	=
	\begin{pmatrix} Q&PW_{E,J}\\0&U\end{pmatrix},
	\]
	where \(Q=I+PW_{E,E}\).
	
	\emph{Statement 1.}
	For every $j\in J$, the definition of $L^F$ gives $(Wp)(j)<0$. Moreover, as $|W_{ji}|\leq1$, we have $(Wp)(j)\geq-1$. Thus $U_{jj}\in[0,1)$.
	To establish the claim for $Q$, let $K=P^{1/2}W_{E,E}P^{1/2}$.
	Then $P^{-1/2}QP^{1/2}=I+K$. Since $K$ is antisymmetric,
	$z^\top Kz=0$ for every $z\in \mathbb{R}^{|E|}$. Hence
	$\norm{(I+K)z}_2^2=\norm{z}_2^2+\norm{Kz}_2^2$ for every $z\in \mathbb{R}^{|E|}$.
	Taking $z=P^{-1/2}h$ gives $\norm{Qh}_P^2=\norm{(I+K)z}_2^2\geq\norm{z}_2^2=\norm{h}_P^2$.
	Thus $Qh=0$ implies $h=0$, so $Q$ is invertible. Applying the
	same inequality to $Q^{-1}h$ gives $\norm{Q^{-1}h}_P\leq\norm{h}_P$.
	
	\emph{Statement 2.}
	If $v\in X_1$, then $A^n(Av)=A^{n+1}v\to0$, so $Av\in X_1$.
	The block form gives $A(h,0)=(Qh,0)$, so $X_2$ is also invariant
	under $A$. If $(h,0)\in X_1\cap X_2$, then $Q^nh\to0$.
	By Statement 1, $\norm{Q^nh}_P\geq\norm{h}_P$ for every $n$,
	so $h=0$. Hence $X_1\cap X_2=\{0\}$.
	
	We next show that every $z\in\mathbb R^{|J|}$ is the vector of
	$J$-coordinates of some $v\in X_1$. Fix $j\in J$, let $e_j$ be
	the corresponding coordinate vector, and write $u_j=U_{jj}$.
	If $(u_jI-Q)h=0$, then
	$\norm{h}_P\leq\norm{Qh}_P=u_j\norm{h}_P$.
	Since $0\leq u_j<1$, this implies $h=0$. Thus $u_jI-Q$ is
	invertible, and we may choose $h^j$ satisfying
	$(u_jI-Q)h^j=PW_{E,J}e_j$. Since $Ue_j=u_je_j$, the block
	form gives $A(h^j,e_j)=u_j(h^j,e_j)$. Consequently,
	$A^n(h^j,e_j)=u_j^n(h^j,e_j)\to0$, so $(h^j,e_j)\in X_1$. For any $z\in\mathbb R^{|J|}$, the vector
	$v=\sum_{j\in J}z(j)(h^j,e_j)$ therefore belongs to $X_1$
	and satisfies $v_J=z$. It is unique because the difference of
	any two such vectors belongs to $X_1\cap X_2=\{0\}$.
	
	Finally, given any $y\in\mathbb R^N$, choose $v\in X_1$ with
	$v_J=y_J$. Then $y-v\in X_2$, so $y=v+(y-v)\in X_1+X_2$.
	Since $X_1\cap X_2=\{0\}$, this proves $\mathbb R^N=X_1\oplus X_2$.
	The lower block row of $A$ gives $(Av)_J=Uv_J$ for every $v\in X_1$.
	
	\emph{Statement 3.}
	For $v\in X_2$, write $T(v)=P^{-1/2}v_E$. Since every vector in
	$X_2$ has the form $(v_E,0)$ and $P$ is invertible, $T$ is a linear
	bijection from $X_2$ onto $\mathbb R^{|E|}$. Moreover,
	$Av=(Qv_E,0)$ and $P^{-1/2}QP^{1/2}=I+K$, so
	$T(Av)=(I+K)T(v)$.
	
	Define $X_2^c=\{v\in X_2:T(v)\in\ker K\}$ and
	$X_2^u=\{v\in X_2:T(v)\in(\ker K)^\perp\}$.
	Since $\mathbb R^{|E|}=\ker K\oplus(\ker K)^\perp$ and $T$ is a
	linear bijection, $X_2=X_2^c\oplus X_2^u$.
	If $v\in X_2^c$, then $KT(v)=0$, so $T(Av)=T(v)$.
	Injectivity of $T$ gives $Av=v$. Thus $X_2^c$ is invariant under
	$A$, and $A$ acts as the identity on $X_2^c$.
	To show that $X_2^u$ is invariant under $A$, fix $v\in X_2^u$
	and write $z=T(v)$. By definition, $z\in(\ker K)^\perp$.
	For every $w\in\ker K$, we have $w^\top z=0$ and $Kw=0$.
	Since $K$ is antisymmetric, $w^\top Kz=-(Kw)^\top z=0$.
	Hence $w^\top T(Av)=w^\top(I+K)z=w^\top z+w^\top Kz=0$.
	This holds for every $w\in\ker K$, so $T(Av)\in(\ker K)^\perp$.
	Since $Av\in X_2$ by Statement 2, it follows that $Av\in X_2^u$.
	
	For the second part of the statement, if $W_{E,E}=0$, then $K=0$. In this case, $(\ker K)^\perp=\{0\}$ and
	$X_2^u=\{0\}$. Suppose instead that $W_{E,E}\neq0$. Since $P$ is
	invertible, $K=P^{1/2}W_{E,E}P^{1/2}$ is a nonzero antisymmetric
	matrix and hence has rank at least two. It follows that
	$\dim X_2^u=\dim(\ker K)^\perp=\operatorname{rank}(K)\geq2$.

	For every $z$ on the unit sphere in $(\ker K)^\perp$, we have
	$Kz\neq0$. Indeed, $Kz=0$ would imply
	$z\in\ker K\cap(\ker K)^\perp=\{0\}$, contradicting $\norm{z}_2=1$.
	Thus $z\mapsto\norm{Kz}_2$ is strictly positive on this sphere.
	Since the function is continuous and the sphere is compact,
	it attains a strictly positive minimum $\beta$. Thus
	$\norm{Kz}_2\geq\beta\norm{z}_2$ for every $z\in(\ker K)^\perp$.
	For $u\in X_2^u$, take $z=T(u)=P^{-1/2}u_E$. We have	$\norm{(Au)_E}_P^2=\norm{(I+K)z}_2^2
	=\norm{z}_2^2+\norm{Kz}_2^2
	\geq(1+\beta^2)\norm{u_E}_P^2$.
	Taking $b=\sqrt{1+\beta^2}>1$ gives the claimed bound.
\end{proof}

The next lemma constructs an initial distribution (that is sufficiently close to $p$) from which the imitation dynamics converge to a given $p\in L^F$.

\begin{lemma}\label{lem:attainability}
	Fix $p\in L^F$ and suppose $J=\mathcal R\setminus E^F$ is nonempty.
	There exists $\delta>0$ such that, for every $z\in(0,\infty)^{|J|}$
	with $\norm{z}_2<\delta$, there is an initial distribution
	$\mu_1\in\Delta_+(\mathcal R)$ with $(\mu_1)_J=z$ whose imitation
	dynamics converge to $p$.
\end{lemma}

\begin{proof}
	Since $\Phi(p)=p$, the map $G(y)=\Phi(p+y)-p$ satisfies $G(0)=0$
	and $DG(0)=A$. Use the decomposition $\mathbb R^N=X_1\oplus X_2$
	from Lemma~\ref{lem:limit-linearization}, and write
	$A_\ell=A|_{X_\ell}$ for $\ell=1,2$.
	Equip $X_1$ with the norm $\norm{w}=\norm{w_J}_2$ and $X_2$ with
	the norm $\norm{v}=\norm{v_E}_P$. By Statement 2 of the lemma,
	every $z\in\mathbb R^{|J|}$ is the $J$-coordinate vector of a
	unique $w\in X_1$, and $(Aw)_J=(A_1w)_J=Uw_J$. Hence, the operator norm $\norm{A_1}=\max_{j\in J}U_{jj}<1$. On $X_2$, we have
	$A_2(h,0)=(Qh,0)$, so Statement 1 gives $\norm{A_2^{-1}}\leq1$.
	
	We apply Theorem~1.2(a) of \citet*{cabre2003parameterization}
	with $L=1$ and $r=2$. The map $G$ is twice continuously
	differentiable. We verify their conditions (0) to (5). For condition~(0), Remark 5 allows us to dispense with the
	invertibility of $A$, provided that $\norm{A_2^{-1}}\norm{A_1}^2<1$
	for $L=1$. This inequality follows from the preceding bounds
	$\norm{A_2^{-1}}\leq1$ and $\norm{A_1}<1$. Condition (1) holds because $X_1$ is invariant under $A$. Condition (2) holds because $\norm{A_1}<1$, so every eigenvalue of $A_1$ has modulus less than one. Condition (3) holds because $A_2$ is represented by the invertible matrix $Q$ on $X_2$. Condition (4) is vacuous for $L=1$. Last, Condition (5) requires $L+1\leq r$, which	holds with equality. Theorem 1.2(a) therefore gives $C^2$ maps $K$ and $R$ defined
	near zero in $X_1$, with values in $\mathbb R^N$ and $X_1$,
	respectively. They satisfy $K(0)=R(0)=0$, $DR(0)=A_1$, and
	$G(K(w))=K(R(w))$. Moreover, the $X_1$-component of $K(w)$
	equals $w$, so $K(w)-w\in X_2$.

	Choose $q\in(\norm{A_1},1)$. By continuity of $DR$, there is a
	ball $B_\delta=\{w\in X_1:\norm{w}<\delta\}$ on which
	$\norm{DR(w)}\leq q$. Since $R(0)=0$, the fundamental theorem of calculus gives $\norm{R(w)}\leq q\norm{w}$ on $B_\delta$. Thus
	$R(B_\delta)\subseteq B_\delta$ and
	$\norm{R^n(w)}\leq q^n\norm{w}\to0$ for every $w\in B_\delta$. 
	
	By the definition of $G$, the identity $G(K(w))=K(R(w))$ gives
	$\Phi(p+K(w))=p+K(R(w))$ for every $w\in B_\delta$. Thus, we have
	$\Phi^n(p+K(w))=p+K(R^n(w))$ for every $n\geq1$.
	Since $R^n(w)\to0$ and $K$ is continuous at zero, we have $\Phi^n(p+K(w))\to p$. Since $K(0)=0$ and $p(i)>0$ for every $i\in E$, we may reduce
	$\delta$ so that $p(i)+K(w)(i)>0$ for every $i\in E$ and every
	$w\in B_\delta$.
	
	Now fix $z\in(0,\infty)^{|J|}$ with $\norm{z}_2<\delta$, and let
	$w\in X_1$ be the unique vector satisfying $w_J=z$. Then
	$\norm{w}=\norm{z}_2<\delta$, so $\mu_1=p+K(w)$ generates a
	trajectory converging to $p$. Since $K(w)-w\in X_2$, we have
	$K(w)_J=w_J$. Together with $p_J=0$, this gives $(\mu_1)_J=z$.
	Thus $\mu_1$ is positive on $J$, and our choice of $\delta$
	ensures that it is positive on $E$. Finally, $\Phi$ preserves the
	sum of coordinates on $\mathbb R^N$, so
	$\sum_i\mu_1(i)=\lim_{n\to\infty}\sum_i\Phi^n(\mu_1)(i)
	=\sum_i p(i)=1$. Hence $\mu_1\in\Delta_+(\mathcal R)$.
\end{proof}

\subsection{A Lemma for the Proof of Theorem~\ref{thm:weak-imitation-convergence}}\label{online:local-null}

\begin{lemma}\label{lem:appendix-local-null}
	Suppose $W_{E,E}\neq0$. For every $p\in L^F$, there exists a bounded open neighborhood $\mathcal U_p$ of $p$ with the following property.
	Let $K_p=\{x\in\Delta(\mathcal R):\Phi^n(x)\in\mathcal U_p
	\text{ for every }n\geq0\}$ be the set of initial distributions
	whose trajectories remain in $\mathcal U_p$. Then, for every integer
	$m\geq0$, the set $\Phi^{-m}(K_p)\cap\Delta_+(\mathcal R)$ has
	zero $(N-1)$-dimensional Lebesgue measure.
\end{lemma}

\begin{proof}
	Fix $p\in L^F$. We first show that there is a neighborhood
	$\mathcal U_p$ of $p$ such that the set of initial distributions
	whose trajectories remain in $\mathcal U_p$ at every date has
	measure zero.
	
	In the coordinates of Lemma~\ref{lem:limit-linearization}, write $y\in\mathbb R^N$ as $y=s+u$, where $s\in X_1\oplus X_2^c$ and $u\in X_2^u$.
	Let $C$ and $B$ denote the restrictions of $A$ to
	$X_1\oplus X_2^c$ and $X_2^u$, respectively. Writing $s=w+v$ with
	$w\in X_1$ and $v\in X_2^c$, define
	$\norm{s}=\max\{\norm{w_J}_2,\norm{v_E}_P\}$ and
	$\norm{u}=\norm{u_E}_P$. By the lemma, $\norm{Cs}\leq\norm{s}$
	and $\norm{Bu}\geq b\norm{u}$ for some $b>1$. Equip the product
	space with the norm $\norm{(s,u)}=\max\{\norm{s},\norm{u}\}$.
	
	In these coordinates, the translated map $G(y)=\Phi(p+y)-p$ takes
	the form $G(s,u)=(Cs+f(s,u),Bu+g(s,u))$. Since $f$ and $g$ are
	continuously differentiable and $Df(0)=Dg(0)=0$, we can fix
	$0<\varepsilon<(b-1)/2$ and choose a bounded open convex
	neighborhood $V$ of zero on which both $f$ and $g$ are
	$\varepsilon$-Lipschitz.
	
	Set $\mathcal U_p=p+V$ and let $H=\{x:\sum_i x(i)=1\}$ be the
	affine hull of the simplex. To show that $K_p$ has measure zero in $H$, we consider lines along which the initial $s$-coordinate is fixed. We show that on each such line, at most one
	initial point can generate a trajectory that stays in $\mathcal U_p$ at every date.
	
	Suppose, toward a contradiction, that two distinct initial points
	in $K_p$ have the same $s$-coordinate. Write them as $p+s+u$ and
	$p+s+\tilde u$, where $u\neq\tilde u$. Since both points belong
	to $K_p$, both trajectories remain in $\mathcal U_p$ at every date.
	After translation by $-p$, they therefore remain in $V$. We show
	that this is impossible because their $u$-coordinates grow
	arbitrarily far apart.
	
	Let $\Delta s_n$ and $\Delta u_n$ denote the differences between
	the respective coordinates after $n$ updates. Initially,
	$\Delta s_0=0$ and $\norm{\Delta u_0}>0$. We claim that
	$\norm{\Delta s_n}\leq\norm{\Delta u_n}$ at every date. This holds
	at $n=0$. If it holds at date $n$, then the distance between the
	two points in the product norm is $\norm{\Delta u_n}$. The Lipschitz
	bounds on $f$ and $g$ therefore give
	$\norm{\Delta s_{n+1}}\leq\norm{\Delta s_n}
	+\varepsilon\norm{\Delta u_n}\leq(1+\varepsilon)\norm{\Delta u_n}$
	and $\norm{\Delta u_{n+1}}\geq(b-\varepsilon)\norm{\Delta u_n}$.
	Since $b-\varepsilon>1+\varepsilon$, we obtain
	$\norm{\Delta s_{n+1}}\leq\norm{\Delta u_{n+1}}$, proving the claim
	by induction. The lower bound on $\norm{\Delta u_{n+1}}$ thus
	applies at every date and gives
	$\norm{\Delta u_n}\geq(b-\varepsilon)^n\norm{\Delta u_0}$.
	Since $b-\varepsilon>1$, this contradicts the boundedness of $V$.
	Hence no two distinct points of $K_p$ have the same initial
	$s$-coordinate.
	
	To apply this conclusion within $H$, choose a nonzero vector
	$h\in X_2^u$ whose coordinates sum to zero. Such a vector exists
	because $\dim X_2^u\geq2$ by Lemma \ref{lem:limit-linearization} and $\sum_i h(i)=0$ imposes only one
	linear restriction. Each line in $H$ parallel to $h$ has the form $\{x+ah:a\in\mathbb R\}$ for some $x\in H$. Since $h\in X_2^u$, all points on this line have the same $s$-coordinate. Thus, each such line meets $K_p$ in at most one point. Since $\Phi$ is continuous, the set
	$K_p=\Delta(\mathcal R)\cap \left(\cap_{n\geq0}\Phi^{-n}(\mathcal U_p)\right)$
	is Borel. Its intersection with each of these lines has
	one-dimensional measure zero. Fubini's theorem therefore implies
	that $K_p$ has $(N-1)$-dimensional Lebesgue measure zero in $H$.

	We have shown that $K_p$ has measure zero. Now fix $m\geq1$.
	We must show that the set of interior initial distributions whose
	trajectories remain in $\mathcal U_p$ from date $m$ onward also has measure zero. By definition of $K_p$, this set is
	$\{x\in\Delta_+(\mathcal R):\Phi^m(x)\in K_p\}=\Phi^{-m}(K_p)\cap\Delta_+(\mathcal R)$.
	
	To apply the inverse function theorem, we first show that
	$D\Phi(x)$ is nonsingular at every interior point $x$. Fix
	$x\in\Delta_+(\mathcal R)$ and let $z\in\mathbb R^N$ be any
	direction. Differentiating
	$\Phi(x+tz)(i)=(x(i)+tz(i))[1+(Wx)(i)+t(Wz)(i)]$
	with respect to $t$ at $t=0$ gives
	$[D\Phi(x)z](i)=[1+(Wx)(i)]z(i)+x(i)(Wz)(i)$. Suppose $D\Phi(x)z=0$. We show that $z=0$. Multiply the $i$-th
	coordinate of this identity by $z(i)/x(i)$ and sum over $i$.
	The terms involving $Wz$ sum to $z^\top Wz=0$ by antisymmetry,
	leaving $\sum_i [1+(Wx)(i)]z(i)^2/x(i)=0$.
	Since $1+(Wx)(i)\geq1-\sum_{j\neq i}x(j)=x(i)>0$, the coefficient
	of each $z(i)^2$ is positive. The sum can therefore vanish only if
	$z(i)=0$ for every $i$. This proves that $D\Phi(x)$ is nonsingular.
	
	Fix $x\in\Delta_+(\mathcal R)$. By
	Lemma~\ref{lem:relative-growth}, every iterate of $x$ remains in
	the simplex interior. Since $D\Phi$ is invertible at every interior point, the chain rule implies that $D\Phi^m(x)$ is invertible. The inverse function theorem therefore gives a neighborhood $U$ of $x$ on which $\Phi^m$ has a $C^1$ inverse on its image. Since $\Phi^m$ preserves the sum of coordinates, $w\in H$ if and
	only if $\Phi^m(w)\in H$. We may therefore restrict $U$ to $H$
	and take it small enough to lie in the simplex interior.
	
	Within this neighborhood $U$ of $x$ in $H$, any set whose image
	under $\Phi^m$ has measure zero in $H$ must itself have measure
	zero in $H$, since the inverse is locally Lipschitz. In particular, $\{w\in U:\Phi^m(w)\in K_p\}$ has
	measure zero because its image is contained in the measure-zero set $K_p$. Since countably many such neighborhoods cover the simplex interior, $\Phi^{-m}(K_p)\cap\Delta_+(\mathcal R)$ has measure zero.
	\end{proof}

\subsection{Proof of Non-convergence in Example \ref{example:largeintensitybreakconvergence}} \label{appendix:largeintensitybreakconvergence}

Let $\Phi$ denote the population update map under $F$, satisfying
$\Phi(x)(i)=x(i)[1+(Wx)(i)]$ for each $i$. Consider the boundary face
$\Gamma=\{x\in\Delta(\mathcal R):x(1)=x(2)=0, x(3)x(4)x(5)=0\}$. This is a compact invariant set of $\Phi$.

On the relative interior of each edge of $\Gamma$, the population share of the rule that defeats the other rule strictly increases. Hence, every such path converges to the corresponding vertex, which implies that the only recurrent points in $\Gamma$ are therefore the fixed points $e_3,e_4,e_5$. By Poincar\'e's recurrence theorem, any invariant
distribution supported on $\Gamma$ is concentrated on these three
points. Since any nondegenerate mixture of their point masses assigns
probability strictly between zero and one to an invariant singleton,
the only ergodic invariant distributions supported on $\Gamma$ are
the point masses on $e_3,e_4,e_5$.

For an interior population distribution $x\in\Delta_+(\mathcal R)$,
define three transformations:
$A(x)=x(1)/[x(4)x(5)^6]$,
$B(x)=x(2)/[x(3)^5x(4)^6]$, and
$C(x)=x(3)x(4)x(5)$. The exponents are chosen so that each
transformation decreases around every recurrent point of $\Gamma$.

Let $h_i(x):=\log[1+(Wx)(i)]$. By the evolution equation, the
logarithmic changes in these functions are
$d_A(x):=\log A(\Phi(x))-\log A(x)=h_1(x)-h_4(x)-6h_5(x)$,
$d_B(x):=\log B(\Phi(x))-\log B(x)
=h_2(x)-5h_3(x)-6h_4(x)$, and
$d_C(x):=\log C(\Phi(x))-\log C(x)
=h_3(x)+h_4(x)+h_5(x)$.
Since every term $1+(Wx)(i)$ is bounded away from zero on $\Delta(\mathcal{R})$, each $d_G$,
$G\in\{A,B,C\}$, extends continuously to $\Delta(\mathcal R)$.
By the entries of $W$, we have $d_G(e_3)<0$, $d_G(e_4)<0$, and
$d_G(e_5)<0$ for every $G\in\{A,B,C\}$.

For each $G\in\{A,B,C\}$, let
$D_G(x,n):=\sum_{r=0}^{n-1}d_G(\Phi^r(x))$ denote the cumulative
logarithmic change in $G$ over $n$ periods. Each $D_G(\cdot,n)$ is
continuous on $\Gamma$ and satisfies
$D_G(x,n+m)=D_G(x,n)+D_G(\Phi^n(x),m)$. Define
$D(x,n):=\max_{G\in\{A,B,C\}}D_G(x,n)$. The function $D$ is
continuous and subadditive.

By \citet[Theorem~1]{schreiber1998growth}, the largest long-run
average change in $D$ over all points in $\Gamma$ equals the largest
growth rate associated with an ergodic invariant distribution
supported on $\Gamma$. In our setting, the only ergodic invariant distributions on $\Gamma$
are the point masses at $e_3,e_4,e_5$. Since each $e_i$ is fixed,
$D_G(e_i,n)=n d_G(e_i)$ for every $G\in\{A,B,C\}$. Hence the growth
rate at $e_i$ is $D(e_i,n)/n=\max_{G\in\{A,B,C\}}d_G(e_i)$ for every
$n\geq1$. The theorem therefore gives
$\inf_{n\geq1}n^{-1}\sup_{x\in\Gamma}D(x,n)
=\max_{i\in\{3,4,5\},\,G\in\{A,B,C\}}d_G(e_i)<0$.

Hence, there exists an integer $K\geq1$
such that $M:=\sup_{x\in\Gamma}D(x,K)<0$. Choose $\rho\in(e^M,1)$.
Then $D_G(x,K)\leq D(x,K)\leq M<\log\rho$ for every $x\in\Gamma$
and every $G\in\{A,B,C\}$. By continuity of $D(\cdot,K)$, the set
$U=\{x\in\Delta(\mathcal R):D(x,K)<\log\rho\}$ is a relatively
open neighborhood of $\Gamma$. For every interior $x\in U$ and
every $G\in\{A,B,C\}$, summing the one-period logarithmic changes
gives $\log G(\Phi^K(x))-\log G(x)=D_G(x,K)<\log\rho$.
Thus $G(\Phi^K(x))\leq\rho G(x)$.

For $\eta>0$, define 
$V_\eta=\{x\in\Delta_+(\mathcal R):
\max\{A(x),B(x),C(x)\}<\eta\}$. For some sufficiently small $\eta>0$, we have $\emptyset\neq V_\eta\subseteq U$. Thus, for every $x\in V_\eta$, we have
$A(\Phi^K(x))\leq\rho A(x)$, $B(\Phi^K(x))\leq\rho B(x)$, and
$C(\Phi^K(x))\leq\rho C(x)$. Therefore,
$\Phi^K(V_\eta)\subseteq V_\eta$, and both $A(\Phi^{nK}(x))$ and
$B(\Phi^{nK}(x))$ converge to zero. Since $x(1)\leq A(x)$ and
$x(2)\leq B(x)$, we obtain that
$(\Phi^{nK}(x))(1)+(\Phi^{nK}(x))(2)$ converges to $0$. Meanwhile, the evolution equation implies that $\Phi(x)(1)\le 1.51 x(1)$ and $\Phi(x)(2)\le 1.51 x(2)$ hold for all $x$. Therefore, the path must satisfy $\mu_t(1)+\mu_t(2)\rightarrow 0.$ If this path converged, Proposition~\ref{prop:stationary-convergence}
would require its limit to belong to $L^F$ and hence to have support
$E^F=\{R_1,R_2\}$. This contradicts the preceding conclusion.
Therefore, every path beginning in $V_\eta$ fails to converge.

\subsection{Proof of Proposition~\ref{prop:boundary-approach}}\label{appendix:boundary-approach}

Let $s_t=\mu_t(S)$ and $\eta_t=1-s_t$ denote the population shares
inside and outside the top cycle, respectively.

\emph{Statement 1.}
If $S=\mathcal R$, then $s_t=1$ and $\eta_t=0$ at every date.
Suppose $S\neq\mathcal R$, and let
$\delta=\min_{i\in S,\,j\notin S}W_{ij}>0$. Summing the evolution
equation over $i\in S$, by the antisymmetry of $W$, we have
\begin{equation}\label{eqn:incrementS}
s_{t+1}-s_t=\sum_{i\in S,\,j\notin S}W_{ij}\mu_t(i)\mu_t(j)
\geq\delta s_t\eta_t>0.
\end{equation}
Thus, $s_t$ is strictly increasing and
$\eta_{t+1}\leq(1-\delta s_t)\eta_t$.

Fix $T\geq1$. Since $s_t\geq s_T$ for every $t\geq T$, iteration
gives $\eta_t\leq\eta_T(1-\delta s_T)^{t-T}$ for every $t\geq T$.
Taking $T=1$ shows that $\eta_t\to0$ and
$\sum_{t\geq1}\eta_t<\infty$. In particular, $s_t\to1$.
For each fixed $T$, the same bound gives
$\limsup_{t\to\infty}t^{-1}\log\eta_t\leq\log(1-\delta s_T)$.
Letting $T\to\infty$ yields
$\limsup_{t\to\infty}t^{-1}\log\eta_t\leq\log(1-\delta)$,
including the case $\delta=1$ under the convention $\log0=-\infty$.

\emph{Statement 2.} Consider any convergent subsequence $\mu_{t_k}\to z$. By Statement 1, every accumulation point assigns
all its mass to $S$. Suppose now that $z(i)>0$ for every $i\in S$.
To prove the statement, we show first that $z$ is stationary and then show that the entire trajectory $(\mu_t)_{t\ge 1}$ must converge to $z$.

For any lottery $v$ supported on $S$, write
$D_t(v)=D_{\mathrm{KL}}(v\|\mu_t)$. The evolution equation gives
$D_{t+1}(v)-D_t(v)
=-\sum_{i\in\supp(v)}v(i)\log[1+(W\mu_t)(i)]$.
Since $\log(1+u)\leq u$, this difference is bounded below by
$-v^\top W\mu_t=(Wv)^\top\mu_t$, where the equality follows from
antisymmetry. The terms outside $S$ contribute at least $-\eta_t$,
since $|(Wv)(j)|\leq1$ and $\mu_t(S^c)=\eta_t$.
Hence, 
\begin{equation}\label{eqn:incrementDonS}
	D_{t+1}(v)-D_t(v)\geq\sum_{i\in S}(Wv)(i)\mu_t(i)-\eta_t
\end{equation}

To show that $z$ is stationary, apply the Farkas argument from
the proof of Statement 1 in Appendix \ref{appendix:stationary-convergence}
with $-W_{S,S}$ in place of $W$. This gives a lottery $q\in\Delta(S)$. We regard $q$ as a lottery on
$\mathcal R$ by setting $q(i)=0$ for $i\notin S$. Then $(Wq)(i)=0$
for $i\in\supp(q)$ and $(Wq)(i)>0$ for $i\in S\setminus\supp(q)$.  In particular, $(Wq)(i)\geq0$ throughout $S$, so the preceding
bound gives $D_{t+1}(q)-D_t(q)\geq-\eta_t$. By Statement 1,
$L_t=\sum_{r=t}^{\infty}\eta_r$ is finite and tends to zero.
Since $L_t-L_{t+1}=\eta_t$, the sequence $D_t(q)-L_t$ is
nondecreasing.

We now show that $\supp(q)=S$. We first prove that
$D_{t+1}(q)-D_t(q)\to0$. Along the subsequence $\mu_{t_k}\to z$,
we have $D_{t_k}(q)\to D_{\mathrm{KL}}(q\|z)$. Because $z(i)>0$
for every $i\in\supp(q)$, this limit is finite, and because $L_{t_k}\to0$, we have $D_{t_k}(q)-L_{t_k}\to D_{\mathrm{KL}}(q\|z)$. Since $D_t(q)-L_t$
is nondecreasing, the entire sequence $D_t(q)-L_t$
converges to the same limit. Since $L_t\to0$, $D_t(q)$ also converges to $D_{\mathrm{KL}}(q\|z)$.
It follows that $D_{t+1}(q)-D_t(q)\to0$. Taking $t=t_k$ in \eqref{eqn:incrementDonS} and using $\mu_{t_k}\to z$ and $\eta_{t_k}\to0$, we obtain $0\geq\sum_{i\in S}(Wq)(i)z(i)$. Every term in the sum is nonnegative. Moreover, if some $i\in S\setminus \supp(q)$, the choice
of $q$ would give $(Wq)(i)>0$. Since $z(i)>0$, that term would
be strictly positive, a contradiction. Hence, we have $\supp(q)=S$.

We now show that $z$ is stationary. The evolution equation \eqref{eqn:evolution} gives
$D_{t+1}(q)-D_t(q)=-\sum_{i\in S}q(i)\log[1+(W\mu_t)(i)]$.
We have shown that the left-hand side tends to zero. Along the
subsequence $\mu_{t_k}\to z$, each logarithm on the right-hand side
converges because $1+(Wz)(i)\geq 1 - \sum_{j\neq i} z_j = z(i)>0$ for every $i\in S$. Taking the limit gives $\sum_{i\in S}q(i)\log[1+(Wz)(i)]=0$.

Since $\supp(q)=S$, the choice of $q$ gives $(Wq)(i)=0$ for every
$i\in S$. As $z$ is supported on $S$, the antisymmetry of $W$ gives
$\sum_{i\in S}q(i)(Wz)(i)=q^\top Wz=-(Wq)^\top z=0$.
For each $i\in S$, we have $\log[1+(Wz)(i)]\leq(Wz)(i)$, with
strict inequality if $(Wz)(i)\neq0$. Recall that $0=\sum_{i\in S}q(i)\log[1+(Wz)(i)]\le -(Wq)^\top z=0$. Thus, $\log[1+(Wz)(i)]= (Wz)(i)$  must hold
for every $i\in S$, which implies $(Wz)(i)=0$ throughout $S$. Consequently, the evolution
equation leaves each share $z(i)$ unchanged on $S$, and
the shares outside $S$ remain zero. Hence,  $z$ is stationary.

It remains to show that the entire trajectory $(\mu_t)_{t\ge 1}$, rather than a subsequence of it, converges to $z$.
Since $(Wz)(i)=0$ for every $i\in S$, applying
\eqref{eqn:incrementDonS} with $v=z$ gives
$D_{t+1}(z)-D_t(z)\geq-\eta_t$. Fix $t$ and take any $t_k>t$.
Summing this bound from $t$ to $t_k-1$ gives
$D_t(z)\leq D_{t_k}(z)+\sum_{r=t}^{t_k-1}\eta_r \le D_{t_k}(z)+ L_t$.
Along the subsequence $\mu_{t_k}\to z$, we have
$D_{t_k}(z)\to D_{\mathrm{KL}}(z\|z)=0$. Letting $k\to\infty$
therefore gives $0\leq D_t(z)\leq L_t$. Since $L_t\to0$,
Pinsker's inequality yields $\norm{\mu_t-z}_1^2\leq2D_t(z)\to0$.

By Statement 1, every accumulation point belongs to $\Delta(S)$.
We have shown that an accumulation point in $\Delta_+(S)$ implies
convergence of the entire trajectory. Thus, if the trajectory does
not converge, every accumulation point belongs to $\partial\Delta(S)$.
If $\operatorname{dist}(\mu_t,\partial\Delta(S))$ did not tend to
zero, some subsequence would stay a positive distance from
$\partial\Delta(S)$. By compactness of the simplex, that subsequence
would have an accumulation point at least that distance from the
boundary, a contradiction. This proves Statement 2.

\subsection{Proof of Proposition \ref{prop:nonconvergencesufficient}}\label{appendix:nonconvergencesufficient}

Since $v$ is the unique maximal lottery and has support $S$, we have $E=S$ and $(Wv)(i)=0$ for every $i\in S$. Write $s_t=\mu_t(S)$ and $\eta_t=1-s_t$. Suppose first that $S\neq\mathcal R$. Put $C=\beta/\delta\geq1$ and $J_t=D_{\mathrm{KL}}(v\|\mu_t)+C\log s_t$. By Jensen's inequality, we have 
\[
D_{\mathrm{KL}}(v\|\mu_{t+1})-D_{\mathrm{KL}}(v\|\mu_t)
=-\sum_{i\in S}v(i)\log[1+(W\mu_t)(i)]
\geq-\log(1+v^\top W\mu_t).\] 
Furthermore, using that $W^\top v = - Wv$ vanishes on $S$, we have
\[
D_{\mathrm{KL}}(v\|\mu_{t+1})-D_{\mathrm{KL}}(v\|\mu_t)
\geq -\log\left(1+\sum_{j\notin S} \mu_t(j)(W^\top v)(j)\right) \ge -\log(1+\beta\eta_t).
\]
Moreover, by \eqref{eqn:incrementS}, we have $s_{t+1}-s_t\geq\delta s_t\eta_t$. Thus, $\log s_{t+1}-\log s_t\geq\log(1+\delta\eta_t)$. Hence, $J_{t+1}-J_t\geq C\log(1+\delta\eta_t)-\log(1+\beta\eta_t)\geq0$, where the last inequality follows from $(1+u)^C\geq1+Cu$ for $u\geq0$.

If $S=\mathcal R$, set $C=1$ and define $J_t$ in the same way. Then $s_t=1$, and Jensen's inequality again implies that $J_t$ is nondecreasing. Thus, in both cases, $J_t$ is nondecreasing. Nevertheless, either stated condition gives $J_1>0$, but the convergence to $v$ would require $J_t\to0$. Contradiction. Thus, the dynamics do not converge.

\subsection{Proof of Proposition \ref{prop:heteroclinic-cycle}}\label{appendix:heteroclinic-cycle}

It is straightforward to verify that $v=(b,c,a)/(a+b+c)$ is the unique probabilistic solution to $Wv=0$. Since $S=\mathcal R$, Proposition~\ref{prop:nonconvergencesufficient} establishes nonconvergence, and Proposition~\ref{prop:boundary-approach} shows that the trajectory approaches the boundary. In this case, every rule survives: if $\mu_t(1)\to0$, then $(W\mu_t)(2)-(W\mu_t)(3)\to b>0$. Then, Lemma~\ref{lem:relative-growth} implies $\mu_t(3)/\mu_t(2)\to0$, hence $\mu_t\to e_2$, a contradiction. The other cases are symmetric.

Let $\Phi$ be the population update map defined in \eqref{eqn:Phi} and $\Omega\subseteq\partial\Delta(\mathcal R)$ be the set of accumulation points of the trajectory $(\mu_t)_{t\ge 1}$. As a closed subset of a compact set, \(\Omega\) is compact and satisfies $\Phi(\Omega)=\Omega$. Fix any rule \(R_i\), and choose \(x\in\Omega\) maximizing $x(i)$. Since $R_i$ survives, $x(i)>0$. 

Now we prove $x(i)=1$. Suppose that $x(i)<1$. Because $x\in \partial\Delta(\mathcal R)$, $\supp(x)=\{i,j\}$ for some $j\neq i$. If $R_i$ defeats $R_j$ (so $W_{ij}>0$), then $\Phi(x)(i)=x(i)(1+W_{ij}x(j))>x(i)$. This contradicts the choice of $x$, as $\Phi(x)\in\Omega$ will be an accumulation point with a larger $i$-th entry. If $R_j$ defeats $R_i$ (so $W_{ij}<0$), take $y\in\Omega$ with $\Phi(y)=x$. Note that a rule has positive share after an update if and only if it had positive share before the update. Therefore, $\supp(y)=\{i,j\}.$ But the share of $R_i$ falls under the update, giving $x(i)=\Phi(y)(i)<y(i)$, again contradicting the choice of $x$. Thus, $x(i)=1.$ Since the choice of $i$ is arbitrary, we have shown $\limsup_{t\to\infty}\mu_t(i)=1$ for every $i$, which in turn implies $\liminf_{t\to\infty}\mu_t(i)=0$ for every $i$. 

Last, the boundary dynamics follow the cycle $e_1\to e_3\to e_2\to e_1$, opposite to the arrows in $G^F$. Since each $e_1,e_2,e_3$ is a stationary distribution, the cycle is a heteroclinic cycle.

\subsection{Proof of Proposition \ref{prop:dominated-rule-extinction}}\label{appendix:dominated-rule-extinction}

By Lemma~\ref{lem:relative-growth}, the population ratio $r_t=\mu_t(j)/\mu_t(i)\ge 0$ is strictly decreasing over time. So $r_t$ converges to some $\bar{r}\ge 0$. The same lemma also gives $[1+(W\mu_t)(i)](r_t-r_{t+1})=r_t[(W\mu_t)(i)-(W\mu_t)(j)]$. By the antisymmetry, $W_{ii}-W_{ji}=W_{ij}$. Moreover, since $R_i$ dominates $R_j$, we have $W_{ik}\ge W_{jk}$ for $k\neq i$. Thus,
\[(W\mu_t)(i)-(W\mu_t)(j)=\sum_k(W_{ik}-W_{jk})\mu_t(k)\geq W_{ij}\mu_t(i).\] 
Multiplying by $r_t$ and using $r_t\mu_t(i)=\mu_t(j)$, we obtain $W_{ij}\mu_t(j)\leq[1+(W\mu_t)(i)](r_t-r_{t+1})\to0$, where the convergence follows because $1+(W\mu_t)(i)$ is bounded by $2$. Since $W_{ij}>0$, we have $\mu_t(j)\to0$. 

To show a dominated rule can be in the top cycle, we consider an environment with four rules, and the pairwise comparison matrix is given by
$$
W^F=\begin{pmatrix}
	0&0.12&-0.08&-0.07\\
	-0.12&0&0.04&0.05\\
	0.08&-0.04&0&0.01\\
	0.07&-0.05&-0.01&0
\end{pmatrix}.
$$
In this environment, $R_4$ is dominated by $R_3$, but the top cycle $S=\mathcal R$, because the induced graph $\mathcal{G}^F$ contains the cycle $R_3\to R_4\to R_1\to R_2\to R_3$. 

\subsection{Strengthening of Proposition~\ref{prop:REwins}}\label{appendix:REsharp}

The following proposition strengthens Proposition~\ref{prop:REwins} by allowing the fixed belief to be misspecified, and characterizes the sharp threshold on permissible misspecification in terms of total variation distance,
$d(\pi^1,\pi^2)=\frac{1}{2}\sum_{k=1}^{K}|\pi^1_k-\pi^2_k|$ for $\pi^1,\pi^2\in\Delta(\mathcal K).$ In particular, setting $\hat\pi^*=\pi$ yields Proposition~\ref{prop:REwins}.

\begin{prop}\label{prop:REsharp}
	Fix an integer $N\ge 2$. For each integer $K\ge 3$ and each pair of full-support distributions $(\pi,\hat\pi^*)\in\Delta_+(\mathcal K)\times \Delta_+(\mathcal K)$, let the belief set $\mathcal R$ consist of $\hat\pi^*$ and $N-1$ other beliefs drawn independently and uniformly from $\Delta_+(\mathcal K)$. There exists a constant $\delta^*\in (0,1)$ such that the following statements hold:
	\begin{enumerate}
		\item For every $\delta\in[0,\delta^*)$ and $\varepsilon>0$, there exists an integer $K_0\ge 3$ such that, for all $K\ge K_0$ and all full-support pairs $(\pi,\hat\pi^*)$ satisfying $d(\hat\pi^*,\pi)\le\delta$, the probability that $\hat\pi^*$ is the Condorcet winner is greater than $1-\varepsilon$.
		
		\item For every $\delta\in(\delta^*,1)$ and $\varepsilon>0$, there exists an integer $K_0\ge 3$ such that, for all $K\ge K_0$, there exists a full-support pair $(\pi,\hat\pi^*)$ satisfying $d(\hat\pi^*,\pi)=\delta$ for which the probability that $\hat\pi^*$ is the Condorcet winner is strictly less than $\varepsilon$.
	\end{enumerate}
	Specifically, $\delta^*$ is the unique solution to $\delta+(1-\delta)e^{-(1-\delta)}=1/2$.
\end{prop}

The proof of this proposition relies on the following moment bounds.

\begin{lemma}\label{lem:uniform-belief-moments}
	Let $K\geq3$, fix $\hat\pi^*\in\Delta_+(\mathcal K)$, and
	let $\hat\pi$ be drawn uniformly from $\Delta_+(\mathcal K)$.
	Write $X_k=\mathbf{1}(\hat\pi_k>\hat\pi_k^*)$ and
	$r_k=(1-\hat\pi_k^*)^{K-1}$. For any nonnegative weight vector
	$b=(b_1,\ldots,b_K)$, let $Z_b=\sum_{k=1}^K b_kX_k$.
	Then, we have the following two statements.
	\begin{enumerate}
		\item We have $\mathbb{E}[Z_b]=\sum_{k=1}^K b_kr_k$ and
		$\mathrm{Var}(Z_b)\leq
		\sum_{k=1}^K b_k^2r_k(1-r_k)
		\leq\frac14\sum_{k=1}^K b_k^2.$
		\item If additionally $b_k\leq\hat\pi_k^*$ for every $k$ and
		$s=\sum_{k=1}^K b_k$, then
		$\mathbb{E}[Z_b]\leq
		s\left(1-\frac{s}{K}\right)^{K-1}$ and 
		$\mathrm{Var}(Z_b)\leq\frac{s}{K}.$
	\end{enumerate}
\end{lemma}

\begin{proof}[Proof of Lemma~\ref{lem:uniform-belief-moments}]
	The uniform distribution of $\hat{\pi}$ gives
	$\mathbb{E}[X_k]=r_k$, as the set $\{q\in\Delta(\mathcal K): q_k\ge\hat{\pi}_k^*\}$ is parametrized by $\{q\in\Delta(\mathcal K): q= \hat{\pi}_k^*e_k+(1-\hat{\pi}_k^*)z, z\in \Delta(\mathcal K)\}$. For the same reason, for $k\neq\ell$, we have
	$\mathbb{E}[X_kX_\ell]
	=(1-\hat\pi_k^*-\hat\pi_\ell^*)^{K-1}$, which further implies
	$\mathbb{E}[X_kX_\ell]
	\leq(1-\hat\pi_k^*)^{K-1}
	(1-\hat\pi_\ell^*)^{K-1}
	=r_kr_\ell.$  Thus, $\mathrm{Cov}(X_k,X_\ell)\leq0$ for $k\neq\ell$, and Statement 1 follows.
	
	To prove Statement 2, we suppose $b_k\leq\hat\pi_k^*$ for every $k$.
	Let $f_K(x)=x(1-x)^{K-1}$ and
	$g_K(x)=f_K(\min\{x,1/K\})$.
	The function $f_K$ is increasing and concave on $[0,1/K]$,
	attains its global maximum at $1/K$, and satisfies
	$f_K'(1/K)=0$.
	Consequently, $g_K$ is concave on $[0,1]$ and
	$g_K\geq f_K$. Since $s=\sum_kb_k\leq1$,
	Jensen's inequality yields
	$$
	\mathbb{E}[Z_b]
	=\sum_{k=1}^K b_k(1-\hat\pi_k^*)^{K-1}
	\leq\sum_{k=1}^K f_K(b_k)
	\leq\sum_{k=1}^K g_K(b_k)
	\leq K g_K(s/K)
	=s(1-s/K)^{K-1}.
	$$
	Finally, note that $b_kr_k\leq f_K(b_k)\leq f_K(1/K)<1/K$. By Statement 1, we have
	$
	\mathrm{Var}(Z_b)
	\leq\sum_{k=1}^K b_k^2r_k
	\leq\frac1K\sum_{k=1}^K b_k
	=\frac{s}{K}.
	$
\end{proof}

\begin{proof}[Proof of Proposition~\ref{prop:REsharp}]
	Let $h(\delta)=\delta+(1-\delta)e^{-(1-\delta)}$.
	This function is continuous and strictly increasing on $[0,1]$,
	with $h(0)=e^{-1}<1/2<h(1)=1$. Hence, there is a unique
	$\delta^*\in(0,1)$ satisfying $h(\delta^*)=1/2$.
	
	For one uniformly drawn competing belief $\hat\pi$, define $X_k=\mathbf{1}(\hat\pi_k>\hat\pi_k^*)$ as in Lemma~\ref{lem:uniform-belief-moments}, and let
	$\phi(\hat\pi)=\sum_{k=1}^K\pi_kX_k$.
	Since $\Pr(\hat\pi_k=\hat\pi_k^*)=0$ for every $k$,
	the fixed belief $\hat\pi^*$ defeats $\hat\pi$ if and only if
	$\phi(\hat\pi)<1/2$, with probability one.\\
	\noindent \emph{Statement 1.}
	Fix $\delta\in[0,\delta^*)$, let $s=1-\delta$, and consider
	any full-support pair $(\pi,\hat\pi^*)$ with
	$d(\hat\pi^*,\pi)\leq\delta$.
	Since 
	\[\sum_{k=1}^K\min\{\pi_k,\hat\pi_k^*\} 
	=\frac12\sum_{k=1}^K \left(\pi_k+\hat\pi_k^*-|\pi_k-\hat\pi_k^*|\right)
	=1-d(\hat\pi^*,\pi)\geq s,\] 
	we can choose weights
	$0\leq m_k\leq\min\{\pi_k,\hat\pi_k^*\}$ with
	$\sum_{k=1}^K m_k=s$. Writing $Y=\sum_{k=1}^Km_kX_k$, we have
	\[
	\phi(\hat\pi)
	=Y+\sum_{k=1}^K(\pi_k-m_k)X_k
	\leq Y+\sum_{k=1}^K(\pi_k-m_k)=Y+1-s=Y+\delta.
	\]
	Define
	$\gamma_K=1/2-\delta-s(1-s/K)^{K-1}$. 
	Recall $\phi(\hat\pi)\leq Y+\delta$. So whenever $\phi(\hat\pi)\geq1/2$, we must have $Y\geq1/2-\delta$. Using $\mathbb E[Y]\leq s(1-s/K)^{K-1}$ in Lemma \ref{lem:uniform-belief-moments}, we have that $\phi(\hat\pi)\geq1/2$ implies
	\[
	Y-\mathbb E[Y]
	\geq\frac12-\delta-\mathbb E[Y]
	\geq\frac12-\delta-s(1-s/K)^{K-1}
	=\gamma_K.
	\]
	Note $\gamma_K$ is positive for all sufficiently large $K$: since $\left(1-\frac{s}{K}\right)^{K-1}\rightarrow e^{-s}$ as $K\rightarrow \infty$, we have $\gamma_K\to 1/2 -\delta- se^{-s}=1/2-h(\delta)>0$ as $K\rightarrow \infty$. Therefore, the variance bound $\mathrm{Var}(Y)\leq\frac{s}{K}$ in Lemma~\ref{lem:uniform-belief-moments} and Chebyshev's inequality give, for all sufficiently large $K$,
	 \[\Pr\left(\phi(\hat\pi)\geq\frac12\right) \le \Pr\left(Y-\mathbb E[Y]\geq\gamma_K\right)
	 \leq\Pr\!\left(|Y-\mathbb E[Y]|\geq\gamma_K\right)
	 \leq\frac{\mathrm{Var}(Y)}{\gamma_K^2}
	 \leq\frac{s}{K\gamma_K^2}.
	 \]
	Taking a union bound over the $N-1$ competing beliefs yields
	\[
	\Pr\left(\hat\pi^*\text{ is not the Condorcet winner}\right)
	\leq\frac{(N-1)s}{K\gamma_K^2}\longrightarrow0.
	\]
	Since this bound does not depend on the fixed pair $(\pi,\hat\pi^*)$, statement 1 follows.\\
	\noindent\emph{Statement 2.}
	Fix $\delta\in(\delta^*,1)$. For sufficiently large $K$, set
	$c_K=(1-\delta-2K^{-2})/(K-2)>0$ and
	$$
	\pi=(\delta+K^{-2},\ K^{-2},\ c_K,\ldots,c_K),
	\qquad
	\hat\pi^*=(K^{-2},\ \delta+K^{-2},\ c_K,\ldots,c_K).
	$$
	Both distributions have full support, and
	$d(\hat\pi^*,\pi)=\delta$.
	For one uniformly drawn competing belief, write
	$\phi(\hat\pi)=(\delta+K^{-2})X_1+K^{-2}X_2+Y_K$,
	where $Y_K=c_K\sum_{k=3}^KX_k$. 
	
	We first claim that the sum of the first two terms converges in probability to $\delta$: when $X_1=1$, we have $
	|(\delta+K^{-2})X_1+K^{-2}X_2 -\delta|=K^{-2}(1+X_2)\rightarrow 0
	$. Thus, for any $\eta>0$, the deviation $|(\delta+K^{-2})X_1+K^{-2}X_2 -\delta|>\eta$ cannot occur when $X_1=1$. Consequently, for any sufficiently large $K$,
	\[
	\Pr(|(\delta+K^{-2})X_1+K^{-2}X_2-\delta|>\eta)
	\leq\Pr(X_1=0)
	=1-(1-K^{-2})^{K-1}
	\leq\frac{K-1}{K^2}
	\longrightarrow0,
	\]
	where the last inequality follows from Bernoulli's inequality.	Moreover, as $K\rightarrow \infty,$ we have $
	(K-2)c_K=1-\delta-2K^{-2}\longrightarrow1-\delta$, and 
	\[
	\log\left((1-c_K)^{K-1}\right)
	=(K-1)c_K\frac{\log(1-c_K)}{c_K}
	\longrightarrow-(1-\delta),
	\]
	where the convergence follows from $\log(1-x)/x\to-1$ as $x\to0$.
	Thus, Lemma~\ref{lem:uniform-belief-moments} gives
	\[
	\mathbb{E}[Y_K]
	=(K-2)c_K(1-c_K)^{K-1}
	\longrightarrow(1-\delta)e^{-(1-\delta)}, \qquad \mathrm{Var}(Y_K)
	\leq\frac{(K-2)c_K^2}{4}\longrightarrow0.
	\]
	For any $\eta>0$, for sufficiently large $K$, $|\mathbb{E}[Y_K]-(1-\delta)e^{-(1-\delta)}|<\eta/2$. Thus, Chebyshev's inequality implies that 
	\[
	\Pr(|Y_K-(1-\delta)e^{-(1-\delta)}|>\eta)
	\leq\Pr\!\left(|Y_K-\mathbb E[Y_K]|>\frac{\eta}{2}\right)
	\leq\frac{4\,\mathrm{Var}(Y_K)}{\eta^2}
	\longrightarrow0.
	\]
	Thus, $Y_K$ converges in probability to $(1-\delta)e^{-(1-\delta)}$ , which implies $\phi(\hat\pi)$
	converges in probability to $h(\delta)>1/2$.
	Since being the Condorcet winner requires defeating the
	first competing belief,
	\[
	\Pr\left(\hat\pi^*\text{ is the Condorcet winner}\right)
	\leq\Pr\left(\phi(\hat\pi^1)<\frac12\right)
	\longrightarrow0.
	\]
	This probability is therefore strictly less than any given
	$\varepsilon>0$ for all sufficiently large $K$, proving statement 2.
\end{proof}

\subsection{Proof of Lemma~\ref{lem:appendix:uniform-defeat}}\label{appendix:uniform-defeat}

Fix $K\geq3$ and a full-support objective distribution $\pi$
satisfying $\max_{1\leq k\leq K}\pi_k<1/2$. Let $\hat\pi^*\in\Delta(\mathcal K)$
be any fixed belief, and let $\hat\pi$ be drawn uniformly
from $\Delta_+(\mathcal K)$.
For each $k\in\mathcal K$, define
$A_k(\hat\pi^*)
=
\left\{
\tilde\pi\in\Delta(\mathcal K):
\tilde\pi_\ell>\hat\pi^*_\ell
\text{ for every }\ell\neq k
\right\}.$ Every belief in $A_k(\hat\pi^*)$ outperforms $\hat\pi^*$
in all states except $k$. Since these states have total
objective probability $1-\pi_k>1/2$, every belief in
$A_k(\hat\pi^*)$ defeats $\hat\pi^*$ in the pairwise comparison. Moreover, we have
$\Pr\bigl(\hat\pi\in A_k(\hat\pi^*)\bigr) = (\hat\pi^*_k)^{K-1}$ and
the sets $A_k(\hat\pi^*)$ are pairwise disjoint. Therefore,
\[
\Pr\bigl(\hat\pi\text{ defeats }\hat\pi^*\bigr)
\geq \sum_{k=1}^K\Pr\bigl(\hat\pi\in A_k(\hat\pi^*)\bigr)=
\sum_{k=1}^{K}(\hat\pi^*_k)^{K-1}
\geq
K\left(\frac1K\right)^{K-1}
=
K^{2-K}=c_K,
\]
where the second inequality follows from Jensen's inequality.
This lower bound does not depend on either $\pi$ or $\hat\pi^*$.

\subsection{Proof of Lemma~\ref{lem:FOSD-local-construction}}
\label{appendix:FOSD-local-construction}

	Define two perturbation vectors with $K$ entries:
	$$
	p=\left(
	2K,
	\underbrace{1,\ldots,1}_{M-2},
	-2,
	\underbrace{1,\ldots,1}_{K-M-1},
	5-3K
	\right),\qquad
	o=\left(
	-2K,
	\underbrace{-1,\ldots,-1}_{M-1},
	\underbrace{2,\ldots,2}_{K-M-1},
	3M+1
	\right),
	$$
	where empty blocks are omitted. Both vectors sum to zero.
	Since $\pi$ has full support, choose $\varepsilon>0$
	sufficiently small that
	$\hat\pi^P=\pi+\varepsilon p$ and
	$\hat\pi^O=\pi+\varepsilon o$ belong to
	$\Delta_+(\mathcal K)$ and satisfy
	$d(\hat\pi^P,\pi)<\delta$ and
	$d(\hat\pi^O,\pi)<\delta$.
	Such a choice is possible because
	$d(\hat\pi^P,\pi)=(\varepsilon/2)\sum_{k=1}^K|p_k|$ and
	$d(\hat\pi^O,\pi)=(\varepsilon/2)\sum_{k=1}^K|o_k|$.
	
	For every $r=1,\ldots,K-1$, we have
	$\sum_{k=1}^{r}p_k\geq2K-2(r-1)>0$ and
	$\sum_{k=1}^{r}o_k\leq-2K+2(r-1)<0$.
	Therefore,
	$\sum_{k=1}^{r}\hat\pi_k^O
	<\sum_{k=1}^{r}\pi_k
	<\sum_{k=1}^{r}\hat\pi_k^P$
	for every $r=1,\ldots,K-1$, which implies
	$\hat\pi^O\succeq_{\mathrm{FOSD}}\pi
	\succeq_{\mathrm{FOSD}}\hat\pi^P$.
	
	Finally, the state-by-state rankings follow from
	$p_k>0>o_k$ for $k<M$,
	$0>o_M=-1>p_M=-2$,
	$o_k=2>p_k=1>0$ for $M<k<K$, and
	$o_K=3M+1>0>p_K=5-3K$.

\subsection{Proof of Proposition \ref{prop:ordinalimitations}} \label{appendix:ordinalimitations}

It is straightforward to verify that a protocol $F$ satisfying $F_k(u,v)=a_k\mathbf{1}(v>u)$ is an improving ordinal protocol. Conversely, suppose $F$ is improving and ordinal. Fix $k\in\mathcal K$ and let $a_k=F_k(0,1)$, $b_k=F_k(1,0)$, and $c_k=F_k(0,0)$. By definition, $a_k,b_k,c_k\in[0,1]$. Ordinality implies that $F_k(u,v)$ equals $a_k$, $b_k$, or $c_k$, respectively, when $v>u$, $v<u$, or $v=u$. That is, $F_k(u,v)=a_k\mathbf{1}(v>u)+b_k\mathbf{1}(v<u)+c_k\mathbf{1}(v=u)$. Since $F$ is improving, $b_k=c_k=0$. Therefore, $F_k(u,v)=a_k\mathbf{1}(v>u)$. As $F_k$ is assumed to be not identically zero, we have $a_k\in(0,1]$.

\subsection{Proof of Lemma~\ref{lem:additivity-cycle}}\label{appendix:additivity-cycle}

Relabel the contexts so that additivity fails in context 1 at some $u<v<w$ in $U$. Write $a=F_1(v,w)$, $b=F_1(u,v)$, $c=F_1(u,w)$, and $\eta=(a+b-c)/3\neq 0$. Because $F_k$ is not identically zero for any $k$, for each $k\in\{2,3\}$, we can choose $u_k<v_k$ in $U$ such that $q_k=F_k(u_k,v_k)>0$. 

We construct an environment with three rules $\mathcal R=\{R_1,R_2,R_3\}$. Throughout the construction, performance triples are listed in the order $(R_1,R_2,R_3)$, and contributions to the margins $(W_{12},W_{23},W_{31})$ are computed using unnormalized context frequencies $\tilde\pi_k$ (which need not sum to one). Our goal is to make all three unnormalized margins equal to $\eta$.

In context 1, assign performances $(w,v,u)$ and set $\tilde\pi_1=1$. The contribution to the margins is $(a,b,-c)$. The third component is negative because $W_{31}$ measures the advantage of $R_3$ over $R_1$, whereas $R_1$ has the higher performance.

Performances in context 2 are designed to contribute $(\eta-a, 0, a-\eta)$ to the margins. We achieve this by tying $R_2$ and $R_3$ and placing $R_1$ above or below them:
\begin{itemize}
	\item If $\eta>a$, assign performances $(v_2,u_2,u_2)$ and set $\tilde\pi_2=(\eta-a)/q_2>0$. Rule $R_1$ outperforms both other rules while $R_2$ and $R_3$ tie, giving an unweighted contribution of $(q_2,0,-q_2)$ and a weighted contribution of $(\eta-a,0,a-\eta)$.
	\item If $\eta<a$, assign performances $(u_2,v_2,v_2)$ and set $\tilde\pi_2=(a-\eta)/q_2>0$. Rule $R_1$ is outperformed by both other rules, which tie, giving an unweighted contribution of $(-q_2,0,q_2)$ and a weighted contribution of $(\eta-a,0,a-\eta)$.
	\item If $\eta=a$, assign the same performance in $U$ to all three rules and set $\tilde\pi_2=1$, yielding a contribution of $(0,0,0)$.
\end{itemize}

Performances in context 3 are designed to contribute $(0, \eta-b, b-\eta)$ to the margins. The construction is analogous to context 2:
\begin{itemize}
	\item If $\eta>b$, assign performances $(v_3,v_3,u_3)$ and set $\tilde\pi_3=(\eta-b)/q_3>0$. Rules $R_1$ and $R_2$ tie and both outperform $R_3$, giving an unweighted contribution of $(0,q_3,-q_3)$ and a weighted contribution of $(0,\eta-b,b-\eta)$.
	\item If $\eta<b$, assign performances $(u_3,u_3,v_3)$ and set $\tilde\pi_3=(b-\eta)/q_3>0$. Rules $R_1$ and $R_2$ tie and are both outperformed by $R_3$, giving an unweighted contribution of $(0,-q_3,q_3)$ and a weighted contribution of $(0,\eta-b,b-\eta)$.
	\item If $\eta=b$, assign the same performance in $U$ to all three rules and set $\tilde\pi_3=1$, yielding a contribution of $(0,0,0)$.
\end{itemize}

In every remaining context $k\notin\{1,2,3\}$, assign performance $\underline u$ to all three rules and set $\tilde\pi_k=1$. These contexts contribute $(0,0,0)$ to the margins.

Summing the contributions across all contexts gives total unnormalized margins
\[
(a,b,-c)+(\eta-a,0,a-\eta)+(0,\eta-b,b-\eta)=(\eta,\eta,\eta),
\]
where the third component equals $\eta$ because $a+b-c=3\eta$. All performances lie in $U$, and all unnormalized frequencies are strictly positive. Normalizing the frequencies therefore yields $\pi\in\Delta_+(\mathcal K)$ with $W_{12}=W_{23}=W_{31}=\eta/\sum_{k\in\mathcal K}\tilde\pi_k\neq 0$. Since $W$ is antisymmetric, the uniform lottery $p=(1/3,1/3,1/3)$ satisfies $Wp=0$ and is a maximal lottery with full support. Hence, we have $E=\mathcal R$.

\subsection{Proof of Proposition \ref{prop:myopic_comparison_loss}} \label{appendix:myopic_comparison_loss}

Let $F^*$ denote imitate-if-better, so that $F_k^*(u,v)=\mathbf{1}\{v>u\}$ for every context $k$. Fix any environment and any population state $\mu$. Because $\ell$ is weakly increasing, $u_k(R_j)>u_k(R)$ implies $L(R_j\mid k,\mu)\leq L(R\mid k,\mu)$. Moreover, since every $F\in\mathcal{F}$ is improving, $F_k(u,v)=F_k^*(u,v)=0$ whenever $v\leq u$. Hence, for any $F\in\mathcal{F}$, any rule $R$, and any context $k$, we have
$L(F\mid R,k,\mu)-L(F^*\mid R,k,\mu)
= \sum_{j:u_k(R_j)>u_k(R)} \mu(j)\bigl[1-F_k(u_k(R),u_k(R_j))\bigr]\bigl[L(R\mid k,\mu)-L(R_j\mid k,\mu)\bigr] \geq 0.$
Thus, $F^*$ minimizes every agent's expected loss. Summing over all agents and contexts yields $L(F^*\mid\mu)\leq L(F\mid\mu)$ in every environment and at every population state $\mu$.

Conversely, suppose $F\in\mathcal{F}$ differs from $F^*$. Since both protocols are improving, there exist a context $k$ and performances $u<v$ such that $F_k(u,v)<1$. Consider an environment with two rules $R_1,R_2$ satisfying $u_k(R_1)=u$ and $u_k(R_2)=v$, and equal performances in every other context. Choose any full-support context distribution $\pi$, and let $\mu(1)=\mu(2)=1/2$. In context $k$, we have $L(R_1\mid k,\mu)=\ell(v-u)/2$ and $L(R_2\mid k,\mu)=0$, while all losses in every other context are zero. Substituting these into the aggregate loss difference gives
\[
L(F\mid\mu)-L(F^*\mid\mu)=\frac{\pi_k}{8}\bigl[1-F_k(u,v)\bigr]\ell(v-u)>0,
\]
since $\pi_k>0$, $F_k(u,v)<1$, and $\ell(v-u)>0$. Thus, $F$ does not minimize aggregate expected loss in this environment, establishing uniqueness.

\subsection{Omitted Proof in Example~\ref{example:robustperturbation}}
\label{appendix:robustperturbation}

Under this resampling perturbation, when the population state is $\mu_t$, an $R_i$-user ends up comparing with a peer using $R_j$ ($j\neq i$) with probability $\mu_t(j)(1+\mu_t(i))$. The perturbed law of motion under $F$ is given by
\[
\begin{aligned}
	\mu_{t+1}(i)
	={}&\mu_t(i)
	+\frac12\sum_{k=1}^{2}\sum_{j\neq i}
	\mu_t(j)\mu_t(i)(1+\mu_t(j))
	F_k(u_k(R_j),u_k(R_i))\\
	&-\frac12\sum_{k=1}^{2}\sum_{j\neq i}
	\mu_t(i)\mu_t(j)(1+\mu_t(i))
	F_k(u_k(R_i),u_k(R_j)).
\end{aligned}
\]
The first sum captures the inflow of agents switching from $R_j$ to $R_i$, and the second captures the outflow of agents switching from $R_i$ to $R_j$. Since an agent retains her rule whenever both sampled peers use the same rule, $\mu_{t+1}(i)\geq\mu_t(i)^3>0$.
Hence, the population distribution remains interior. We
write $G_t(i)=\mu_{t+1}(i)/\mu_t(i)$ for the population growth rate of using $R_i$. Then, the law of motion becomes
$$G_t(i)-1 = \frac12\sum_{j\neq i}\mu_t(j)\Big[(1+\mu_t(j))\tilde{F}_{ji} - (1+\mu_t(i))\tilde{F}_{ij}\Big],$$
where  $\tilde{F}_{ji} := F_1\big(u_1(R_j), u_1(R_i)\big)+F_2\big(u_2(R_j), u_2(R_i)\big)$ is the sum of switching probabilities from $R_j$ to $R_i$ across both contexts. Under proportional imitation with $\sigma=1/2$, we have 
$$\tilde{F}_{12}=\frac{1}{2}, \tilde{F}_{21}=\frac{1}{2}, \tilde{F}_{13}=\frac{1-\delta}{2}, \tilde{F}_{31}=\frac{1+\delta}{2}, \tilde{F}_{23}=0,\tilde{F}_{32}=\delta.$$
In contrast, under imitate-if-better, 
$$\tilde{F}_{12}=\tilde{F}_{21}=\tilde{F}_{13}=\tilde{F}_{31}=1, \tilde{F}_{32}=2, \tilde{F}_{23}=0.$$

\noindent \emph{Survival under perturbed proportional imitation.}
Substituting the values of $\tilde{F}_{ji}$ for proportional imitation ($\sigma=1/2$) into the growth-rate equation yields
\begin{align*}
	G_t(1)-1
	&=
	\frac12\mu_t(2)\left[\frac{1+\mu_t(2)}{2} -\frac{1+\mu_t(1)}{2} \right]
	+\frac12\mu_t(3)\left[(1+\mu_t(3))\frac{1+\delta}{2} - (1+\mu_t(1))\frac{1-\delta}{2}\right] \\
	&=
	\frac14\Big[\mu_t(2)(\mu_t(2)-\mu_t(1))+\mu_t(3)(\mu_t(3)-\mu_t(1))\Big]
	+\frac{\delta}{4}\mu_t(3)\big[2+\mu_t(1)+\mu_t(3)\big],\\[6pt]
	G_t(3)-1
	&=
	\frac12\mu_t(1)\left[(1+\mu_t(1))\frac{1-\delta}{2} - (1+\mu_t(3))\frac{1+\delta}{2}\right]
	-\frac12\mu_t(2)\big(1+\mu_t(3)\big)\delta \\
	&=
	\frac14\mu_t(1)(\mu_t(1)-\mu_t(3))
	-\frac{\delta}{4}\mu_t(1)\big[2+\mu_t(1)+\mu_t(3)\big]
	-\frac{\delta}{2}\mu_t(2)\big(1+\mu_t(3)\big).
\end{align*}

First, we show that $\mu_t(1)\geq 1/4$ for all sufficiently large $t$. By the Cauchy--Schwarz inequality, $\mu_t(2)^2+\mu_t(3)^2\geq [\mu_t(2)+\mu_t(3)]^2/2=(1-\mu_t(1))^2/2$. Dropping the nonnegative $\delta$-term from the growth equation for $R_1$ gives
\[
G_t(1)-1
\geq
\frac14\Big[\mu_t(2)^2+\mu_t(3)^2-\mu_t(1)\big(1-\mu_t(1)\big)\Big]
\geq
\frac18\big[1-\mu_t(1)\big]\big[1-3\mu_t(1)\big].
\]
Whenever $\mu_t(1)\leq 1/4$, the right-hand side is at least $3/128>0$. Thus, while $\mu_t(1)$ remains below $1/4$, it grows at a rate bounded away from zero and must reach $1/4$ in finite time. Once $\mu_t(1)\geq 1/4$, it cannot fall below $1/4$: if $1/4\leq \mu_t(1)\leq 1/3$, the bound above is nonnegative, so $\mu_{t+1}(1)\geq \mu_t(1)\geq 1/4$; if $\mu_t(1)>1/3$, dropping all positive terms in the equation for $G_t(1)$ gives $G_t(1)\geq 1-\mu_t(1)(1-\mu_t(1))/4\geq 15/16$, whence $\mu_{t+1}(1)\ge 5/16>1/4$. Hence, there exists $T$ such that $\mu_t(1)\geq 1/4$ for all $t\geq T$.

Now we show $R_3$ survives. Since $2+\mu_t(1)+\mu_t(3)\leq 4$ and $1+\mu_t(3)\leq 2$, the two negative terms in the growth equation for $R_3$ sum to at least $-\delta[\mu_t(1)+\mu_t(2)]\geq -\delta$. For any $t\geq T$ with $\mu_t(3)\leq 1/8$, we have $\mu_t(1)\geq 1/4$ and $\mu_t(1)-\mu_t(3)\geq 1/8$, so
\[
G_t(3)-1
\geq
\frac14\mu_t(1)\big(\mu_t(1)-\mu_t(3)\big)-\delta
\geq
\frac{1}{128}-\delta>0.
\]
Thus, for all sufficiently large $t$, $\mu_t(3)$ increases whenever $\mu_t(3)\leq 1/8$. Thus, $\mu_t(3)$ cannot go to zero. That is, $R_3$ survives.

\noindent \emph{Extinction under perturbed imitate-if-better.}
Substituting the values of $\tilde{F}_{ji}$ for imitate-if-better into the growth-rate equation yields
\begin{align*}
	G_t(2)-1
	&=
	\frac12\mu_t(1)\Big[(1+\mu_t(1)) - (1+\mu_t(2))\Big]
	+\frac12\mu_t(3)\Big[2(1+\mu_t(3)) - 0\Big] \\
	&=
	\frac12\mu_t(1)\big(\mu_t(1)-\mu_t(2)\big)
	+\mu_t(3)\big(1+\mu_t(3)\big),\\
	G_t(3)-1
	&=
	\frac12\mu_t(1)\Big[(1+\mu_t(1)) - (1+\mu_t(3))\Big]
	+\frac12\mu_t(2)\Big[0 - 2(1+\mu_t(3))\Big] \\
	&=
	\frac12\mu_t(1)\big(\mu_t(1)-\mu_t(3)\big)
	-\mu_t(2)\big(1+\mu_t(3)\big).
\end{align*}
Subtracting the second equation from the first yields
\begin{align*}
	G_t(2)-G_t(3)
	&=
	\frac12\mu_t(1)\big(\mu_t(3)-\mu_t(2)\big)+\big(\mu_t(2)+\mu_t(3)\big)\big(1+\mu_t(3)\big)\\
	&\geq
	-\frac12\mu_t(1)\mu_t(2)+\mu_t(2)+\mu_t(3)\geq
	\frac12\big(\mu_t(2)+\mu_t(3)\big),
\end{align*}
where the final inequality uses $\mu_t(1)\mu_t(2)\leq \mu_t(2)+\mu_t(3)$. Moreover, dropping the negative term in the expression for $G_t(2)$, we have $G_t(2)\leq 1+\frac12\mu_t(1)^2+\mu_t(3)+\mu_t(3)^2\leq 3,$ as
 $\mu_t(3)\leq 1$ and $\frac12\mu_t(1)^2+\mu_t(3)^2\leq (\mu_t(1)+\mu_t(3))^2\leq 1$.

We conclude by using the ratio argument. Set $r_t=\mu_t(3)/\mu_t(2)$. Note that
\[r_{t+1}
=\frac{\mu_{t+1}(3)}{\mu_{t+1}(2)}
=\frac{\mu_t(3)G_t(3)}{\mu_t(2)G_t(2)}
=r_t\frac{G_t(3)}{G_t(2)}.\]
Together with the bounds above, this gives
\[
r_t-r_{t+1}
=
r_t\frac{G_t(2)-G_t(3)}{G_t(2)}
\geq
\frac16 r_t\big(\mu_t(2)+\mu_t(3)\big)
\geq
\frac16\mu_t(3)>0.
\]
Hence, $r_t\ge 0$ is decreasing. Thus, it converges, implying $r_t-r_{t+1}\to 0$. It further implies $\mu_t(3)\to 0$, showing that $R_3$ goes extinct under perturbed imitate-if-better.

\renewcommand{\refname}{References for the Online Appendix}
\begin{singlespace}
	\putbib[RS]
\end{singlespace}

\end{bibunit}

\end{document}